\documentstyle[12pt]{article}
  \advance\oddsidemargin  by -1in
  \advance\evensidemargin by -1in
\def\lsim{\mathrel{\rlap{\lower4pt\hbox{\hskip1pt$\sim$}}
    \raise1pt\hbox{$<$}}}         
\def\gsim{\mathrel{\rlap{\lower4pt\hbox{\hskip1pt$\sim$}}
    \raise1pt\hbox{$>$}}}         

\def\Pom{{\bf I\!P}}
\def\beq{\begin{equation}}
\def\endeq{\end{equation}}
\def\arr{\begin{eqnarray}}
\def\endarr{\end{eqnarray}}
\makeindex
\begin{document}

\vspace{0.7cm}
\phantom{.}\hspace{10cm}{\sl \large Landau-9-93} \\
\phantom{.}\hspace{9.8cm}{\sl \large 15 February 1993} \\
\begin{center}
{\bf \LARGE Colour transparency:\\
a novel test of QCD in nuclear interactions}
\vspace{0.5cm}\\
{\Large N.N.Nikolaev}
\medskip\\
{\sl
L.D.Landau Institute for Theoretical Physics\\
GSP-1, 117940, ul.Kosygina 2,
V-334 Moscow, Russia\\
{\Large E-mail: kph154@zam001.zam.kfa-juelich.de}\\}
\vspace{0.5cm}
{\bf \large A b s t r a c t \smallskip\\}
\end{center}
Color transparency (CT) is a cute and indispensable property
of QCD as a gauge theory. CT tests of QCD consist of production
of the {\sl perturbative} small-sized hadronic state and
measuring the strength of its {\sl non-perturbative} diffraction
interaction in a nuclear matter. The energy dependence of
the final-state interaction in a nuclear matter probes a
dynamical evolution from the {\sl perturbative} small-sized
state to the full-sized {\sl nonperturbative} hadron. QCD
observables of CT experiments correspond to a novel mechanism
of {\sl scanning of hadronic wave functions} from the large
nonperturbative to the small perturbative size.
In these lectures, which are addressed to experimentalists and
theorists, I discuss the principle ideas of CT physics and
the physics potential of the hadron and electron facilities in
the $\gsim 10\, GeV$ energy range. The special effort was made
to present the material in the pedagogical and self-consistent
way, with an emphasis on the underlying rich quantum-mechanical
interference phenomena. \medskip
\vspace{0.3cm}\\

{ \bf \sl Lectures delivered at:
\begin{enumerate}
\item
XXVII Winter School on Nuclear and Particle Physics
of the St-Petersburg Nuclear Physics Institute,
18-23 February 1993,
Zelenogorsk near St-Petersburg.
\item
XX Winter  School on
Elementary  Particle Physics,
of the Institute for Theoretical and Experimental Physics,
25 February - 2 March 1993,
 Beliye Stolby near Moscow.
 \end{enumerate}
}
{\large To be published in: {\sl Surveys in High Energy Physics}, Harwood
Academic Publishers. }

\newpage
\begin{center}
{\bf \sl \large The contents:}
\end{center}
\begin{enumerate}


\item     
Introduction


\item                  
Exponential attenuation, Gribov's inelastic shadowing
and formation length.


\item                  
Introduction into colour transparency and
diffraction scattering in QCD.


\item                  
Colour transparency, nonexponential attenuation and nuclear
filtering.

\subitem       
4.1. Nonexponential nuclear attenuation of hadrons

\subitem       
4.2. Nuclear filtering of small size Fock states

\subitem       
4.3. Nonexponential attenuation and hadron-nuclei cross sections


\item                  
Why weak nuclear attenuation of hadrons or colour transparency
sum rules

\subitem      
5.1. Quark-hadron duality and colour transparency sum rules

\subitem      
5.2. Filtering small size Fock components in the diffraction matrix

\subitem      
5.3. Spatial expansion of wave packets and the formation (coherence) length


\item                  
Colour transparency, the coherency constraint and the diffraction
operator.

\subitem      
6.1. Overview of Glauber's multiple-scattering theory

\subitem      
6.2. Gribov's inelastic shadowing and the diffraction operator

\subitem      
6.3. Coherency constraint and the effective diffraction operator

\item                 
Nonexponential attenuation of ultrarelativistic positronium
      in den\-se medium.

\item                 
Colour transparency experiments: the candidate reactions.

\item                 
Colour transparency in the diffractive
deep inelastic scattering: Other
men's flowers.

\item                 
Colour transparency and QCD scanning of hadronic wave functions

\subitem     
10.1. Novel feature of colour transparency
experiments: scanning hadronic wave functions

\subitem     
10.2. Scanning and the node effect. Perils of the vector dominance model.

\subitem     
10.3. The two scenarios of scanning the wave function of
light vector mesons: anomalous $Q^{2}$ dependence

\subitem     
10.4. Anomalous $A$-dependence of photoproduction of the $\rho'$-meson
\item                 
Quantum evolution, the coherency constraint and energy dependence
of final state interaction

\subitem     
11.1. Energy dependence: from small to large coherence length

\subitem     
11.2. Spatial expansion of the ejectile: the hadronic basis description

\subitem     
11.3. Spatial expansion of the ejectile:
the path integral in the basis of the quark Fock states

\subitem     
11.4. Interplay of the spatial expansion and variable scanning radius

\item                 
Colour transparency and nuclear shadowing in  deep inelastic
scattering.

\subitem     
12.1. Scaling nuclear shadowing

\subitem     
12.2. Colour transparency and $R=\sigma_{L}/\sigma_{T}$ in the diffractive
deep inelastic scattering

\subitem     
12.3. Nuclear shadowing and fusion of partons

\subitem     
12.4. Colour transparency
and diffraction dissociation into jets on nucleons and nuclei.

\item                 
QCD observables of colour transparency $(e,e'p)$ experiments

\subitem     
13.1. The ejectile state has a transverse size identical to the free
proton size

\subitem     
13.2. Strength of final state interaction: the nonrelativistic model

\subitem     
13.3. The effective size of the ejectile

\subitem     
13.4. Strength of final state interaction: the relativistic model

\subitem     
13.5. Strength of final state interaction: the model estimates

\subitem     
13.6. Filtering the small size ejectile: the Sudakov form factor.

\item                   
Multiple-scattering theory of final state interaction in
$(e,e'p)$ scattering

\subitem    
14.1. Final state interaction and the semiexclusive cross section.

\subitem    
14.2. The Glauber model prediction for nuclear transparency

\subitem    
14.3. Gribov's inelastic shadowing and nuclear transparency in
$(e,e'p)$ scattering

\item                   
The coherency constraint, the observable $\Sigma_{ep}$ and
the signal of co\-lo\-ur trans\-pa\-ren\-cy  in $(e,e'p)$
scattering

\subitem    
15.1.The coherency condition and the effective ejectile state

\subitem    
15.2. Final state interaction at asymptotic $Q^{2}$ and the
observable $\Sigma_{ep}$

\subitem    
15.3. The observable $\Sigma_{ep}$ and the onset of colour transparency
at moderate $Q^{2}$

\subitem    
15.4. Realistic estimate of the signal of colour transparency

\subitem    
15.5. Conspiration of the two sates does not produce colour transparency

\item                   
QCD motivated diffraction operator

\subitem    
16.1. Overview of diffraction dissociation of protons

\subitem    
16.2. QCD diffraction operator in the quark basis

\subitem    
16.3. The hybrid pion-nucleon model: size of the $3q$-core.

\subitem    
16.4. How stripping of pions off nucleons contributes to the signal
of colour transparency

\item                   
Predictions for nuclear transparency $Tr_{A}$ in $(e,e'p)$ scattering

\subitem    
17.1. It is not easy to shrink the proton.

\subitem    
17.2. Quantitative predictions for the transmission $(e,e'p)$ experiment

\subitem    
17.3. Electroproduction of resonances and continuum states

\subitem    
17.4. Scaling properties of the nuclear transparency in
$(e,e'p)$ scattering

\subitem    
17.5. Scaling
properties of the nuclear transparency in virtual photoproduction
of vector mesons

\subitem    
17.6. Testing QCD diffraction operator against the data on the
diffraction dissociation of protons

\item                    
Effects of the Fermi motion

\subitem    
18.1. The Fermi-motion bias in nuclear transparency

\subitem    
18.2. Rescattering broadening of angular distribution

\item                    
Search for colour transparency in the quasielastic $(p,p'p)$
scattering on nuclei

\item                    
Colour transparency and colour opacity in diffractive
hadron-nucleus scattering.

\subitem   
20.1. Anomalous nuclear attenuation in diffraction dissociation of
hadrons on nuclei

\subitem    
20.2. Signal of colour transparency in the charge-exchange scattering on
nuclei

\subitem   
20.3. Colour transparency and diffractive production of jets on nuclei

\item                    
Beyond the dilute-gas approximation: correlation effect
in the nuclear transparency

\subitem    
21.1. Correlations and Glauber's optical potential.

\subitem    
21.2. The hole and spectator effects in the quasielastic
$(e,e'p)$ scattering.

\item                    
Use and misuse of the multiple scattering theory.

\item
Conclusions

\end{enumerate}
\newpage

\section{Introduction.}

The subject of Colour Transparency (CT), which originated from
the seminal papers by Zamolodchikov, Kopeliovich and Lapidus [1],
Bertsch, Brodsky, Goldhaber and Gunion [2], Mueller [3] and
Brodsky [4], is approaching its maturity. CT
 is an indispensable
property of QCD as the gauge theory, and hadronic and
electronuclear facilities
of the next generation could well become dedicated to the study of
CT.

CT predicts that in the hard-scattering
exclusive reactions on nuclear targets the produced hadrons
may escape a nucleus without being absorbed, so that the
so-called nuclear transmission coefficient or the nuclear
transparency
\beq
Tr_{A}= {d\sigma_{A} \over Ad\sigma_{N}} \approx 1 \, ,
\label{eq:1.1}
\endeq
although with the typical hadronic cross section
a thickness of nuclei is of the order of few absorption
lengths. CT phenomenon has the two facets:
\begin{itemize}
\item
Hard scattering selects small-sized components of interacting
hadrons and is tractable in the {\sl perturbative} QCD. The
size $\rho_{Q}$ of the {\sl ejectile state} emerging from the hard
scattering process is controlled by the reaction kinematics
and dynamics.

\item
A strength of the {\sl final-state interaction} (FSI) of the
small size ejectile is characterized by the {\sl size-dependent
interaction cross section} $\sigma(\rho_{Q})$. This cross
section is measured by nuclear attenuation produced by the
{\sl nonperturbative} forward diffractive scattering of
the ejectile in the nuclear matter.
\end{itemize}

QCD as a theory
of strong interactions gives a strong prediction that
a strength of this diffractive scattering vanishes as
$\rho_{Q}\rightarrow 0$ and gives rise to a rather
nontrivial CT sum rules [5-8], which relate the {\sl
perturbative
and nonperturbative QCD observables}.

CT physics is the nuclear QCD physics in which intranuclear
interactions do emerge not as an undue complication, but
as an analyzer of CT effect. CT physics is a
novel physics of
QCD diffractive scattering. Concerning the diffractive
scattering aspect of CT, the second pillar of CT is the
Glauber-Gribov multiple-scattering theory, in particular
Gribov's theory of the inelastic shadowing [9] and its extensions
in 70's, found mostly in the Russian literature [10-17].
Gribov's theory complemented by the quark-hadron duality
becomes a powerful tool for analyzing the onset of CT
regime. Manifestations of the quark-hadron duality
in CT processes are unparalelled in other hadronic or
leptonic interactions.

Besides the colour gauge invariance and the Glauber-Gribov
multiple scattering theory, another pillar of CT is
the relativistic rise of the formation (coherence) length
$l_{f}$ in high energy diffractive interactions, introduced
in 1953 by Landau, Pomeranchuck and Feinberg [18]. It is
the formation length which controls the quantum evolution
of the perturbative small size ejectile into the
full size nonperturbative final-state hadrons.

{}From the classical point of view, CT is precisely
the QCD counterpart of the QED Chudakov
effect: weak ionization of medium by the Bethe-Heitler produced
ultrarelativistic $e^{+}e^{-}$ pairs at short distances from
the production vertex (1949, published in 1955 [19]). When
(anti)quarks of the colourless hadronic state sit on top of
each other, co\-lour\-ed  gluons decouple from such a
quasineutral state, resulting in a small interaction cross
section [1,2] by virtue of the colour gauge invariance.
This is a useful heart of the CT physics.

There are important differences between the Chudakov effect and CT
in their quantum aspects. The latter is very rich in the
nontrivial observable phenomena, which will be a major topic in
these lectures. The lectures are based on a review of traditional
sources [9-32], selected original works on
CT [1-4,33-37] and recent work by V.Barone, O.Benhar,
B.Jennings, M.Genovese,
B.Z.Kopeliovich, C.Mariotti, J.Nemchik, E.Predazzi,
A.Szczurek, J.Speth, B.G.Zakharov, V.Zoller and myself
[5-8,38-51].
The subject of CT is being developed for more than one decade,
but the consistent quantum theory of CT phenomena has emerged
only very recently. The purpose of these lectures is to
present the self-contained overview of CT physics. The
major emphasis will be on which aspects of QCD are tested
in CT experiments.

The onset of CT and principle effects of scanning the hadronic
wave functions take place in the 10-50 GeV energy range. CT
is a physics of the new phenomena on the boundary
between the perturbative and nonperturbative QCD, and is a
good case for the high-luminosity hadron and electron
facilities in the 10-50 GeV energy range. The unique potential of
the electron facilities is particularly noteworthy. CT tests
of QCD can not be performed at other facilities like LEP, HERA,
LHC and SSC, which can be classified as the perturbative QCD
facilities.

The contents of these lectures can be grouped into few
larger topics:
\begin{itemize}
\item
How is it possible that hadrons which have large free-nucleon
cross section might have weak intranuclear interaction?
\item
How QCD is tested and what are QCD observables in CT
experiments?
\item
How the nonperturbative wave function of hadrons is scanned
in CT experiments?
\item
What aspects of the quark-hadron duality are tested in CT
experiments?
\end{itemize}

For the early reviews on the subject see [37,7,8].


\section{Exponential attenuation, Gribov's in\-elas\-tic sha\-dow\-ing
and formation length.}

The interaction strength is measured by the total cross section,
which is determined via the exponential attenuation
of a beam in the absorber of
thickness $d$,
\beq
N(d)=N(0)\exp\left(-{d/l}\right)\, ,
\label{eq:2.1}
\endeq
where  $l$ is the mean free path $l= 1/ n\sigma_{tot}$ and
$n$ is the density of scatterers.
Textbooks  seldom mention that (\ref{eq:2.1}) is derived under
the implicit assumption of {\sl uncorrelated consecutive
interactions}. Rescattering of  atomic and nuclear particles
is followed by their excitation. The consequtive interactions are
uncorrelated only provided the Lorentz-dilated
relaxation (de-exciatiton) time
$\tau_{R} \sim R/v \sim 1/\Delta m$  is short, i.e.,
the so-called formation length
\beq
l_{f} =\gamma c\tau_{R} = \gamma {1 \over \Delta m} \ll l
\label{eq:2.2}
\endeq
Here $R$ is a size of the projectile, $v$ is a typical
velocity of its constituents and $\Delta m$ is a typical
excitation energy.
At high-energy $\nu$ the Lorentz factor
$\gamma =\nu/m >>1$
and one easily gets $l_{f}$ exceeding the size of nuclei $R_{A}$.

This observation, and
introduction of the formation (coherence) length $l_{f}$ into
the particle
physics, is due to Landau, Pomeranchuck and Feinberg [18],
who in 1953-54 have discovered a novel mechanism of the coherent
diffractive production on nuclei at $l_{f} \gg R_{A}$.
Pomeranchuck's paper on the high energy photoproduction on nuclei
is particularly noteworthy as the precursor of the vector
dominace model (VDM) and as
the first demonstration of the nonexponential attenuation:
despite the weak interaction with free nucleons, photoproduction
on nuclei has diffractive properties typical of the strongly
absorbed hadrons.

The novel physics, which emerges at $l_{f} \gg l$, is the so-called
Gribov's inelastic shadowing [9] or regeneration. In a collision
with an isolated atom or a free nucleon the
excitation of the projectile
is an inelastic scattering and contributes to absorption. In a
dense medium  (nucleus) the consequtive rescatterings result in
regeneration of the projectile from the excited states.
In high-energy interactions of hadrons diffraction excitation
of the projectile is a quasi-two-body reaction with the
(approximately) constant cross section, very much alike the elastic
scattering. For this reason the regeneration effects persist
at high energy.
In the elastic scattering this regeneration reduces the overall
nuclear attenuation and the particle-nucleus cross section,
making nuclei less opaque. In transitions from the ground
state of the projectile to the excited states regeneration may
either enhance or diminish the overall attenuation. In certain
cases the rate of production on nuclei is enhanced
({\sl antishadowed})
rather than attenuated ({\sl shadowed}).
The statement that Gribov's inelastic shadowing
may {\sl completely wipe out nuclear attenuation} is a very nontrivial
one. In QCD it is a consequence of the fundamental property of
colour gauge invariance, as it will be explained below.

      Technically, a description of Gribov's
inelastic shadowing effects is very much similar to description
of propagation of the elliptically polarized photons in the
optically active medium, of the regeneration of the short-lived
$K_{S}$ and attenuation of the long-lived $K_{L}$ mesons
in an  absorber, or of the neutrino oscillations in a medium.
These are basically the two-channel problems, the inelastic
shadowing in nuclear interactions is a many-channel problem.
The unifying feature is that in all the cases the off-diagonal
transitions  between the states having unequal free-nucleon
cross section lead to a departure from the simple exponential
attenuation law (\ref{eq:2.1}).

Physics of CT is, in fact, a {\sl physics of the nonexponential
attenuation.} The issue at stake is why and when attenuation
would be anomalously strong or anomalously weak, how does the
weak attenuation of hadrons in a nuclear matter emerge
and how it can be quantified. Hence I proceed first with
introduction into diffraction scattering in QCD.


\section{Introduction into colour transparency and
dif\-frac\-ti\-on scat\-ter\-ing in QCD}

Hadrons are made of the constituent quarks. The high energy
interactions are followed by transition of the projectile
into the higher orbital or angular excitations. These transitions
between the mass-eigenstates are similar
to rotation of the polarization plane in an
optically active medium.
The condition $l_{f} \gg R_{N}$ corresponds to freezing of the
intrinsic
motion of quarks in colliding hadrons, so that the transverse
separation of quarks $\vec{\rho}$ is conserved in the scattering
process. Fixed-$\vec{\rho}$ states are counterparts of the circular
polarization states of the photon.

Conservation of $\vec{\rho}$ in the high-energy limit
allows casting the $aN$ scattering cross section
in the quantum mechanical form (here for the sake of simplicity
$a$ is a $q\bar{q}$ meson) [1,2,39,44]:
\beq
\sigma_{tot}(aN)=\langle a| \sigma(\rho)|a \rangle =
\int dz\, d^{2}\vec{\rho}\sigma(\rho)
|\Psi_{a}(z,\vec{\rho})|^{2}                     \, ,
\label{eq:3.1}
\endeq
Here $\sigma(\rho)$ is the total cross section for a $q\bar{q}$
pair of the transverse size $\vec{\rho}$. To the lowest
order in perturbative QCD, in the Low-Nussinov two-gluon
approximation [24,25], fig.1, one finds:
\beq
\sigma(\rho)=
{ 16\alpha_{S}(1/\rho) \over 9 }
\int { d^{2}\vec{\kappa}
V(\vec{\kappa})[1-\exp(-i\vec{\kappa}\cdot \vec{\rho})]
\over
(\vec{\kappa}\,^{2} + \mu_{G}^{2})^{2} }  \alpha_{S}(\kappa)   \, .
\label{eq:3.2}
\endeq
where $\vec{\kappa}$ is a momentum of the exchanged gluons and
for the nucleon target
\beq
V(\vec{\kappa})=
1-\left<N|\exp[\vec{\kappa}(\vec{r}_{1}-\vec{r}_{2})]
|N\right> = 1 - G_{em}(3\vec{\kappa}^{2})
  \,  .
\label{eq:3.3}
\endeq

In (\ref{eq:3.2}) the effective mass of gluons
$\mu_{G}$ is included not to have the colour
forces propagating beyond the
confinement radius $R_{c}=1/\mu_{G} \sim 1/\mu_{\pi}$.
However, since by virtue of the colour gauge
invariance the both  gluon-nucleon vertex function
$V(\vec{\kappa})$ and gluon-$q\bar{q}$ vertex
$\left[ 1-\exp(i\vec{\kappa}\vec{\rho})\right]$
vanish at $\kappa^{2}\rightarrow 0$ ( soft gluons with
the wavelength $\lambda_{g} \gsim R_{a},\rho$ decouple
from the colourless hadrons), the above cross section
is infrared convergent even at $\mu_{G} \rightarrow 0$.
At small $\rho$ the virtuality of the exchanged gluons is
large, $\kappa^{2} \sim 1/\rho^{2}$, and
for the charmonium or bottonium scattering
the Low-Nussinov
approximation is quantitaively good.

The quantum mechanical representation
(\ref{eq:3.1}) has a few salient features:
\begin{enumerate}
\item  It emphasizes the transverse separation of quarks $\vec{\rho}$
is a diagonal parameter in the scattering matrix.
\item The cross section (\ref{eq:3.2}) only
depends on the target and is universal for all projectiles.
\item At $\rho << R_{h}\sim R_{c}$ one can expand
the factor $[1-\exp(i\vec{\kappa}\cdot \vec{\rho})]$ in (\ref{eq:3.2})
and
\beq
\sigma(\rho) \propto \rho^{2}\alpha_{S}(1/\rho)
\log\left[1/\alpha_{S}(1/\rho)\right] \propto \rho^{2}
\label{eq:3.4}
\endeq
\item In the opposite limit of
$\rho \gg R_{c}$ the cross section (\ref{eq:3.2})
saturates, $\sigma(\rho) \sim R_{c}^{2}$,
since the colour forces can not propagate beyond the confinement radius
$R_{c}$.
\end{enumerate}
Of the above, CT property (\ref{eq:3.4}) is the most important one.
The way is was derived above makes clear that
 it is a direct consequence of
color gauge invariance and of the confinement, i.e., the
colour neutrality of hadrons.
Formula (\ref{eq:3.2}) for the cross section $\sigma(\rho)$
and eq.(\ref{eq:3.1})
quantify the property of CT.
How $\sigma(\rho)$ behaves in the nonperturbative domain
$\rho \gsim R_{c}$, is a matter of
educated guess. I only emphasize that
the behaviour of $\sigma(\rho)$,
shown in Fig.2 for the nucleon
target, is obviously protected against the
higher order QCD corrections.
The representation (\ref{eq:3.1}) has a broader region of
applicability than the Low-Nussinov approximation it was derived
from. It can easily be extended to the higher order ($q\bar{q}g,....$)
Fock states of interacting hadrons.

The universality of $\sigma(\rho)$ must be emphasized [39,44]:
Different projectiles probe the same universal $\sigma(\rho)$
at different size $\rho$, depending on the wave function
$\Psi_{a}(\rho)$: the $\Upsilon N$ scattering
would have probed $\sigma(\rho)$ at $\rho \sim r_{\Upsilon}$,
the $(J/\Psi)N$ and $\Psi'N$ scattering probe
$\sigma(\rho)$ at $\rho \sim R_{J/\Psi},R_{\Psi'}$ with the result
$\sigma_{tot}((J/\Psi)N) \approx 5 mb$ and
$\sigma_{tot}(\Psi'N) \approx 12 mb$ [42]. The $\pi N$ total
cross section receives its leading contribution from
$\rho \sim 0.7-1\,fm$. With realistic pion wave function
one finds $\sigma_{tot}(\pi N) \approx 25 mb$ [25,39].

     I shall comment more on implications of
CT property (\ref{eq:3.1}) for the hadronic interactions
in section 5. Now I shall proceed with derivation of one
of the most spectacular consequences of CT: the nonexponential
attenuation of hadrons in nuclear medium.


\section{Color transparency, nonexponential nuclear attenuation
and nuclear filtering [{\sl Ref.1}].}


\subsection{Nonexponential nuclear attenuation of hadrons.}

Consider now the charmonium-nucleus scattering in the high
energy limit of $l_{f} \gg R_{A}$. Since $\vec{\rho}$ is
conserved in the scattering process,  for a $q\bar{q}$
pair of size $\vec{\rho}$ the nuclear cross section is given
by the standard Glauber's formula [20] (see also below, section 7)
\beq
\sigma_{A}(\rho)=2\int d^{2}\vec{b}\left\{
1-\exp\left[-{1 \over 2} \sigma(\rho)T(\vec{b})\right]\right\}    \, .
\label{eq:4.1.1}
\endeq
Here $T(\vec{b})$ is the optical thickness of a nucleus,
$T(\vec{b})=\int_{-\infty}^{+\infty}dz n_{A}(z,\vec{b})$,
so that
\beq
\exp\left[-\sigma(\rho)T(\vec{b})/2 \right]=\exp(-d/2l)
\label{eq:4.1.2}
\endeq
is precisely the {\sl exponential attenuation amplitude} of the
fixed-$\vec{\rho}$ state.

Since the size of the charmonium is small, one can write
down
\beq
\sigma(\rho)= {\rho^{2} \over R_{J/\Psi}^{2}}
\sigma_{tot}(J/\Psi N)
\label{eq:4.1.3}
\endeq
Applying eq.(\ref{eq:3.1}) with the Gaussian wave function
$\langle \rho|J/\Psi\rangle\propto \exp[-\rho^{2}/2R_{J/\Psi}^{2}]$,
I find a dramatic
change from the {\sl exponential} to the
{\sl inverse thickness} attenuation law [1]:
\beq
\langle J/\Psi|\exp\left[-{1 \over 2} \sigma(\rho)T(\vec{b})\right]
|J/\Psi\rangle =
{1 \over 1 +{1 \over 2} \sigma_{tot}(J/\Psi N)T(\vec{b})}    \, .
\label{eq:4.1.4}
\endeq


\subsection{Nuclear filtering of small size Fock states}

I emphasize that for the heavy nuclei one finds the $1/T(b)$ law
irrespective of the detailed form of the wave function $|J/\Psi\rangle$.
Indeed, the attenuation can be neglected for the Fock components with
\beq
\rho^{2} \lsim R_{J/\Psi}^{2}
{2 \over \sigma_{tot}(J/\Psi N)T(b)}
\label{eq:4.2.1}
\endeq
and the law (\ref{eq:4.1.4}) is recovered for the sufficiently
thick nuclei for {\sl all mesons}, albeit with different
absolute normalization (for baryons $B$ one finds the
$[\sigma_{tot}(BN)T(b)]^{-2}$ law).

Eq.(\ref{eq:4.2.1}) demonstrates, how the sensitivity to the
small-sized Fock component of
the wave function, eventually reaching the {\sl perturbative}
region of $\rho$, is developed in a typical {\sl nonperturbative}
soft diffraction scattering, and this sensistivity has its origin
in the CT law (\ref{eq:3.4}), (\ref{eq:4.1.3}).
This property of the {\sl nuclear
filtering} is one of the principal phenomena in CT physics.


\subsection{Nonexponential attenuation and hadron-nuclei cross
sec\-tions [{\sl Refs.10,11,14,17}]}

The nonexponential attenuation law by Zamolodchikov et al.
[1] is a very cute effect, but difficult to test experimentally.
In the case of nucleons, pions or kaons the dominant
contribution to the attenuation factor comes from
 $\sigma(\rho)$ close to the saturation regime, fig.2.
 In this case it
is more convenient to describe a departure from the exponential
attenuation in terms of moments
$\langle h | \sigma(\rho)^{n} |h \rangle $ ([14-16], for
evaluation of these moments see [16,17]):
\arr
\langle h     |\exp\left[-{1 \over 2} \sigma(\rho)T(\vec{b})\right]
|h     \rangle = ~~~~~~~~~~~~~~~~~~~~~~~~~~\nonumber \\
\exp\left[-{1 \over 2}
\langle h | \sigma(\rho) |h \rangle T(\vec{b})\right]
\left\{1+ {1 \over 8}
\left[\langle h | \sigma(\rho)^{2} |h \rangle -
\langle h | \sigma(\rho) |h \rangle ^{2}\right] T(\vec{b})^{2}+
.... \right\}
\label{eq:4.3.1}
\endarr
Gribov's inelastic shadowing correction in the curly brackets
in (\ref{eq:4.3.1}) is numerically very significant
and brings the theory to a much better agreement with the experiment
[22,23], Fig.3, (for the review see [15,17]). For the more
detailed discussion of Gribov's inelastic
shadowing as a pillar of CT physics see below.


\section{Why  weak nuclear attenuation of hadrons or colour
transparency sum rules [{\sl Refs.5-8}]}


\subsection{Quark-hadron duality and colour transparency sum rules}

The above QCD formalism is deceptively simple and the origin
of CT could easily be overlooked. The principal issue is that
what propagates in a nuclear matter is a superposition of the
mass-eigenstates, i.e., physical hadrons, all having large
cross sections. Where have these large cross sections gone?

In order to set a frame of reference, let us consider
the quantum-mechanical $\bar{q}q$ system with the
$\vec{\rho}$ being a size in the
plane normal to ejectile's momentum, $\vec{r}=(\vec{\rho},z)$.
The ejectile can be described in terms of the
two complete sets of wave functions: the mass eigenstates
$|i\rangle$ and the fixed-$\vec{r}$ states
$|\vec{r}\rangle$. The {\sl quark-hadron duality} states that
the {\sl hadronic basis} and the {\sl quark-gluon-Fock-state
basis} descriptions of the same process are complementary to
each other and must yield the identical results. CT property
(\ref{eq:3.4}) is readily formulated in terms of the quark
Fock states. It is interesting to see what constraint on the
hadronic interactions  is imposed by this CT property.

The mass-eigenstates $|i\rangle$ can
be expanded in the basis $|\vec{r}\rangle$ yielding
the stationary wave function
\beq
\Psi_{i}(\vec{r})=\left<\vec{r}|i\right> \, .
\label{eq:5.1.1}
\endeq
Vice versa, the $\vec{r}$-eigenstates expand as
\beq
\left.|\vec{r}\right>=
\sum_{i}\Psi_{i}(\vec{r})^{*}\left.|i\right>     \, .
\label{eq:5.1.2}
\endeq
Let $\hat{\sigma}$ be the cross
section or diffraction operator. In the $\vec{r}$-representation
$\hat{\sigma}$ is a diagonal operator:
 $\hat{\sigma} = \sigma(\rho)$.
In the mass-eigenstate basis the diagonal matrix elements give
the total cross section:
$\sigma_{tot}(iN)=\sigma_{ii}= \left<i|\hat{\sigma}|i\right>$ and
the differential cross section of the forward elastic scattering at
$t=0$ equals
$d\sigma_{el}(iN)/dt\bigr\vert _{t=0}=\sigma_{ii}^{2}/16\pi$.
The off-diagonal matrix elements
$\sigma_{ik}= \left<i|\hat{\sigma}|k\right>$
describe  the differential cross section of the forward diffraction
excitation $k\, N \rightarrow i\, N$:
\beq
\left.{d \sigma_{D}(k N \rightarrow i N) \over dt}\right\vert _{t=0}=
{\sigma_{ik}^{2} \over 16\pi}                            \, .
\label{eq:5.1.3}
\endeq
The mass-eigenstate  expansion for $\sigma(\rho)$ reads  as
\beq
\sigma(\rho)=\langle \vec{r}|\hat{\sigma}|\vec{r}\rangle =
\sum_{i,k}
\left<\vec{r}|k\right>
\left<k|\hat{\sigma}|i\right>\left<i|\vec{r}\right>=
\sum_{i,k} \Psi^{*}_{i}(\vec{r})\Psi_{k}(\vec{r})
\sigma_{ki}
\label{eq:5.1.4}
\endeq
and relates the non-perturbative diffraction amplitudes $\sigma_{ik}$
with $\sigma(\rho)$ which is perturbatively calculable at small
$\rho$. The special case of $\vec{r}=0$ yields the relationship
between the wave functions at the origin $\Psi_{i}(0)$
and the diffraction transition amplitudes
$\sigma_{ik}$, which may be called the "CT sum rule" [5-8]:
\beq
\sum_{i,k} \Psi_{k}(0)^{*}\Psi_{i}(0)\sigma_{ki} = 0      \, .
\label{eq:5.1.5}
\endeq

The CT sum rule is a very nontrivial one.
Experimentally, the diffraction excitation rate is much smaller
than the elastic scattering rate,
$\sigma_{ik} \ll \sigma_{ii},\sigma_{kk}$ (see below), so that one
might be tempted to neglect the off-diagonal terms
$\propto \sigma_{ik}$ in (\ref{eq:5.1.4}.\ref{eq:5.1.5}).
In doing so one runs into contradiction with CT as formulated
in eq.(\ref{eq:3.4}):
\beq
\sigma(\rho) \approx \sum_{i}
\Psi^{*}_{i}(\vec{r})\Psi_{i}(\vec{r}) \sigma_{tot}(i\,N)\sim
\sigma_{tot}(NN)\sum_{i} |\Psi_{i}(\vec{r})|^{2}=
\sigma_{tot}(NN)  \,\,  ,
\label{eq:5.1.6}
\endeq
with the $l.h.s$ expected to vanish at $\rho \rightarrow 0$,
whereas the $r.h.s$ does not depend on $\rho$.
This makes it obvious that in the mass-eigenstate basis
CT has its origin in the strong cancellations between the diagonal
and off-diagonal diffractive transitions [5-8].
Considering the matrix elements $\langle \vec{r}|\hat{\sigma}|k\rangle$
at $\vec{r} \rightarrow 0$, one readily finds a whole family of CT
sum rules [6], which generalize the sum rule (\ref{eq:5.1.5}) and which
will be of much use in our discussion of FSI in $(e,e'p)$ scattering:
\beq
\sum_{i} \Psi_{i}(0)\sigma_{ik} = 0      \, .
\label{eq:5.1.7}
\endeq
When considered on the basis of a
finite number $N_{eff}$ of states
$|i\rangle, |k\rangle$, the sum rules (\ref{eq:5.1.5}), (\ref{eq:5.1.7})
are evidently better saturated the larger is a number of
contributing states. Similar sum rules, which will be derivatives
of the family (\ref{eq:5.1.7}), can be written down
for matrix elements of $\hat{\sigma}^{n}$ or functions
$F(\hat{\sigma})$ which start from $\hat{\sigma}^{n}$ with $n \geq 1$.


\subsection{Filtering small
size Fock components in the diffraction
matrix}

Since
the proton $p$ and its excitations $p^{*}$ have a large size
$\gsim R_{p}$, the unification of the nonperturbative and
perturbative aspects of QCD in the CT sum rule might look
surprizing. However,
inspect more carefully the forward
diffraction transition amplitude
\beq
\sigma_{ik}=\langle i|\hat{\sigma}|k\rangle =
\int d^{3}\vec{r} \Psi_{i}^{*}(\vec{r})\Psi_{k}(\vec{r})
\sigma(\rho)  \, .
\label{eq:5.2.1}
\endeq
By the orthogonality of wave functions, it
would have vanished if $\sigma(\rho) =const$. Because of
the saturation of $\sigma(\rho)$ at large $\rho$, the matrix
element (\ref{eq:5.2.1}) is dominated by $\rho \lsim R_{p}$.
Moreover, the heavier the state $|i\rangle$, the more nodes has
$\Psi_{i}(\vec{r})$, and the smaller becomes the region of
$\rho$ the nonvanishing contribution to the matrix element
(\ref{eq:5.2.1}) comes from. As we shall see later on, heavy
states do indeed play an important role in the CT phenomenon.

Eqs.(\ref{eq:5.1.4},\ref{eq:5.1.5}) can be given a still more
practical formulation: since $\sigma(\rho)$  is
an eigenvalue  of the cross section operator $\hat{\sigma}$,
CT sum rule as a property of QCD simply states that on the basis
of hadronic states $\hat{\sigma}$  has a {\sl vanishing eigenvalue.}
Incidentally, a possibility of vanishing cross section eigenvalues
has been discussed for quite a time from a very different
perspective ([52,53], for the review see [15]).


\subsection{Spatial expansion of wave packets and the forma\-ti\-on
(co\-he\-ren\-ce) length}

Let the small-size state $|\rho\rangle$ be produced in the hadronic
interaction. The proton, as the ground state of the mass operator,
has the smallest size of all the nucleonic states. In order to form
the wave packet $|\rho\rangle$ of the size $\rho \ll R_{p}$,
where $R_{p}$ is the proton radius, one has to mix the proton with the
higher excitations which have an  ever growing size.
By virtue of the unceratinty relation, the transverse momentum
of quarks in the state of size $\rho$ is $k_{\perp}\sim 1/\rho$,
and the mass of states with this transverse momentum can
be estimated as $m \sim 4k_{\perp} \sim 4/\rho$.

The time evolution of the nonstationary wave packet  (\ref{eq:5.1.2})
is given by
\beq
\left.|\vec{r},t\right>=
\sum_{i}\Psi_{i}(\vec{r})^{*}\left.|i\right> \exp(-im_{i}t)
\label{eq:5.3.1}
\endeq
and
\beq
\left<\vec{r},t|\hat{\sigma}|\vec{r},t\right> = \sum_{i,k}
\left<\vec{r}|k\right>\left<k|\hat{\sigma}|i\right>
\left<i|\vec{r}\right>
\exp[i(m_{k}-m_{i})t]  \, .
\label{eq:5.3.2}
\endeq
The proper time $t$ is related to the distance $z$ from the
production vertex as  $t=z/\gamma$. As soon as large phases
$(m_{k}-m_{i})z/\gamma \sim z/l_{f} \gsim 1$  emerge  in
the phase factors of (\ref{eq:5.3.2}),
they will destroy the delicate cancellations neccessary
for CT. In a propagation of the wave packet in a nucleus
$z \sim R_{A}$. Then, a coherence of the components $|i\rangle$
and $|k\rangle$ of the wave packet requires that
\beq
{\gamma  \over |m_{k}-m_{i}|} > R_{A}  \,  .
\label{eq:5.3.3}
\endeq
The higher the excitation energy, the faster the corresponding
admixture to the expanding wave packet
$|\vec{\rho},t\rangle$ becomes out of phase,
so that only a finite number of excited
states $N_{eff}$ can interfere coherently at finite energy [5].
The diffraction operator $\hat{\sigma}$
which should now be considered on the truncated set of $N_{eff}$
coherently interfering states, will no longer have the vanishing
eigenvalue. We see, that the formation length $l_{f}$ controls the
coherence and expansion properties of the small-sized states.
The crucial issue is how rapidly CT sum rules
(\ref{eq:5.1.5}),\,(\ref{eq:5.1.7}) are
saturated by the lowest lying excitations
of nucleons?

Intranuclear absorption of the ejectile sate is described by
diffractive interactions of the ejectile wave packet. In order
to proceed further we need fundamentals of the Glauber-Gribov
theory of diffractive scattering.


\section{Colour transparency, the coherency con\-straint and the
dif\-fraction operator}


\subsection{Overview of Glauber's multiple scattering theory
[{\sl Ref.20}]}

We are interested in the intranuclear diffraction interactions
at energy $\nu \gsim 10 GeV$. The appropriate technique
is the Glauber-Gribov multiple-scattering theory (MST), which
to a certain extent we
have already used in section 4.

In order to set up the formalism,
let us start with the proton-nucleus total cross section.
The starting point
is the Glauber's relationship between the nuclear
and free-nucleon profile functions [20]
\arr
\Gamma_{A}(\vec{b},\vec{c}_{A},...,\vec{c}_{1})=
1- \prod_{i=1}^{A} \left[1-\Gamma(\vec{b}-\vec{c}_{i})\right]
{}~~~~~~~~~~~~~~~\nonumber\\
= \sum_{i}^{A}\Gamma(\vec{b}-\vec{c}_{i})
-\sum_{k>i}^{A}\Gamma(\vec{b}-\vec{c}_{k})\Gamma(\vec{b}-\vec{c}_{i})+...+
(-1)^{A-1}\prod_{i=1}^{A}\Gamma(\vec{b}-\vec{c}_{i})
\label{eq:6.1.1}
\endarr
Here $\vec{b}$ is the impact parameter,
$\vec{r}_{i}=(\vec{c}_{i},z_{i})$ are
positions of nucleons in the target nucleus, and the profile functions
are normalized as
\beq
f(\vec{q})=2i\int d^{2}\vec{b}\,\Gamma(\vec{b})
\exp(-i\vec{q}\,\vec{b})
\label{eq:6.1.2}
\endeq
\beq
\sigma_{tot}(hN)=Im f(0)=2Re\int d^{2}\vec{b}\,\Gamma(b)
\label{eq:6.1.3}
\endeq
Experimentally, to a good approximation,
\beq
f(\vec{q})=i\sigma_{tot}(pN)
\exp\left(-{1\over 2}B\vec{q}\,^{2}\right)  \,  ,
\label{eq:6.1.4}
\endeq
so that
\beq
\Gamma(\vec{b})={\sigma_{tot}(pN) \over 4\pi B}
\exp\left(-{b^{2} \over 2B}\right)
\label{eq:6.1.5}
\endeq
At energies above $10 GeV$ the proton-nucleon diffraction slope
$B \approx 0.5\, f^{2}$ and only slowly rises with energy [54].

For the sake of simplicity, in the sequel I neglect a small real part
of the high-energy scattering amplitude, so that $\Gamma(\vec{b})$ is
real; corrections for finite $Re f(0)/Im f(0)$ can easily be included.

The nuclear matrix element
\beq
\Gamma_{A}(\vec{b})=
\langle A|\Gamma_{A}(\vec{b},\vec{c}_{A},...,\vec{c}_{1})|A\rangle=
1-\langle A| \prod_{i=1}^{A} \left[1-\Gamma(\vec{b}-\vec{c}_{i})\right]
|A\rangle
\label{eq:6.1.6}
\endeq
can readily be
calculated in the dilute-gas or
independent particle model (IPM), when
(here we neglect the center-of-mass correlations)
\beq
|\Psi(\vec{r}_{A},...,\vec{r}_{1})|^{2}=
\prod_{j=1}^{A} n(\vec{r}_{j}) \,   ,
\label{eq:6.1.7}
\endeq
where $n(\vec{r})=n_{A}(\vec{r})/A$ and $n_{A}(\vec{r})$ is the
target matter density, and $A$ is the nuclear mass number.
The result is [20]
\arr
\langle A| \prod_{i=1}^{A} \left[1-\Gamma(\vec{b}-\vec{c}_{i})\right]
|A\rangle
=\left[1-{1 \over A}\int d^{2}\vec{c}\,\Gamma(\vec{c})
\int dz n_{A}(\vec{b}-\vec{c}, z)\right]^{A}  \nonumber\\
=
\left[1-{1 \over 2A} \sigma_{tot}(hN)T(\vec{b}) \right]^{A}=
\exp\left[-{1 \over 2} \sigma_{tot}(hN)T(\vec{b}) \right]~~~~~~~~~~~~~~~
\label{eq:6.1.8}
\endarr
and
\beq
\sigma_{tot}(pA)=2\int d^{2}\vec{b}\,
\left\{ 1 -\exp\left[-{1\over 2}\sigma_{tot}(pN)T(b)\right]\right\}
\label{eq:6.1.9}
\endeq
When deriving eq.(\ref{eq:6.1.8}) I have neglected the $hN$ interaction
radius (the diffraction slope $B$)
compared to the nuclear radius. Incorporation of the
finite diffraction slope $B$
is trivial:
\beq
\int dz \int d^{2}\vec{c}\,\Gamma(\vec{c})
n_{A}(\vec{b}-\vec{c}, z)=
{1\over 2}\sigma_{tot}(pN)\int {d^{2}\vec{c}\over 2B}
T(\vec{b}-\vec{c})
\exp\left[-{\vec{c}^{2} \over 2B}\right]
\approx {1\over 2}\sigma_{tot}(pN)
T(b)
\label{eq:6.1.10}
\endeq
 Since the $\nu$-fold scatterings with
\beq
\nu \sim \sigma_{tot}(pN)T(0)/2 \ll A
\label{eq:6.1.11}
\endeq
 dominate in (\ref{eq:6.1.8}), for all
the practical purposes exponentiation is accurate enough.


\subsection{Gribov's inelastic shadowing
and the diffraction operator [{\sl Refs.9,11,14}]}

In Gribov's field-theoretic language  Glauber's result
(\ref{eq:6.1.8}), (\ref{eq:6.1.9})
corresponds to a   sum of all the
mutiple-scattering diagrams of Fig.4a
with the free-nucleon cross section $\sigma_{tot}(hN)$.
It must be corrected for the
off-diagonal diffractive transitions (regeneration) of Fig.4b
(Gribov's inelastic shadowing [9]).

Let us start with the high-energy limit of $l_{f} >> R_{A}$,
when the time-dependent
phases in eq.(\ref{eq:5.3.2}) can be neglected. In this
limit the inelastic shadowing  corresponds to the
coupled-channel
generalization of (\ref{eq:6.1.9}) [11,14]
\beq
\sigma_{tot}(pA)=2\int d^{2}\vec{b}\,
\left\{ 1 -\langle p|
\exp\left[-{1\over 2}\hat{\sigma}T(b)\right]|p\rangle\right\}
\label{eq:6.2.1}     \,  .
\endeq

A convenient way to calculate the nuclear cross
section is to diagonalize the diffraction matrix $\hat{\sigma}$.
Let the corresponding eigenstates be $|\alpha\rangle$ and the
corresponding eigenvalues of $\hat{\sigma}$ be
$\sigma_{\alpha}=\langle \alpha |\hat{\sigma}|\alpha\rangle=
\sigma_{tot}(\alpha N)$. These diffraction eigenstates $|\alpha\rangle$
make a complete set, and one can
expand the proton wave function as
\beq
|p\rangle = \sum_{\alpha}a_{\alpha}|\alpha\rangle  \,   .
\label{eq:6.2.2}
\endeq
Then, one can compute a matrix element
of the exponential in eq.(\ref{eq:6.2.1})
as
\arr
\langle p|
\exp\left[-{1\over 2}\hat{\sigma}T(b)\right]|p\rangle  =
{}~~~~~~~~~~~~~~~~~~~~~~~~~~~~~ \nonumber\\
\sum_{\alpha,\beta}\langle p|\alpha\rangle \langle \alpha|
\exp\left[-{1\over 2}\hat{\sigma}T(b)\right]
|\beta\rangle \langle\beta|p\rangle
=\sum_{\alpha}a_{\alpha}^{*}a_{\alpha}
\exp\left[-{1\over 2}\sigma_{tot}(\alpha N)T(b)\right]  \,  ,
\label{eq:6.2.3}
\endarr
and calculate the cross section (\ref{eq:6.2.1}) as  [11,14]
\beq
\sigma_{tot}(pA)=2\sum_{\alpha}a_{\alpha}^{*}a_{\alpha}
\int d^{2}\vec{b}\,
\left\{ 1 -
\exp\left[-{1\over 2}\sigma_{tot}(\alpha N)T(b)\right]\right\}    \,  .
\label{eq:6.2.4}
\endeq
Here the Glauber formula (\ref{eq:6.1.9}) holds for the
separate diffraction
eigenstates, which undergo only elastic rescatterings.
The fixed-$\vec{\rho}$ states are precisely the diffraction
eigenstates of the QCD theory of diffractive scattering,
which we have already used in sections 3 and 4.
The original formulation of the diffraction-eigenstate
formalism goes back to 1955 paper by Feinberg on
the diffraction dissociation of deuterons ([55], see also [56,57]).

The least possible attenuation depends on the minimal
eigenvalue $\sigma_{min}$  of the diffraction operator.
QCD predicts $\sigma_{min}=0$, which produces the
nonexponential attenuation discussed in section 4.


\subsection{Coherency constraint and the effective diffraction
operator [{\sl Refs.5,6}]}

At moderate energies the spatial
 expansion of the wave packet (\ref{eq:5.3.1})
becomes significant. The source of expansion is an emergence of
relative phases between different components of the wave packet.
In MST language, the origin of these phases is a change of
the longitudinal momentum
in the off-diagonal transition $i\, N \rightarrow k\, N$
when the wave packet propagates inside the nucleus [9,10]:
\beq
\kappa_{ik}={ m_{k}^{2} - m_{i}^{2} \over 2\nu   }
\sim {1 \over l_{f}}\,.
\label{eq:6.3.1}
\endeq

In order to see how the high-energy formula
(\ref{eq:6.2.1}) should be modified at moderate energy,
consider the $\nu$-fold scattering contribution to (\ref{eq:6.2.1}):
\beq
{1\over \nu!}
T(b)^{\nu}\langle p|\hat{\sigma}^{\nu}
|p\rangle                   =
\sum_{i,j...k}\sigma_{pi}\sigma_{ij}...\sigma_{kp}
\int_{-\infty}^{+\infty}dz_{\nu}n_{A}(b,z_{\nu})....
\int_{-\infty}^{z_{2}}dz_{1}n_{A}(b,z_{1})
\label{eq:6.3.2}
\endeq
The finite longitudinal momentum transfer (\ref{eq:6.3.1})
gives rise to the phase factor
\beq
\exp\left[i\kappa_{pi}z_{\nu}+ i\kappa_{ij}z_{\nu-1}
+...+i\kappa_{kp}z_{1}\right]
\label{eq:6.3.3}
\endeq
in the integrand of the $r.h.s.$ of
(\ref{eq:6.3.2}).
(In the time-ordered perturbation theory with the conserved
momentum $\vec{p}$ one finds the same phase factor expanding
the energy of the relativistic intermediate state as
$\sqrt{p^{2}+m_{i}^{2}} \approx p + m_{i}^{2}/2\nu$ and
factoring out the common  $t$-dependence in
eq.(\ref{eq:5.3.1})).

The result of the $z$ integrations with the
phase factor (\ref{eq:6.3.3})
in the $r.h.s.$ of eq.(\ref{eq:6.3.2})  will be that the
high-energy formula in the
$r.h.s$ of eq.(\ref{eq:6.3.2}) is multiplied by a certain
longitudinal form factor
$F_{\nu}(b,\kappa_{pi},\kappa_{ij},...,\kappa_{kp})$.
This form factor can most easily be evaluated using
the momentum space representation of MST
(for instance, see
considerations in [16]). Here the $\nu$-fold
scattering amplitude can be written as (recall our normalization
(\ref{eq:7.3}))
\arr
f_{A}^{(\nu)}(\vec{q})=\left({i \over 8\pi^{2}}\right)^{\nu-1}
{A!\over \nu!(A-\nu)!} ~~~~~~~~~~~~~~~~~~~~\nonumber\\
\int
\prod_{i=1}^{\nu}d^{2}\vec{\kappa}_{\perp i}
\Phi_{\nu}(\vec{\kappa}_{pi},\vec{\kappa}_{ij},...
,\vec{\kappa}_{kp})
f_{pi}(\vec{\kappa}_{pi})f_{ij}(\vec{\kappa}_{ij})...
f_{kp}(\vec{\kappa}_{kp})
\,
\delta\left(\sum_{i=1}^{\nu}\vec{\kappa}_{i}-\vec{q}\right)
\label{eq:6.3.4}
\endarr
The nuclear form factor
$\Phi_{\nu}(\vec{\kappa}_{pi},\vec{\kappa}_{ij},...
,\vec{\kappa}_{kp})$
is the Fourier-transform of the $\nu$-body density
$n^{(\nu)}(\vec{r}_{\nu},...,\vec{r}_{1})$ :
\beq
\Phi_{\nu}(\vec{\kappa}_{\nu},...,\vec{\kappa}_{1})=\int
d^{3}\vec{r}_{\nu}...d^{3}\vec{r}_{1}
n^{(\nu)}(\vec{r}_{\nu},...,\vec{r}_{1})
\exp(+i\vec{\kappa}_{\nu}\vec{r}_{\nu}+...
+i\vec{\kappa}_{1}\vec{r}_{1})  \,  .
\label{eq:6.3.5}
\endeq
The normalization is such that
\beq
\Phi_{\nu}(0,...,0)=1
\label{eq:6.3.6}
\endeq
Still another representation of the form factor
$\Phi_{\nu}(\vec{\kappa}_{\nu},...,\vec{\kappa}_{1})$ in terms
of the momentum space
wave function of the ground state of the nucleus is as follows:
\beq
\Phi_{\nu}(\vec{\kappa}_{\nu},...,\vec{\kappa}_{1})=\int
\prod_{i=1}^{A}d^{3}\vec{k}_{i}
\Psi^{*}(\vec{k}_{A},...,\vec{k}_{\nu}+\vec{\kappa}_{\nu},...,
\vec{k}_{1}+\vec{\kappa}_{1})
\Psi(\vec{k}_{A},...,\vec{k}_{1})
\label{eq:6.3.7}
\endeq
The center-of-mass correlation improved
independent particle model:
\beq
|\Psi(\vec{r}_{A},...,\vec{r}_{1})|^{2}=(2\pi)^{3}\rho_{c}
\delta\left(\sum_{i=1}^{A}\vec{r}_{i}\right)
\prod_{i=1}^{A}n(\vec{r}_{i})
=\rho_{c} \int d^{3}\vec{k}
\exp\left(-i\vec{k}\sum_{i=1}^{A}\vec{r}_{i}\right)
\prod_{i=1}^{A}n(\vec{r}_{i})   \,  ,
\label{eq:6.3.8}
\endeq
gives the $\nu$-body form factor
\beq
\Phi_{\nu}(\vec{\kappa}_{\nu},....,\vec{\kappa}_{1})
=\rho_{c}\int d^{3}\vec{k}\,[\phi(\vec{k})]^{A-\nu}
\prod_{i=1}^{\nu}\phi(\vec{k}-\vec{\kappa}_{i})
\label{eq:6.3.9}
\endeq
where $\phi(\vec{k})$ is the Fourier transform of $n(\vec{r})$.
The constant $\rho_{c}$ is so chosen that the normalization
(\ref{eq:6.3.6}) is satisfied with $\phi(0)=1$.

Now notice, that at $\nu \ll A$  the $\vec{k}$ integration in
(\ref{eq:6.3.9}) is dominated by a rapid decrease of
$\phi(\vec{k})^{A}$. The steepest descent estimation
$\vec{k}=-\vec{q}/A$ shows that to a good accuracy
the charge form factor of  nucleus
$G_{A}(\vec{q})\approx \phi(\vec{q})$. Similarly,
at $\nu \ll A$ of the interest ( see (\ref{eq:6.1.11}))
\beq
\Phi_{\nu}(\vec{\kappa}_{\nu},....,\vec{\kappa}_{1})
\approx
\prod_{i=1}^{\nu}\phi(\vec{\kappa}_{i})\approx
\prod_{i=1}^{\nu}G_{A}(\vec{\kappa}_{i})
\label{eq:6.3.10}
\endeq
Finally, at moderate values of $\kappa_{i}\lsim 1/R_{A}$
the charge form factor
is close to the Gaussian and
\beq
G_{A}(\vec{\kappa})\approx G_{A}(\vec{\kappa}_{\perp})
G_{A}(\kappa_{L})
\label{eq:6.3.11}
\endeq
Performing
in (\ref{eq:6.3.4})
the transverse momentum integrations,
which are
dual to the impact parameter integration in (\ref{eq:6.2.1}), one
is left with the longitudinal form factor [5,6]
\beq
F_{\nu}(b,\kappa_{pi},\kappa_{ij},...,\kappa_{kp})
\approx
G_{A}(\kappa_{pi})G_{A}(\kappa_{ij})...G_{A}(\kappa_{kp})\,  .
\label{eq:6.3.12}
\endeq
By virtue of the factorization (\ref{eq:6.3.12}),
at each rescattering vertex the form factor
$G_{A}(\kappa_{ik})$ can be absorbed into the corresponding
rescattering amplitude, giving rise to the effective
diffraction operator [5]
\beq
\sigma_{ik}^{eff}=\sigma_{ik}G_{A}(\kappa_{ik})\, .
\label{eq:6.3.13}
\endeq
Notice, that
for the elastic rescatterings $\kappa_{ii}=0$ and
$G_{A}(\kappa_{ii})=1$.

Therefore, at moderate energy, one has to
solve for the eigenvalues and eigenstates
of the effective energy-dependent diffraction operator
$\hat{\sigma}^{eff}$ and calculate the nuclear cross section
applying eqs.(\ref{eq:6.2.2}),(\ref{eq:6.2.4}).
The nuclear form factor $G_{A}(\kappa_{ik})$ quantitatively
describes the freezing out of
new coherently interefering channels $|i\rangle$ with the mass
(notice a slight change $\gamma \rightarrow 2\nu/(m_{i}+m_{p})$
from eq.(\ref{eq:5.3.3}))
\beq
m_{i}^{2} -m_{p}^{2} \lsim {\nu \over R_{A}}
\label{eq:6.3.14}
\endeq
as the incident energy $\nu$ is gradually increased and
quantifies the coherency constraint (\ref{eq:6.3.14}).

At lower energies, when
$\kappa_{ik}R_{A} \sim R_{A}/l_{f} \gg 1$, all the off-diagonal
matrix elements of $\hat{\sigma}^{eff}$ vanish,
Gribov's inelastic shadowing is frozen,
only the
elastic rescattering, $p\,N \rightarrow p\,N$ , contribution
survives in the $l.h.s.$ of (\ref{eq:6.3.2} and Glauber's
formula (\ref{eq:6.1.9}) with the free-nucleon
cross section $\sigma_{tot}(pN)$ is recovered.


\section{Nonexponential
attenuation of ultrarela\-ti\-vis\-tic
po\-si\-tronium in medium [{\sl Refs.26,27}]}

In section 2 I have argued that whenever the formation length
becomes large, the nonexponential attenuation does emerge.
The specific law (\ref{eq:4.1.4}) derived in section 4 is
a property of QCD with its CT cross section (\ref{eq:4.1.3}).
Another beautiful example is propagation of the ultrarelativistic
positronium in a medium, which emphasizes rather a regeneration
and large formation length phenomena.

The velocity of the positron and electron in the
positronium $v \sim \alpha_{em}$, and
in this case $l_{f} \approx \gamma a_{B}/\alpha_{em}$,
where $a_{B}$ is
the Bohr radius and $\alpha_{em} \approx 1/137 $.
For the 10 $GeV$ positronium $l_{f} \sim 0.03cm$ compared to the
mean free path in graphite $l_{C^{12}} \approx 0.1 \mu m$
[26,30].

If the formation length exceeds the target thickness, then
the $e^{+}e^{-}$ separation $\vec{\rho}$ can be considered
frozen.
If $\Psi_{0}(\vec{\rho})$ is the ground state wave function,
downstream the target we have
\beq
\Psi(\vec{\rho})=\Psi_{0}(\vec{\rho})
\exp\left[-{i\over 2}\vec{Q}\cdot\vec{\rho}\right]              \, ,
\label{eq:7.1}
\endeq
where $\vec{Q}=\sum\vec{q}_{i}~^{(+)}-\sum\vec{q}_{i}~^{(-)}$
is the total momentum transfer to the positronium.
A probability to find the positronium in the ground state downstream
the target is given by
\beq
w_{0}=\Biggr\vert \int d^{3}\vec{\rho}\Psi_{0}(\vec{\rho})^{*}
\Psi(\vec{\rho})\Biggr\vert^{2}=\Phi_{1}(\vec{Q})^{2}
=\left[{1\over 1+Q^{2}a_{B}^{2}/16}\right]^{4}                \, .
\label{eq:7.2}
\endeq
The size of positronium $a_{B}$ is about twice the atomic size,
the atomic electric field is strong only inside the inner shells
$r \lsim a_{B}/Z \ll a_{B}$, and
only positron or only electron gets the transverse
kick when flying by a separate atom. Hence,  $\vec{q}_{i}~^{(+)}$ and
$\vec{q}_{j}~^{(-)}$ are uncorrelated, for the thick target
$\vec{Q}$ will have the
Gaussian distribution
with $\langle Q^{2}\rangle=2Q^{2}_{0}d/l
 \gg 1/a_{B}^{2}$, and [27]
\beq
\langle w_{0} \rangle = \int d^{2}\vec{Q}
{1 \over \pi \langle Q^{2}\rangle }
\exp\left(- {\vec{Q}^{2} \over \langle Q^{2}\rangle }\right)
\Phi_{1}(\vec{Q})^{2}
= {8 \over 3} {1 \over a_{B}^{2}Q_{0}^{2}}
{l \over d}
\label{eq:7.3}
\endeq
We have obtained the inverse thickness law for the {\sl survival
probability}, not the {\sl amplitude} like in the charmonium-nuclei
interaction, eq.(\ref{eq:4.1.4}). Corrections to the
inverse thickness law at
$l_{f}$ comparable to or smaller than
the target thickness $l$ are discussed in
[28,29].

I emphasize that the nonexponential attenuation of positronium
derives from a large coherence length, since in view of
$a_{B} \gg r_{a}$ the
color (charge) neutrality of positronium is not relevant here.
The nonexponential attenuation of positronium was
observed experimentally in an ingenuous experiment performed at
the Serpukhov accelerator in a positronium beam formed
in the decay $\pi^{0}\rightarrow \gamma (e^{+}e^{-})_{atom}$ [30].


\section{Colour transparency experiments: the candidate re\-ac\-tions}

Having set up the necessary formalism, we shall present a
brief list of possible CT experiments. Theoretically, the
photo- and electroproduction reactions are the best understood
ones. Apart from much theoretical appeal and rich CT phenomena,
these reactions are of the practical interest too, as could
easily be studied at SLAC or EEF (European Electron Facility).

The electron accelerators can be viewed as a high-luminosity
source of the real and virtual photons. By changing the
virtuality $Q^{2}=-q^{2}$ of the photon or the Compton
wavelength of the produced (anti)quarks, one can control
the intitial size $\rho_{Q}$ of the ejectile. By changing
the energy transfer $\nu$, one can control the formation
length $l_{f}$ and the onset of CT signal.
Here we list some of the interesting processes, which can be
studied at SLAC or the $15-30\, GeV$ EEF and
for which the quantitative QCD predictions are available.

\begin{itemize}

\item
The quasielastic $(e,e'p)$ scattering on nuclei at large momentum
transfer [3,4].

\item
Photoproduction of charmonium on nuclei, where
the ejectile's size is given by a short Compton wavelength
of heavy quarks [35]
\beq
\rho_{Q} \approx {1 \over m_{c}}
\label{eq:8.1}
\endeq

\item
Large-angle photoproduction of light vector mesons with
the large momentum transfer $Q=\sqrt{|t|}$, where [58] (for
the related discussion of the quasielastic pion charge-excahnge
on nuclei see [34,41])
\beq
\rho_{Q} \sim {1\over \sqrt{|t|} }
\label{eq:8.2}
\endeq

\item
Electroproduction of vector mesons on nuclei, where [48]
\beq
\rho_{Q} \approx {2 \over \sqrt{m_{V}^{2} +Q^{2}} }
\label{eq:8.3}
\endeq

\item
Nuclear shadowing in the deep inelastic scattering (DIS)
[38-40,46].
This is the textbook reaction, and in textbooks it is often stated
that $\rho_{Q} \sim 1/Q$.

\item
Diffraction dissociation of pions or (virtual) photons into
pairs of the high transverse momentum jets on nuclei [59,60]~
$\pi+ A\, , ~~\gamma^{*} A \rightarrow jet(q)+jet(\bar{q})+A^{*}$.
Here
\beq
\rho_{Q} \sim 1/k_{\perp}  \,\,  ,
\label{eq:8.4}
\endeq
where $k_{\perp}$ is the transverse momentum of the two jets
which in the comoving frame are produced back-to-back.

\end{itemize}

The QCD scanning of hadronic wave functions in all these
reactions has its own specifics. From the theoretical and,
considering the cross section and counting rates, the experimental
point of view too, the virtual photoproduction of the vector
mesons is an ideal ground test for CT ideas. In this case
the ejectile wave function is QCD calculable, which allows to
predict quantitatively possible CT signals.

I shall demonstrate
that the
quantal evolution of the relativistic wave packets in the
nuclear medium is much richer than the naive attenuation: in
certain cases CT {\sl enhances} the nuclear attenuation, in
other cases it may produce {\sl antishadowing}, i.e., the nuclear
enhancement [42,47,7,8,48]. But before I give a quantitative
definition of the scanning, I need to introduce one more concept:
the wave function of the quark-antiquark Fock state of
virtual photons [39].


\section{Colour transparency in the diffractive
deep in\-el\-as\-tic scattering:
Other men's flowers [{\sl Refs.38-40,46}]}

It is  too late to ask for a Nobel prize
for the discovery of CT. It has already been given to
Freedman, Kendall and Taylor, although CT is never mentioned
in their Nobel prize lectures [61]. There is a certain truth
in this joke.

In the QCD improved parton model higher order
QCD corrections to the Compton scattering on valence quarks
at high $Q^{2}$ are reinterpreted  as a radiative generation of the
glue and quark-antiquark sea in nucleons.
The lowest order perturbative QCD diagram for DIS
on the radiatively generated sea  is shown in fig.5.
Altoghether there are 4 Feynman diagrams with different
coupling of the two gluons to the quark and antiquark, fig.1
(in fact 36 if different couplings of gluons to quarks
in the nucleon are counted).

At $x \ll 1$, the same perturbative QCD diagram
can be treated as a scattering on a nucleon of the
quark-antiquark pairs the virtual photon transforms into at a
large distance upstream the target [62] (called the coherence
length):
\beq
l_{c} \approx {2\nu \over Q^{2}+M^{2}} \sim 1/m_{N}x \ll R_{N} \, .
\label{eq:9.1}
\endeq
Here I assumed excited masses $M^{2} \lsim Q^{2}$, which
can be justified by a direct QCD calculation of the mass
spectrum [40], $x$ is the Bjorken scaling variable, $x=Q^{2}/2(pq)$,
$p$ and $q$ are the target nucleon's and the virtual photon's
4-momenta, $q=(\nu,0,0,\sqrt{\nu^{2}+Q^{2}})$, $Q^{2}=-q^{2}$,
$2(pq)=2m_{N}\nu$.
The photoabsortion cross section can be represented as
\beq
\sigma_{tot}(\gamma^{*}N)=\int_{0}^{1}dz\int d^{2}\vec{\rho}
|\Psi_{\gamma^{*}}(z,\vec{\rho})|^{2}\sigma(\rho)\,\, ,
\label{eq:9.2}
\endeq
where $\sigma(\rho)$  is the same universal cross section
(\ref{eq:3.2}) shown in fig.2.
Wave functions of
the $q\bar{q}$ Fock states of (T) transverse and (L) longitudinal
photons were derived in [39]:
\beq
\vert\Psi_{T}(z,\rho)\vert^{2}={6\alpha_{em} \over (2\pi)^{2}}
\sum_{1}^{N_{f}}Z_{f}^{2}
\{[z^{2}+(1-z)^{2}]\varepsilon^{2}K_{1}(\varepsilon\rho)^{2}+
m_{f}^{2}K_{0}(\varepsilon\rho)^{2}\}                     \, \,,
\label{eq:9.3}
\endeq
\beq
\vert\Psi_{L}(z,\rho)\vert^{2}={6\alpha_{em} \over (2\pi)^{2}}
\sum_{1}^{N_{f}}Z_{f}^{2}
4Q^{2}z^{2}(1-z)^{2}K_{0}(\varepsilon\rho)^{2} \, \, .
\label{eq:9.4}
\endeq
Here $K_{\nu}(x)$
is the Bessel function, $\varepsilon^{2}=z(1-z)Q^{2}+m_{f}^{2}$
and $z$ is the Sudakov variable - fraction of the (light-cone)
momentum of the photon carried
by a quark of the $q\bar{q}$ pair, $0 < z < 1$.
These wave functions describe, in fact, the relativistically boosted
vacuum polarization charge distribution.

Unlike the hadronic wave functions, which are smooth and finite at
small $\rho$, in (\ref{eq:9.3})
$\varepsilon^{2}K_{1}(\varepsilon \rho)^{2} \propto 1/\rho^{2}$.
Only this singular component of the wave function is important in
$\vert\Psi_{T,L}(z,\rho)\vert^{2}$ at large $Q^{2}$.
Upon the $z$ integration I find
\beq
\sigma_{T}(\gamma^{*}N) \propto {1 \over Q^{2}}
\int_{Q^{-2}}^{m_{q}^{-2}}{d\rho^{2} \over \rho^{4}}\sigma(\rho)    \, .
\label{eq:9.5}
\endeq
Here the scaling cross section $1/Q^{2}$ comes, in fact, from a
probability of forming the $q\bar{q}$-fluctuation of the virtual
photon.
The $\rho$ integration  with the CT cross section (\ref{eq:3.4})
produces the
standard $\propto \log[1/\alpha_{S}(Q^{2})]^{2}$ scaling violation.
It is not surprizing, since I am doing the same perturbative QCD
calculation as one does with the conventional
Gribov-Lipatov-Dokshitzer-Altarelli-Parisi evolution equations.

The corollary of eq.(\ref{eq:9.5}) is that at large $1/x$ the
structure function receives contributions from
\beq
{1 \over Q^{2}} < \rho^{2} < {1 \over m_{q}^{2}}
\label{eq:9.6}
\endeq
and the QCD scaling violations probe the universal $\sigma(\rho)$
at $\rho^{2} \sim 1/Q^{2}$, see fig.2.
The small-x structure function predicted [39,46]
by eq.(\ref{eq:9.2}) agrees
well with the deep inelastic scattering
(DIS) data, fig.6. This agreement shows
that CT property of
$\sigma(\rho)$ has already been probed in DIS
 in a broad range of $\rho$. The first qualitative
discussion of DIS in terms of the colour
dipole cross sections appeared in [32].

The charm excitation cross section probes the universal $\sigma(\rho)$
at $\rho^{2} \lsim 1/m_{c}^{2}$, see fig.2, i.e., in the perturbative
QCD region even in the real photoproduction limit of $Q^{2}=0$. The
predicted photoproduction cross section equals [40,44]
$\sigma(\gamma\, N \rightarrow c\bar{c}\, X) \approx 1.1\, \mu b$, in
good agreement with the experiment [63].
When photoproduction of light quarks is considered, it
is natural to ask that quarks in the photon wave function do
not propagate beyond the confinement radius. Then, with the
natural effective mass $m_{u,d} \sim m_{\pi}$,
i.e., with $q\bar{q}$ pairs having a transverse
size of the $\rho^{0}$ meson,
eq.(\ref{eq:9.2})
gives [40,44,46] the total photoabsorption cross section
$\sigma_{tot}(\gamma N) \approx 100 \mu b$, which agrees with the
experiment [64]  and shows that our educated
guess for the large-$\rho$ behaviour of $\sigma(\rho)$ was sensible.


\section{Colour transparency and
QCD scanning of had\-ronic wave functions  [\sl{Refs.7,8,42,47,48}]}


\subsection{Novel feature of colour transparency
experiments: scanning hadronic wave functions}

The quantum-mechanical description of the novel phenomenon
of QCD scanning goes as follows [48]:
In the high-energy limit, to be specified below,
the amplitude of the forward
photoproduction on the free nucleon $M_{N}$ and the nuclear
transmission coefficient or the nuclear transparency
$Tr_{A}=d\sigma_{A} /Ad\sigma_{N}$ in the quasielastic
photoproduction $\gamma^{*}A\rightarrow VA$ read [14,7,8,42,47]:
\begin{equation}
M_{N}=\langle V|\sigma(\rho)|\gamma^{*}\rangle =\langle V|E\rangle
\label{eq:10.1.1}
\end{equation}
\arr
Tr_{A}={1\over A}
\int d^{2}\vec{b} T(b)
{\langle V |\sigma(\rho)
\exp\left[-{1\over 2} \sigma(\rho)T(b)\right] |\gamma^{*}
\rangle^{2} \over
\langle V|\sigma(\rho)|\gamma^{*}\rangle^{2} } \nonumber\\
={1\over A}
\int d^{2}\vec{b} T(b)
{\langle V |
\exp\left[-{1\over 2} \hat{\sigma}T(b)\right] |E\rangle^{2} \over
\langle V|E\rangle^{2} }
\label{eq:10.1.2}
\endarr
where $|E\rangle$ is the ejectile state $|E\rangle =
\hat{\sigma}|\gamma^{*}\rangle$, which is explicitly
known in the $\vec{\rho}$-representation:
\beq
\Psi_{E}(z,\rho)= \sigma(\rho)\Psi_{\gamma^{*}}(z,\rho) \,\, .
\label{eq:10.1.3}
\endeq

The wave function $\Psi_{\gamma^{*}}(z,\rho) $ was discussed
in the preceeding section.
The most important feature of the photon wave functions
(\ref{eq:9.3}),\,(\ref{eq:9.4})
is an exponential decrease
at large distances
\begin{equation}
\Psi_{\gamma^{*}}(z,\rho)
\propto \exp(-\varepsilon \rho)
\label{eq:10.1.4}
\end{equation}
where
\begin{equation}
\varepsilon^{2} = m_{q}^{2}+z(1-z)Q^{2}
\label{eq:10.1.5}
\end{equation}
Calculation of the matrix elements in
(\ref{eq:10.1.1}),(\ref{eq:10.1.2}) involves the
$d^{2}\vec{\rho} dz$ integration. In the nonrelativistic
quarkonium $z \approx 1/2$, so that the relevant $q\bar{q}$
fluctuations have a size [48]
\begin{equation}
\rho \sim \rho_{Q} = {1 \over \sqrt{m_{q}^{2}+{1\over 4}Q^{2}}}
\approx {2 \over \sqrt{m_{V}^{2}+Q^{2}}}
\label{eq:10.1.6}
\end{equation}

What enters eqs.(\ref{eq:10.1.1}),(\ref{eq:10.1.2})
is the ejectile wave function (\ref{eq:10.1.3}).
Becasue of CT property,
$\sigma(\rho) \propto \rho^{2}$ at small $\rho$, this
ejectile wave function will be sharply peaked at
$\rho \approx \rho_{2}=2\rho_{Q}$ with the
width $\Delta \rho \sim 2\rho_{Q}$, which leads naturally to
the idea of the {\sl QCD scanning} [48]: The transition
matrix elements  (\ref{eq:10.1.1})
probe the wave function of vector
mesons at $\rho \sim 2\rho_{Q}$, and varying $\rho_{Q}$
by changing $Q^{2}$, one can
scan the wave function $|V\rangle$ from large to small
distances. In fig.7 we demonstrate qualitatively how the
scanning works. We also show the $z$-integrated
wave functions of the ground state $|V\rangle$ and
of the radial excitation $|V'\rangle$.

General description of diffraction production in terms of
moments $\langle f|\hat{\sigma}^{n}|in\rangle$ was developed
in [14]. Following this technique, the nuclear matrix element in
(\ref{eq:10.1.2}) can be expanded in the moments
$\langle V| \sigma(\rho)^{n}|\gamma^{*}\rangle $.
 These moments probe the
wave function $|V\rangle$ at different values of $\rho
\sim \rho_{2n} \sim 2n\rho_{Q}$ and are the
{\sl true  CT observables} of the
virtual photoproduction process. To the leading
order in FSI [7,8,47],
\begin{equation}
Tr_{A}=1-\Sigma_{V} {1\over A}\int d^{2}\vec{b}T(b)^{2} \,  ,
\label{eq:10.1.7}
\end{equation}
where the CT observable
\begin{equation}
\Sigma_{V}=
{\langle V|\sigma(\rho)^{2}|\gamma^{*}\rangle
\over \langle V|\sigma(\rho)|\gamma^{*}\rangle }
={\langle V|\hat{\sigma}|E\rangle
\over \langle V|E\rangle }
\label{eq:10.1.8}
\end{equation}
measures a strength of intranuclear FSI.
 This expansion works well when $1-Tr_{A} \ll 1$, and is
convenient to explain how the scanning proceeds.
Notice, that $\sigma(\rho)^{2}|\gamma^{*}\rangle$
peaks at $\rho \sim \rho_{4}=4\rho_{Q}$.

The standard VDM [65] gives a similar
expansion for the nuclear transparency, but with $\Sigma_{V}$
substituted for the free-nucleon cross section [7,8]:
\beq
\Sigma_{V} \Leftrightarrow \sigma_{tot}(VN) \,  ,
\label{eq:10.1.9}
\endeq
so that naively one might hope to determine
the free-nucleon cross section
from the photoproduction on nuclei.


\subsection{Scanning and the node effect [{\sl Refs.7,8,47,48}].
Perils of the vector dominance model.}

Start with the real photoproduction ($Q^{2}=0$) of the
ground-state mesons
($\Upsilon$,$\, J/\Psi$,$\,\rho^{0},...$) and take the case
of the charmonium, which is an ideal testing ground
of CT ideas. In this case $\rho_{Q} =
1/m_{c} \ll R_{J/\Psi}$, \,\, $\rho_{2}$ is
still smaller than $R_{J/\Psi}$, but $\rho_{4}\sim
R_{J/\Psi}$. For the latter numerical reason, one finds
$\Sigma_{J/\Psi} \approx \sigma_{tot}(J/\Psi\,N)$, i.e.,
the predicted nuclear shadowing will be
{\sl marginally} similar to the Vector Dominance Model
(VDM) prediction [42,7,8,47]. This holds to much extent for the
$\Upsilon$ and the light vector mesons
$\rho^{o},\,\omega^{o},\,\phi^{o}$ as well [48].
At larger $Q^{2}$, one has $\rho_{2},\,\rho_{4}
\ll R_{V}$ and
$\Sigma_{V} \sim \sigma(\rho_{Q}) \propto \rho_{Q}^{2}$
with the calculable logarithmic corrections [38,6,7], so that
the above marginal similarity with the VDM disappears.
The nuclear transparency will tend to unity {\sl from below}:
\begin{equation}
1-Tr_{A} \propto {A \over R_{A}^{2}}\rho_{Q}^{2}
\label{eq:10.2.1}
\end{equation}

The case of the radial excitation  $V'$ is more interesting.
Radial excitations have larger radius and larger
free nucleon cross section: e.g., $\sigma_{tot} (\Psi' N)
\approx 2.5 \sigma_{tot}(J/\Psi \,N)$ [40], which classically
would
suggest much stronger FSI for the $\Psi'$
than for the $J/\Psi$. Presence of the node in the $V'$ wave
function leads to a rather complex pattern of
the shadowing and antishadowing. In the photoproduction
limit, because of $\rho_{2} \sim R_{\Psi'}$, there are
rather strong cancellations between the contributions
to the amplitude (\ref{eq:10.1.1}) from $\rho$ below and above the
node (the {\sl node effect} [7,8], see fig.7).
For this reason, in photoproduction on the free
nucleons, one predicts the ratio of
differential cross sections of the forward production
$r(Q^{2}=0) =d\sigma(\gamma N\rightarrow V'N)/
d\sigma(\gamma N \rightarrow VN)|_{t=0} < 1$.
For the $\Psi'/(J/\Psi)$ ratio the CT prediction is
$r(0) = 0.17$ [42], which agrees perfectly
with the NMC collaboration result
$r(0)=0.20\pm 0.05(stat)\pm 0.07(syst)$ [66].

In the $\Upsilon'$ photoproduction, the initial size
$\rho_{Q}$ scales as $1/m_{q}$, whereas the radius of
the $\bar{b}b$ bound states decreases with $m_{q}$ less
rapidly. For this reason $\rho_{4} < R_{\Upsilon}$,
for the $\Upsilon'$ the node
effect is much weaker and the
$\Upsilon'/\Upsilon$ ratio $r(0)=0.84$ [48].

The larger is $Q^{2}$, the smaller size $\rho_{2}$
is scanned. A contribution
from the region above the node becomes negligible,
and $r(Q^{2})$ will increase with $Q^{2}$ up to
$r(Q^{2} \gg m_{V}^{2}) \sim 1$ (the exact limiting
ratio depends on the $V$ and $V'$
wave function's at the origin).
Wave function of the second radial excitation
($\Psi''(3770),\,\Upsilon''(10355)$,...)
has two nodes, hence strong suppression
of $r''(Q^{2})=d\sigma(V'')/d\sigma(V)$ at small $Q^{2}$
and  rise to $r''(Q^{2})\sim 1$ at large $Q^{2}$.

Since $\rho_{4}$ is closer to the node position, the
node effect is still stronger in the matrix element
$\langle V'|\sigma(\rho)^{2} |\gamma\rangle$.
In fact, for the $\Psi'$ one finds [42,7,8,47]
$\langle \Psi'|\sigma(\rho)^{2} |\gamma\rangle <0$, so that
$\Sigma_{\Psi'} <0$ and
\beq
Tr_{A}(\Psi') >1
\label{eq:10.2.2}
\endeq
despite the larger free-nucleon cross section,
defying the naive VDM reasoning.
In the $\Upsilon'$ production,
the scanning radius compared to the bottonium radius is
relatively smaller than in the charmonium case,
the node effect is weaker and FSI produces the shadowing
of $\Upsilon'$. Still, in spite of
 $\sigma_{tot}(\Upsilon'N)\gg  \sigma_{tot}(\Upsilon N)$,
shadowing of the $\Upsilon'$ is weaker than
shadowing of the $\Upsilon$.

The $Q^{2}$-dependent scanning changes the
node effect significantly. The larger is
$Q^{2}$, the smaller is the scanning radius
$\rho_{Q}$ and the weaker are the cancellations. In the $\Psi'$
case $\langle V'|\sigma(\rho)^{2}|\gamma^{*}\rangle$
becomes positive valued, so that the antishadowing changes
to the shadowing, which first rises with $Q^{2}$, then
saturates and is followed by the onset of asymptotic
decrease (\ref{eq:10.2.1}).
In the $\Upsilon'$ production the node effect is weaker,
still it affects the $Q^{2}$-dependence of the shadowing
making it different from that of the $\Upsilon$.

The predictions for the heavy quarkonium photoproduction
[48] are
shown in figs.8,9. Because of the small size of heavy
quarkonium, the results are numerically reliable, as they are
dominated by the perturbative QCD domain.


\subsection{The two scenarios of scanning the wave func\-ti\-on of
light vector mesons: anomalous $Q^{2}$ dependence
[{\sl Ref.48}]}

The case of the light vector mesons is particularly interesting,
as similar (anti)shadowing effects occur in the energy range
accessible at SLAC and EEF.
Here at moderate $Q^{2}\lsim m_{\rho}^{2}$ the scanning radius
is large, in the nonperturbative domain of
$\rho_{Q} \sim R_{V}$, so that the
predicted [48] shadowing of the $\rho^{0}$ , shown in fig.10, is
numerically not very accurate. However,
the accuracy increases gradually with $Q^{2}$ as
the scanning radius decreases into the perturbative domain
$\rho_{Q} \ll R_{V}$. Since the wave function of the
$\rho^{0}$ does not have a node, I believe we describe correctly
a smooth transition from the real to virtual photoproduction.

In the photoproduction limit we find a strong node
effect and strong suppression of the
photoproduction of the radial excitation $\rho'$  on nucleons,
by more than one order of magnitude
compared to the $\rho^{0}$ production, which broadly agrees
with the scanty experimental data [65,67]. As here
$\rho_{Q} \sim R_{V}$, we can not give a reliable numerical
estimate of how small the $\rho'/\rho^{0}$ cross section
ratio is. Still, we can describe the two
possible scenarios [48] of scanning the $\rho'$
wave function, experimental tests of which can shed light on
the spectroscopy and identification of the radial
excitations of the light vector mesons:\medskip\\
{\bf
(1)
The
undercompensated
free-nucleon amplitude}:
$\langle \rho'|\sigma(\rho)|\gamma\rangle  >0$.\smallskip\\
In this scenario the $\rho'$ case will be similar
to the $\Psi'$ case, apart from the possibility of
anomalously strong nuclear enhancement $Tr_{A} > 1$,
which might show strong atomic number dependence.
Indeed, because of the  larger relevant
values of $\rho$ the cross section $\sigma(\rho)$ is
larger, and the
attenuation factor $\exp[-{1 \over 2}\sigma(\rho)T(b)]$
in the nuclear matrix element, eq.(\ref{eq:10.1.2}), will suppress
the large $\rho$ region, effectively decreasing the
scanning radius $\rho_{Q}$ and diminishing the
node effect. Detailed description of
the atomic number dependence is presented below.

The $Q^{2}$-dependent scanning too will follow the $\Psi'$-scenario:
a change from the antishadowing to the shadowing with increasing
$Q^{2}$, followed by the saturation and then decrease of the
shadowing according to eq.(\ref{eq:10.2.1}). The range of $Q^{2}$ at
which the major effects should occur corresponds to a
change of the scanning radius
$\rho_{Q}$ by the factor $\sim 2$, i.e., to
$Q^{2} \sim 4m_{q}^{2} \sim m_{\rho}^{2}$. \medskip\\
{\bf
(2)
The overcompensated free-nucleon
amplitude}: $\langle \rho'|\sigma(\rho)|\gamma\rangle < 0$.\smallskip\\
This is the most likely scenario.
It is preferred in the crude oscillator model
used in [40] (see also section 20.1)
and fits the pattern of the node effect becoming
stronger for the lighter flavours.
In this case $\rho_{2} \gsim R_{V}$ and
all the higher moments too will be negative
valued:
$\langle \rho'|\sigma(\rho)^{n}|\gamma\rangle < 0$, so that
in the photoproduction limit one starts with the
conventional shadowing: $Tr_{A} < 1$. With decreasing scanning
radius $\rho_{Q}$, at $\rho_{2} \approx R_{V}$, the first
moment will change the sign,
$ \langle \rho'|\sigma(\rho)|\gamma\rangle >0$, while
higher moments are still negative valued,
$\langle \rho'|\sigma(\rho)|\gamma\rangle < 0$. This will
lead to the antishadowing.

In the $Q^{2}$ depending scanning  the striking
effect is bringing the free-nucleon am\-pli\-tude
$\langle V'| \sigma(\rho)|\gamma^{*}\rangle$ down to
the exact compensation at certain moderate $Q^{2}$,
when the decreasing $\rho_{2}$ intercepts $R_{V}$
(strictly speaking, because of relativistic
corrections and different quark
helicity states, the compensation is unlikely to be
exact). As a result, we find [48] a spike in $Tr_{A}$,
fig.9.
With the further increase of $Q^{2}$ one enters the
undercompensation regime, and the further pattern
of scanning will be essentially
the same as in the undercompensation or the $\Psi'$
scenario. \medskip\\

The above described $Q^{2}$-dependent scanning of the
wave function of light vector
mesons offers a unique possibility of identifying the
radial excitation of light vector mesons (for
the detailed discussion of the spectroscopy of light
vector mesons see [67]). The corresponding
experiments could easily be performed at SLAC and
EEF.

\subsection{Anomalous $A$ dependence of photoproduction of
the $\rho'$ meson [Ref.68]}

The nuclear attenuation factor in the reduced nuclear matrix element
\beq
M(T)={ \langle \rho'|\hat{\sigma}\exp[-{1\over 2}\hat{\sigma}T(b)]|
\gamma^{*}\rangle
\over \langle \rho'|\hat{\sigma}|\gamma^{*}\rangle }
\label{eq:10.4.1}
\endeq
suppresses the large-$\rho$  contribution, which effectively
amounts to a decrease of the scanning radius with the atomic
number $A$. This gives rise to an anomalous $A$ dependence of
diffractive production of the $\rho'$ on nuclei.

The overcompensation scenario is the most interesting
one: $M(T)$ which is $\approx
1$ at small $T$ may change a sign at certain $T$, which
corresponds to the effective scanning radius $\approx R_{V}$,
 and then becomes negative
valued: one starts with the overcompensated nuclear matrix
element in the nominator for the light nuclei and ends up
with the undercompensation regime for the heavy nuclei,
whereas the nucleonic matrix element in the denominator
stays in the overcompenstaion regime.
 If the compensation for the hydrogen was strong enough, then
breaking of compensation of contributions from $\rho >R_{V}$ and
$\rho <R_{V}$ may be more important than the overall nuclear
attenuation. For this reason $M(T)$ could stay numerically large
even for the central interactions with heavy nuclei.

In Fig.11 I show the impact parameter dependence of $M(T)$ for the
photoproduction of the $\rho'$, which shows the anticipated change
from the overcompensation to the undercompensation regime [68].
Since compensation in the free-nucleon amplitude is very strong,
\beq
{d\sigma(\gamma N \rightarrow \rho' N) \over
d\sigma(\gamma N \rightarrow \rho N)} \sim {1 \over 30}\,\, ,
\label{eq:10.4.2}
\endeq
$M(T)$ changes the sign already on light nuclei, in fact,
in the diffuse edge of a nucleus. Since at small $Q^{2}$
the scanning radius $\rho_{Q} \sim R_{V}$, the situation
when this change of sign takes place at small impact parameter,
in the strong attenuation region, is rather
 unlikely. In Fig.12 I show
the $A$ dependence at different values of $Q^{2}$:

In the real photoproduction $Tr_{A}(\rho')$ has a rather deep
minimum for $He-Li-Be$ targets. The integrand of $Tr_{A}$ is
proportional to $T(b)$ and for the heavier nuclei a contribution of
small impact parameters, where $M(T)\sim -1$, very soon takes over.
As a result, $Tr_{A}(\rho')$ rises with $A$, flattens off, and
starts decreasing when the overall nuclear attenuation takes over.

Because of the strong compensation (\ref{eq:10.4.2}) in the free-nucleon
amplitude the $A$ dependence is extremely sensitive to small
variations of the scanning radius, i.e., to $Q^{2}$. The dip on
light nuclei disappears, and at $Q^{2}$ which correspond to the
spike in Fig.9 one starts with the antishadowing in light nuclei,
$Tr_{A}(\rho') >1$, which increases with $A$, flattens and then
decreases for heavier nuclei. At still
larger values of $Q^{2}$, i.e., at still smaller scanning radius,
nuclear transparency stays flat, $Tr_{A}(\rho')\approx 1$, in a
quite broad range of $A$, which is followed by the onset of
shadowing on heavy nuclei.

Of course, at small $Q^{2}$ we can not be truly quantitative.
For instance, the dip in $Tr_{A}(\rho')$ could well take place
at somewhat
larger value of $A$. Likewise, the dip region may be absent,
and the antishadowing regime may start already in light nuclei.
None the less, the likelihood of the anomalous $A$ dependence
of photoproduction of the radial excitations of light vector
mesons is very high. The anomalous $A$ and $Q^{2}$ dependence
of the $\rho'$ production must be compared with the smooth
and structurless $A$ dependence of (virtual) photoproduction
of the $\rho^{0}$. The light nuclei are the most likely
candidates for
anomaluus $Tr_{A}(\rho')$. Search for the above anomalies could
be done at SLAC and EEF.


\section{Quantum evolution, the coherency constraint
and energy dependence of final state interaction}


\subsection{Energy dependence: from small to large coherence length
[\sl{Ref.47}]}

In the photoproduction ($Q^{2}=0$) of charmonium one should
distinguish [35,42] between  the
coherence length in the $\bar{c}c$ photoproduction
\beq
l_{c} = {2 \nu \over Q^{2}+(2m_{c})^{2} }
\approx 0.04\left({\nu \over 1 GeV}\right)f                \, .
\label{eq:11.11}
\endeq
and the length of formation (recombination)
of the charmonium states
\beq
l_{f}= \gamma {1 \over \Delta m_{c\bar{c}} } \approx
0.2\left(\nu \over 1 GeV\right)f                              \, .
\label{eq:11.12}
\endeq
Notice the strong inequality $l_{f} \gg l_{c}$.

By changing the energy $\nu$, one can vary the formation length
from $l_{f} << R_{A}$ (the quasi-instantaneous formation of the
final-state hadron) to $l_{f} > R_{A}$ (the frozen-size limit),
and thus study the dynamical evolution of the small size
{\sl perturbative} $\bar{q}q$ pair to the normal size
{\sl nonperturbative} vector meson.

In the specific case of photoproduction, the
r\^ole of the $\sl coherence$ length $l_{c}$ is somewhat different
from that of the $\sl formation$ length $l_{f}$.
At $l_{c} \ll R_{A}$ the rates of photoproduction on different
nucleons at the same impact parameter add up incoherently.
In the opposite limit of $l_{c} >R_{A}$ the amplitudes of
production on different nucleons at the same impact parameter
add up coherently, and nuclear effects are generally stronger
[47].

I start with the limit of $l_{c} \ll R_{A}$.
If $l_{f} \gg l_{c}$ and $l_{f} > R_{A}$, the transverse
size of the $q\bar{q}$ pair is still frozen
and the nuclear transparency
is given by a simple formula [14,42,7,8]
\begin{equation}
Tr_{A}={1 \over A}\int d^{2}\vec{b}dz n_{A}(b,z)
{ \langle V|\sigma(\rho)\exp[-{1 \over 2}\sigma(\rho)t(b,z)]
|\gamma^{*}\rangle ^{2}
\over
\langle V|\sigma(\rho)|\gamma^{*}\rangle^{2} }
\label{eq:11.13}
\end{equation}
where $t(b,z)=\int_{z}^{\infty}dz'n_{A}(b,z')$.
Notice, that compared to the high-energy limit (\ref{eq:10.1.2}),
here the
attenuation effect in the nuclear matrix element is weaker.

At higher energy, when $l_{c} \sim R_{A}$, the amplitudes of
$\gamma N \rightarrow VN$ transition
on different nucleons of the nucleus add-up with
the $z$-dependent relative phase
[9,10,47]
\beq
M_{A} = \sum_{i}M_{i}\exp\left(iz_{i}/l_{c}\right)
\label{eq:11.14}
\endeq
The quasielastic cross section is $\propto |M_{A}|^{2}$.
After the $z_{i}$ integrations, the interference terms
will enter with the factor $G_{A}(\kappa)^{2}$,
where $G_{A}(\kappa)$ is the charge form factor of the target
nucleus, and $\kappa =1/l_{c}$ (recall
a derivation of the effective diffraction
matrix (\ref{eq:6.3.13}) in section 6.3). In the charmonium or bottonium
photoproduction the shadowing term in (\ref{eq:10.1.7})
will be proportional to the factor [47]
\beq
{1\over 2}[1+G_{A}(\kappa)^{2}]
\label{eq:11.15}
\endeq
 This predicts a {\sl rise} of the
shadowing of the $J/\Psi$ by a factor 2 from the intermediate
to high energy, fig.13, in excellent agreement with the experimental
results of the NMC collaboration [69].
These data reject the erroneous
semiclassical attenuation model by Farrar, Frankfurt, Liu and
Strikman [70], which is extensively used in the current
literature (see also below, section 22).

Derivation of the corresponding energy dependence for the light
vector mesons is more complicated, but still to a good
approximation [48]
\begin{equation}
Tr_{A}(\nu)\approx Tr(l_{c} << R_{A})+G_{A}(\kappa)^{2}
[Tr_{A}(l_{c} >R_{A})-Tr_{A}(l_{c} <<R_{A})]
\label{eq:11.16}
\end{equation}


\subsection{Spatial expansion of the ejectile: the had\-ron\-ic
basis descrip\-tion
[{\sl Refs.48,49}]}

Above I have treated FSI in the
{\sl QCD quark basis}.
If $l_{f} < R_{A}$, the spatial expansion of the $q\bar{q}$ pair
becomes important, and $\rho$ is no longer a diagonal,
conserved variable.
As a nice application of the {\sl quark-hadron duality}, I
first comment on the spatial expansion in terms of
Gribov's inelastic shadowing, i.e., in the {\sl hadronic
basis}. Following to, and extending,
an analysis of sections 6 and 8,
consider the leading term of the final state
interaction in eq.(\ref{eq:10.1.7}).
Inserting a complete set of the
intermediate states, one can write down
\begin{equation}
\langle V|\sigma(\rho)^{2}|\gamma^{*}\rangle =
\sum _{i} \langle V|\sigma(\rho)|V_{i}\rangle
\langle V_{i} |\sigma(\rho)|\gamma^{*}\rangle
\label{eq:11.2.1}
\end{equation}
In terms of the interemediate states in the
{\sl r.h.s} of eq.(\ref{eq:11.2.1}),
antishadowing of the $\Psi'$ comes [48,49] from the destructive
interference of the diagonal VDM-like rescattering
\begin{equation}
\gamma^{*} \rightarrow \Psi' \rightarrow \Psi'
\label{eq:11.2.2}
\end{equation}
and the off-diagonal rescattering
\begin{equation}
\gamma^{*} \rightarrow J/\Psi \rightarrow \Psi'
\label{eq:11.2.3}
\end{equation}
(there is a small contribution from
other intermediate states too).
The reason for strong cancellation is that
the $\gamma \rightarrow \Psi'$ transition is weak
compared to the $\gamma \rightarrow J/\Psi$ transition,
whereas the $J/\Psi \rightarrow \Psi'$ transition
has an amplitude of opposite sign  and smaller, than
the $\Psi' \rightarrow \Psi'$ elastic scattering amplitude.
To the contrary, in the $J/\Psi$ photoproduction both
amplitudes in the off-diagonal transitions like
$\gamma \rightarrow \Psi' \rightarrow J/\Psi$ are small,
the diagonal
rescattering $\gamma \rightarrow \Psi \rightarrow \Psi$
is the predominant one,
and this explains why one finds a marginal similarity
to VDM in the photoproduction.
At larger $Q^{2}$ the node effect is no longer effective
in suppressing the $\gamma^{*} \rightarrow \Psi'$ transition,
the off-diagonal amplitudes become significant for the
$J/\Psi$ photoproduction too and one finds strong
departure from VDM for the $J/\Psi$ too.

At moderate energy, one should repeat the considerations
of section 6, eqs.(\ref{eq:6.3.2}), (\ref{eq:6.3.3}) and
(\ref{eq:6.3.13}). Only those intermediate states which satisfy
the coherence condition
will contribute
to the {\sl r.h.s.} of eq.(\ref{eq:11.2.1}), so that the strength
of the final state interaction will change rapidly
from the near-threshold energy of $l_{f} \ll R_{A}$ to
a higher energy of $l_{f} > R_{A}$. This energy dependence
can be described using the effective diffraction operator
technique [5] described in section 6.


\subsection{Spatial expansion of the ejectile:
the path integral in the basis of quark
Fock states
[{\sl Ref.42}]}

The alternate, and very elegant, path integral
description of the spatial expansion directly in the QCD
quark basis
was developed by Kopeliovich and Zakharov [42].
By virtue of CT (\ref{eq:3.4}), interaction of the
$c\bar{c}$ system with nuclear medium is the absorbing harmonic
oscillator potential
\beq
V_{opt}(\rho,t) = -{i \over 2}\gamma c \sigma(\rho)n_{A}[\vec{r}(t)] \, .
\label{eq:11.3.1}
\endeq
The harmonic oscillator model,
\beq
V(\rho)={1 \over 4}m_{c}\omega^{2}\rho^{2}\, ,
\label{eq:11.3.2}
\endeq
with $\omega =(M_{\psi'}-M_{J/\Psi})/2$  is a good approximation to
the spectroscopy of the low-lying levels of the charmonium.
Then, the net effect of the nuclear attenuation will be the
nuclear-density dependent shift of the oscillator frequency:
\beq
\omega^{2} \rightarrow \omega^{2}-in_{A}(\vec{r})
\gamma c\omega \sigma_{tot}(J/\Psi N)                   \, .
\label{eq:11.3.3}
\endeq

The expansion regime corresponds to  $l_{f} \lsim R_{A}$, i.e.,
to $l_{c} \ll R_{A}$, so that the
low-energy counterpart of eq.(\ref{eq:11.13}) is [42]
\beq
Tr={1 \over A }
\int d^{2}\vec{b} \int dz n_{A}(\rho,z)
{\left<f|\hat{U}\sigma(\rho)|\gamma\right>|^{2} \over
|\left<f|\sigma(\rho)|\gamma\right>|^{2}}            \, .
\label{eq:11.3.4}
\endeq
The evolution operator for the harmonic oscillator is known in an
explicit form [71]:
\beq
\left<y|\hat{U}|x\right>=
\left[{ m\omega \over 2\pi i \sin(\omega t) }\right]^{1/2}
\exp\left\{ { i m\omega \over 2 \sin(\omega t) }
\left[(y^{2}+x^{2})\cos(\omega t) - 2xy\right]\right\}        \, .
\label{eq:11.3.5}
\endeq
For propagation of $c\bar{c}$ one can apply the straight path
approximation, with $z$ playing a r\^ole of time $t$.
Evidently, the evolution operator can numerically be computed
for an arbitrary quark-antiquark potential and more involved
QCD diffraction interaction.
By the {\sl quark-hadron duality},
the effective diffraction operator
formalism in the hadronic basis
and the path-integral formalism
in the QCD quark basis yield the same results.


\subsection{Interplay of the spatial expansion and
of the vari\-able scan\-ning ra\-di\-us [{\sl Refs.48,49}]}

At $l_{f}<<R_{A}$ the amplitudes of transitions (\ref{eq:11.2.2}) and
(\ref{eq:11.2.3}) do not interfere. However, since they are of
comparable magnitude, the incoherent contribution of
the off-diagonal transition (\ref{eq:11.2.3}) enhances
nuclear production of the $\Psi'$. As a
result, for the $\Psi'$ the VDM prediction
for the nuclear shadowing breaks down even near the production
threshold: $Tr_{A}$ is larger than the Glauber model
prediction [48].

For the $J/\Psi$ a similar incoherent
contribution is small at small $Q^{2}$, rises with $Q^{2}$
as described above, and the near-threshold value of $Tr_{A}$
rises too. In the $\Psi'$ case the dominant effect of the
$Q^{2}$-dependent scanning is that the
$\gamma^{*}\rightarrow \Psi'$ transition amplitude increases
with $Q^{2}$ relative to the $\gamma^{*} \rightarrow
J/\Psi$ amplitude, which enhances the diagonal (shadowing)
rescattering contribution compared to the off-diagonal
(antishadowing) rescattering contribution. As a result of
this competition the near-threshold value of $Tr_{A}(\Psi')$
first decreases significantly with $Q^{2}$, which is followed by
the $J/\Psi$-like behaviour at larger $Q^{2}$ , shown in fig.8.
For the both $J/\Psi$ and $\Psi'$ the near threshold
behaviour of $Tr_{A}$ is partially due to the kinematical
rise of the threshold energy with $Q^{2}$, so that at larger
$Q^{2}$ the coherence effects become important relatively
closer to the threshold energy.
In the bottonium the intial size $\rho_{Q}$ compared to
the bottonium radius is relatively smaller, than in the
charmonium, and $Tr_{A}(\Upsilon')$ exhibits a monotonous
$Q^{2}$ and $\nu$ dependence, fig.8.

With the rising energy the destructive interference
of transitions (14) and (15) leads to a rapid rise of
the nuclear transparency $Tr_{A}$ and the onset of
the antishadowing of the $\Psi'$ in the photoproduction limit.
At the larger $Q^{2}$ the smaller $\rho$ are scanned, the
node effect is weaker,
and similar rise of $Tr_{A}(\Psi')$ with energy ends up in the
shadowing region.

In the virtual photoproduction of light vector mesons
$l_{f} \gsim l_{c}$ and eq.(\ref{eq:11.3.4}) is applicable
at $Q^{2} \gsim m_{V}^{2}$. Since the wave function of
the $\rho^{0}$ does not have a node, the $Q^{2}$
dependence of the $\rho^{0}$ production is smooth,
and eq.(\ref{eq:11.3.4}) is a good approximation for
the real photoproduction too. The predicted $Q^{2}$
dependence is shown in fig.10. At large $Q^{2}$, when for all
vector mesons $l_{f} \gg l_{c}$, the nuclear
transparency for $\rho^{0},J/\Psi,\Upsilon$  exhibits
a similar $Q^{2}$- and $\nu$-dependence.

For the $J/\Psi$  weak energy dependence
of the transparency is predicted, fig.8. For this reason,
and the above explained marginal similarity with VDM
in the $J/\Psi$ photoproduction,
the  SLAC determination [72] of the $J/\Psi$ nucleon total cross
section
$\sigma_{tot}(J/\Psi\,N)= 3.5 \pm 0.8 \, mb$ from $Ta/Be$ quasielastic
cross section ratio at $\nu = 20 GeV$ can be regarded as
reliable one. QCD calculations predict
$\sigma_{tot}(J/\Psi\,N) \approx 5.5 \, mb$ [42].


\section{Colour Transparency and nuclear shadowing in
deep inelastic scattering [39,46]}


\subsection{Scaling nuclear shadowing [{\sl Refs.38,39}]}

Consequences of CT for the nuclear shadowing in DIS are particularly
instructive [39,39,44,46].
Textbooks often state that DIS probes the small-size structure of
nucleons: $\rho_{Q} \sim 1/Q$. An analysis of the origin of
the Bjorken scaling in section 9 has already shown that this
simple-minded interpretation of DIS is not quite correct.

In DIS the role of FSI is played by the nuclear shadowing.
The above naive interpretation of DIS, when combined with
the generic CT law (\ref{eq:3.4}),
would have implied that nuclear shadowing
diasappears as $1/Q^{2}$, which was a prevailing point of view
for quite a time.

The standard
Glauber's formula for the hadron-deuteron
scattering reads [73]:
\beq
\sigma_{D}=2\sigma_{N}-\sigma_{N}^{2}
\left< {1 \over 4\pi r_{D}^{2}}\right>             \, .
\label{eq:12.1.1}
\endeq
Here $r_{D}$ is the internucleon distance in deuterium and
the average is taken with the deuterium wave function. In the
case of DIS, with  $\sigma_{N} \propto 1/Q^{2}$, one
can readily jump into the erroneous conclusion that the
nuclear shadowing term should vanish $\propto 1/Q^{4}$.

A correct calculation of the shadowing
goes as follows [38,39,44,46]:
Eq.(\ref{eq:12.1.1}) holds for the $q\bar{q}$ pair of fixed size
$\vec{\rho}$ with $\sigma_{N} \rightarrow \sigma(\rho)$.
Then, one has to average $\sigma_{D}(\rho)$
over $\rho$ with the wave functions
(\ref{eq:9.3},\ref{eq:9.4}), see eqs.(\ref{eq:3.1}),
(\ref{eq:9.2}). For the single scattering
contribution see eq.(\ref{eq:9.5}), here I concentrate
on the shadowing term:
\beq
\int dz d^{2}\vec{\rho}\, |\Psi_{T}(z,\rho)|^{2}\sigma(\rho)^{2}
\propto  {1 \over Q^{2}}
\int_{Q^{-2}}^{m_{q}^{-2}}{d\rho^{2} \over \rho^{4}}\sigma(\rho)^{2}
\propto {1 \over Q^{2}}
\int_{Q^{-2}}^{m_{q}^{-2}}d\rho^{2}                    \, .
\label{eq:12.1.2}
\endeq
Few important conclusions, all based on CT,
 can be drawn from eq.(\ref{eq:12.1.2}):
\begin{enumerate}
\item Shadowing
term scales $\propto 1/Q^{2}$ like the total cross section,
and this $1/Q^{2}$ comes from a probability of
$\gamma^{*} \rightarrow q\bar{q}$ conversion.
\item Shadowing is dominated by the large, hadronic size
$q\bar{q}$ pairs.
\item Shadowing does not contain the logarithmic scaling violations
like in (\ref{eq:9.5}), and at moderately large
$1/x$ and moderately large $Q^{2}$, by the scaling
violations in the free-nucleon cross section,
the nuclear shadowing should
slowly, $\propto log(1/\alpha_{S}(Q))^{-2}$, decrease with $Q^{2}$.
\item Shadowing is $\propto 1/m_{q}^{2}$ and is negligible in the
charm structure function of nuclei.
\end{enumerate}

A comparison of the theoretical predictions [46] for
$\mu He$ and
$\mu Ca$ scattering  with the recent NMC data [74] is shown in
fig.14. We find very good agreement with the experiment.
The rise of shadowing with $1/x$ is a consequnce of freezing
out the large mass intermediate states.


\subsection{Colour transparency and $R=\sigma_{L}/\sigma_{T}$
in the diffractive deep inelastic scattering [{\sl Refs.39,51}]}

A remarkable observation is that by virtue of CT nuclear shadowing
of longitudinal photons vanishes at large $Q^{2}$. The origin
of this phenomenon is in a difference between $|\Psi_{T}|^{2}$
and $|\Psi_{L}|^{2}$, eqs.(\ref{eq:9.3}),\,(\ref{eq:9.4}).
Upon the $z$ integration,
\beq
\int dz d^{2}\vec{\rho}\, |\Psi_{L}(z,\rho)|^{2}\sigma(\rho)
\propto  {1 \over Q^{4}}
\int_{Q^{-2}}^{m_{q}^{-2}}{d\rho^{2} \over \rho^{6}}\sigma(\rho)
\,\, ,
\label{eq:12.2.1}
\endeq
which gives the scaling longitudinal structure function
dominated by $\rho \sim 1/Q$. Notice, this corresponds to
the noncollinear $q\bar{q}$ pairs with $k_{\perp} \sim Q$ -
by virtue of the Callan-Gross relation the collinear
quarks with  $k_{\perp} \ll Q$ can not
contribute to scaling $\sigma_{L}$.

Consider now the shadowing term
\beq
  {1 \over Q^{4}}
\int_{Q^{-2}}^{m_{q}^{-2}}{d\rho^{2} \over \rho^{6}}\sigma(\rho)^{2}
\propto {1 \over Q^{2}} \sigma_{L}
\,\, ,
\label{eq:12.2.2}
\endeq
which vanishes in the scaling limit. Therefore, CT predicts
nuclear enhancement of $R=\sigma_{L}/\sigma_{T}$, fig.15, the principle
source of which is nuclear shadowing of $\sigma_{T}$ [51].
The available experimental data [75,76]
refer to large values of $x$, where
the predicted effect is too small.


\subsection{Nuclear shadowing and fusion of partons
[{\sl Refs.31,38,39}]}

One more comment on the nuclear shadowing is in order.
In the $\nu$-fold
scattering diagrams of fig.16  the $q\bar{q}$ pair can be
attributed to none of the $\nu$ nucleons of the nucleus,
it is shared by them all. This
corresponds precisely to the classic parton fusion mechanism of
nuclear shadowing formulated in 1975 by V.I.Zakharov and
myself [31]: In the Breit
frame the relativistic  nuclei are Lorentz contracted,
\beq
(\Delta z)_{A} \propto R_{A}m_{N}/p \,\, ,
\label{eq:12.3.1}
\endeq
 whereas the sea quarks
with the momentum $k=xp$ always have a longitidunal
localization
\beq
(\Delta z)_{sea} \propto 1/xp \,\, .
\label{eq:12.3.2}
\endeq
 Sea quarks
of nucleons at the same impact parameter will overlap
spatially if $x \lsim (R_{A}m_{N})^{-1}$ and can fuse, cf.
eq.(\ref{eq:9.1}).
Above I have explicitly proven the shadowing is dominated by
the large, hadronic, size $q\bar{q}$ pairs, so that the
fusing partons
have {\sl small} transverse momenta $k_{\perp} \sim m$, not
$k_{\perp} \propto \sqrt{Q^{2}}$.
This is a direct consequence of CT. An important implication
is that the conventional probabilistic assumption [77]
that the fusion contribution to the structure function probed
at the virtuality $Q^{2}$ is proportional to the product of
the parton densities $q_{1}(x,Q^{2})*q_{2}(x,Q^{2})$ at the
same virtuality $Q^{2}$ multiplied by the small fusion
probability $1/Q^{2}$, is not born out by the
quantum-mechanical treatment of the fusion process.
Such a contribution is present in our shadowing correction too,
but it is the $\propto 1/Q^{2}$ correction to the driving
term of the scaling  nuclear shadowing.


\subsection{Colour transparency
and diffractive dissociation into jets on nucleons and
nuclei [Refs.40,44,59,60]}

The quantity closely related to the nuclear shadowing in
deep inelastic scattering is the inclusive forward  cross
section of diffraction dissociation of the photon into
the two jets
\beq
\gamma^{*} \rightarrow q+\bar{q} \rightarrow jet(q)+jet(\bar{q})\, \, ,
\label{eq:12.4.1}
\endeq
which equals
\beq
\left.{d \sigma_{D}(\gamma^{*} \rightarrow X) \over dt }\right|_{t=0}=
\int dM^{2}
\left.{d \sigma_{D}(\gamma^{*} \rightarrow X) \over dt dM^{2} }
\right|_{t=0}=
{1\over 16\pi}
\int dz d^{2}\vec{\rho}\, |\Psi_{T}(z,\rho)|^{2}\sigma(\rho)^{2}
\label{eq:12.4.2}
\endeq
and is reminiscent of the shadowing term.
The mass $M$ of the diffractively produced state is given by
\beq
M^{2} = {\vec{k}^{2} + m_{q}^{2} \over z(1-z) }
\label{eq:12.4.3}
\endeq
where $\vec{k}$ is the transverse momentum of the produced
quark.

The nuclear shadowing and the diffraction dissocation cross
section are proportional to each other [9].
As we have seen above,
eq.(\ref{eq:12.2.2}), the total diffraction dissociation cross
section is dominated by $\rho \sim 1/m_{q}$, i.e., by
production of jets aligned along the photon's momentum,
$\vec{k}^{2} \sim 1/\rho^{2} \sim m_{q}^{2}$, which leads to
the mass spectrum [40,44]
\beq
\left. {d\sigma_{D} \over dM^{2}dt}\right|_{t=0}
\propto {1 \over m_{q}^{2}}{1 \over ( M^{2}+Q^{2})^{2} }
\label{eq:12.4.5}
\endeq
The strong flavour
dependence of the diffraction cross section (\ref{eq:12.4.5})
is particularly noteworthy: because of CT the QCD pomeron does not
factorize [40,44]. The law (\ref{eq:12.4.5}) holds for
excitation of the $q\bar{q}$ Fock components of the photon.
At very large masses $M^{2} \gg Q^{2}$ diffraction excitation
of the $q\bar{q}g$ states with the (triple-Pomeron) mass
spectrum $\propto 1/M^{2}$ takes over [40].

Consider now diffraction dissociation into the two high-$\vec{k}$
jets. Very crude, but quick derivation of the $M^{2}$ and $\vec{k}^{2}$
dependence is making the following substitutions directly
in the integrand of (\ref{eq:12.4.2}):
\beq
\rho \sim {1/k}
\label{eq:12.4.6}
\endeq
and
\beq
dz\,d\rho^{2} \sim {d M^{2} \over (M^{2} + Q^{2})^{2}}
{dk^{2} \over k^{2}}\,\, .
\label{eq:12.4.7}
\endeq
This yields the mass and $\vec{k}^{2}$ spectrum of diffraction
dissociation into the two jets with the total invariant mass
$M^{2} \gsim Q^{2}$ (detailed derivation and discussion of the
region of $M^{2} \lsim Q^{2}$ is given in [40])
\beq
\left.{d \sigma_{D}(\gamma^{*} \rightarrow jet(q)+jet(\bar{q}))
\over dt dM^{2} dk^{2}}\right|_{t=0}
\propto
{1 \over (M^{2} + Q^{2})^{2}} {dk^{2} \over k^{4}} \,\, .
\label{eq:12.4.8}
\endeq

Notice, that diffraction dissociation of
photons can be treated as production
of jets in the photon-pomeron collision
\beq
\gamma^{*}+\Pom \rightarrow jet(q)+ jet(\bar{q})
\label{eq:12.4.9}
\endeq
The result (\ref{eq:12.4.8}) shows that in the quasi-two-body
reaction (\ref{eq:12.4.9}) the pomeron behaves almost
like an elementary particle [40]. This elementary-particle
like behaviour of the pomeron is quite misleading, though.
Its origin is quite subtle: To the lowest order in
the perturbative QCD the pomeron-exchange is an exchange by the
two {\sl uncorrelated} gluons. Large-$\vec{k}$ jets are produced
when these two gluons are exchanged between the two separate quarks of
the $q\bar{q}$ Fock state and the same quark of the
target [40]. The observed momentum of high-$\vec{k}$
jets comes from the large transverse momentum of the exchanged
gluons. Diffraction dissociation of photons is a
somewhat special case,
since for the singular behavior of the photon wave function,
$\varepsilon^{2}K_{1}(\varepsilon\rho)^{2} \propto 1/\rho^{2}$,
the intrinsic momentum $\kappa$ of quarks is large too,
although the dominant contribution to comes from
$\kappa \lsim k$. This justifies crude substitutions
 (\ref{eq:12.4.5}), (\ref{eq:12.4.6}).

Generalization of (\ref{eq:12.4.2}) to the incoherent diffraction
dissociation on nuclear targets reads [59] (cf. eq.(\ref{eq:10.1.2}))
\arr
\left.{d \sigma_{D}(\gamma^{*}A \rightarrow jet(q)+jet(\bar{q})+A^{*})
\over dt }\right|_{t=0}=~~~~~~~~~~~~~~~~~~~~~\nonumber\\
{1\over 16\pi}
\int d^{2}\vec{b}T(b)
\int dz d^{2}\vec{\rho}\, |\Psi_{T}(z,\rho)|^{2}\sigma(\rho)^{2}
\exp\left[-\sigma(\rho)T(b)\right]
\label{eq:12.4.10}
\endarr
Leading term of the nuclear shadowing in (\ref{eq:12.4.10}) is
proportional to the $q\bar{q}$ expectation value of
$\sigma(\rho)^{3}$. Evidently, by virtue of eq.(\ref{eq:12.4.5}),
the nuclear shadowing vanishes
for the large $k^{2}$ jets irrespective of the mass $M$:
the predicted $A^{1}$ dependence for the incoherent
diffraction dissociation
into two high-$\vec{k}$ jets is a direct consequence of CT
(this observation was made in 1987 by Kopeliovich in the
context of diffraction dissociation of hadrons on nuclei [60]).
Nuclear shadowing vanishes also in the coherent diffraction dissociation
into the two high-$\vec{k}$ jets (for the dicussion of CT in
the coherent photoproduction of charmonium see [49]).

Vanishing nuclear attenuation of diffractive nuclear production
of high-$\vec{k}$ jets only holds for the two jet reactions
(\ref{eq:12.4.1}). In the triple-Pomeron region of very large
mass states, diffractive excitation of the $q\bar{q}g$ Fock
states by exchange of the two hard gluons gives rise to
the two sideways jets, which will typically be accompanied
by the soft forward jet. Whenever the forward jet is present
in the final state, it will have the conventional
nuclear attenuation. However, if all the three jets are produced
sideways, then shadowing should vanish again.

Diffractive excitation of the open charm is also an interesting
reaction. According to [40], nuclear shadowing in the
photo- and electroexcitation of charm is very weak, see
section 12.1, and
precisely the same mechanism suppresses the nuclear
shadowing in diffraction dissociation into the jets initiated
by charmed quarks. An interesting prediction [40] is that excitation
of charm contributes $\sim 10 \%$ of the high-$\vec{k}$ jet
production compared to $\sim 1.5\%$ in the total diffraction
dissociation rate: the strong flavour dependence (\ref{eq:12.4.5})
of the $k^{2}$ intergated cross section disappears at
$k^{2} \gg m_{q}^{2}$.
\medskip\\

Now I change the subject and turn to other candidate
reaction: quasileastic $(e,e'p)$ scattering on nuclei.


\section{QCD observables of colour transparency
 $(e,e'p)$ ex\-pe\-ri\-ment [{\sl Refs.5,6,45}]}


\subsection{The ejectile state has a transverse size
identical to the free proton's size [{\sl Ref.45}]}

Firstly, I demonstrate that the transverse size of the
ejectile state probed at short time after its production is
exactly equal to the target proton size.
Indeed, in NRQM the ejectile wave function equals
(here $\vec{r}=(\vec{\rho},z)$,
the $\vec{\rho}$-plane is normal to the momentum
transfer $\vec{Q}$)
\beq
\Psi_{E}(\vec{\rho},z)=
\exp\left({i\over 2}Qz\right)\Psi_{p}(\vec{\rho},z)
\label{eq:13.1.1}
\endeq
The major idea of the CT experiments is to probe the ejectile
state by its interaction with another nucleon of the nucleus
at a time scale small compared to the wave packet
evolution
time. Under these conditions the struck quark acquires large
longitudinal momentum but retains its transverse coordinate.
Consequently, the wave packet  (\ref{eq:13.1.1}) which hits the
second nucleon has the transverse (and longitudinal) size
distribution identical to that of the initial proton:
$|\Psi_{E}(\vec{\rho},z)|^{2}= |\Psi_{p}(\vec{\rho},z)|^{2}$.
As a result,
$\rho_{E}^{2}=\langle E|\rho^{2}|E\rangle=\langle p|\rho^{2}|p\rangle$,
and
\beq
\sigma_{tot}(EN)=\langle E|\hat{\sigma}|E\rangle=
\langle p|\hat{\sigma}|p\rangle=\sigma_{tot}(pN)\,.
\label{eq:13.1.2}
\endeq
The ejectile state probed at the production vertex has exactly the
same interaction cross section as the free proton !
This casts a shadow on predictions of CT effects which explicitly
use an assumption of vanishing size and vanishing
cross section $\sigma_{tot}(EN)=0$ of the ejectile state [70,78-81].
For instance, Jennings and Miller assume $\hat{\sigma}|E\rangle=0$
[78,80,81].


\subsection{Strength of final state interaction:
the nonrelativis\-tic mo\-del [{\sl Ref.45,6}]}


In the absence of
FSI the reaction amplitude ( impulse approximation (IA) amplitude)
is given by the electromagnetic
form factor $G_{em}(Q)=\langle p|E\rangle=
\langle p|J_{em}(Q)|p\rangle$. The driving term of FSI
is the single-rescattering amplitude
which is proportional to the matrix element
$\langle p|\hat{\sigma}J_{em}|p\rangle =
\langle p|\hat{\sigma}|E\rangle$. Therefore, the strength of
FSI can conveniently be characterized by the observable
$\Sigma_{ep}$  defined following  eq.(\ref{eq:10.1.8}):
$\Sigma_{ep}=\langle p|\hat{\sigma}|E\rangle/\langle p|E\rangle$.
In order to elucidate the origin of weak FSI and the
relationship between weak FSI and CT sum rules, we
shall demonstrate major ideas within the nonrelativistic
quantum mechanical (NRQM) model of the $\bar{q}q$ state.

As we have proven above, the ejectile state probed at the
production vertex has a transverse size identical to the proton's
size.
The emergence of weak FSI is a property of semiexclusive reaction
when the full-size ejectile state (\ref{eq:13.1.1})
is projected onto the observed full-size proton.
This projection gives the  electromagnetic form factor (FF)
\beq
G_{em}(\vec{Q})=\int d^{2}\vec{\rho} \int dz |\Psi (z,\vec{\rho})|^{2}
\exp\left(-{i\over 2}Qz\right)
=\int {d^{3}\vec{k} \over (2\pi)^{3} }
\varphi^{*}(\vec{k}+{1\over 2}\vec{Q})\varphi(\vec{k})        \, .
\label{eq:13.2.1}
\endeq
where $\varphi(\vec{k})$ is the momentum-space wave function.
Since $\hat{\sigma} \propto \rho^{2}$, the alternate measure of
the strength of FSI is  $\left<\rho^{2}\right>
=\langle p|\rho^{2}|E\rangle/\langle p|E\rangle$ (which should not
be confused with $\rho_{E}^{2}$)
\arr
\left<\rho^{2}\right>= {1 \over G_{em}(Q)}
\int d^{2}\vec{\rho} \rho^{2}\int dz |\Psi (z,\vec{\rho})|^{2}
\exp(-i{1 \over 2}Qz) \nonumber\\
= {1 \over F_{em}(Q)} \int {d^{3}\vec{k} \over (2\pi)^{3} }
\left[ \partial_{\vec{k}_{\perp}}
\varphi^{*}(\vec{k}+{1\over 2}\vec{Q})\right]
\left[ \partial_{\vec{k}_{\perp}}\varphi(\vec{k})\right]           \, .
\label{eq:13.2.2}
\endarr
With the Gaussian Ansatz for the NRQM wave function
the $d^{2}\vec{\rho}$ and
$dz$ integrations in eqs.(\ref{eq:13.2.1}),(\ref{eq:13.2.2}) do factorize,
so that $\langle \rho^{2}\rangle=R^{2}$ and does not depend on $Q$:
the CT law
$\sigma(\rho)\propto \rho^{2}$ does not guarantee vanishing FSI !

In the second thought, neither $Q$ independent nor rising with $Q$
strength of FSI contradicts any general principles. Indeed,
in the expansion over the intermediate states
\beq
\langle p|\hat{\sigma}|E\rangle =
\sum_{i}\langle p|\hat{\sigma}|i\rangle
\langle i|J_{em}(Q)|p\rangle
\label{eq:13.2.3}
\endeq
the matrix elements of the diffractive operator
$ \langle p|\hat{\sigma}|i\rangle $
do not depend on energy and/or $Q$. Furthermore, the larger
is $Q$ the heavier are the electroproduced
intermediate states $|i\rangle$, which have the ever growing
radius and interaction cross section.

Now I shall demonstrate that, none the less, QCD
predicts that $\Sigma_{ep}(Q)$  vanishes
at large $Q^{2}$. I shall illustrate the basic idea
of the proof starting with the Schr\"odinger equation in the
momentum-representation:
\beq
\varphi(\vec{Q})= {1 \over \varepsilon - \vec{Q}^{2}/ 2m}
\int {d^{3}\vec{k} \over (2\pi)^{3}} V(\vec{Q}-\vec{k})\varphi(\vec{k}) \, .
\label{eq:13.2.4}
\endeq
On the hadronic size, the dominant feature of $\bar{q}-q$ interaction
is the confining potential (funnel potential, harmonic oscillator potential
etc.). At short distances  (anti)quarks interact
via one-gluon exchange Coulomb
interaction, which  can be treated as weak perturbation.
Let $\psi(\vec{r})$ and $\phi(\vec{k})$ be wave functions in the
confining potential.
Then, $\phi(\vec{k})$ is a steep function of $\vec{k}^{2}$,
which vanishes rapidly at $R_{p}^{2}k^{2} \gsim 1$.
On the other hand, the asymptotics of the Fourier transform of
the potential $V(\vec{k})$ is dominated by the QCD one-gluon exchange
contribution $V(\vec{k}) \propto 1/\vec{k}^{2}$ and is a slow
function of $\vec{k}^{2}$
(the logarithmic factor $\alpha_{S}(k)$  is not significant
for our purposes). Consequently, at $R_{p}^{2}Q^{2} \gg 1$, one can
neglect $\vec{k}$ compared to $\vec{Q}$ in $V(\vec{Q}-\vec{k})$
in the {\sl r.h.s.} of (\ref{eq:13.2.4}), and [82]
\beq
\varphi(\vec{Q})\propto {V(\vec{Q}) \over \vec{Q}^{2}}\psi(0)
\label{eq:13.2.5}
\endeq
In the same region of $Q$, the asymptotics of FF will be dominated by
contributions when one of the wave functions
in eq.(\ref{eq:13.2.1}) will be in the Gaussian-like
(confining) regime and the second in Coulomb tail regime
(\ref{eq:13.2.5}).
Then, in (\ref{eq:13.2.1}) I can
neglect $\vec{k}$ compared to $\vec{Q}$ in
$\varphi(\vec{k}+{1\over 2}\vec{Q})$ (this corresponds to
the hard scattering condition that the recoil energy of the struck
quark is much larger than its kinetic energy),
factor out $\varphi({1\over 2}\vec{Q})$,  and will find
\beq
G_{em}(\vec{Q})
\propto {V(\vec{Q}) \over \vec{Q}^{2}}|\psi(0)|^{2}
\label{eq:13.2.6}
\endeq
Generlization of (\ref{eq:13.2.6}) to the transition form factor
is straightforward (notice that $\psi(0)\approx \Psi(0)$):
\beq
G_{ik}(\vec{Q})
\propto {V(\vec{Q}) \over \vec{Q}^{2}}\Psi_{i}^{*}(0)\Psi_{k}(0)
\label{eq:13.2.7}
\endeq
The asymptotics (\ref{eq:13.2.6}), (\ref{eq:13.2.7})
 holds provided that
\beq
Q^{2} \gg 2m\langle T_{i,k} \rangle
\label{eq:13.2.8}
\endeq
where $\langle T_{i} \rangle$ is the mean kinetic energy of quarks
in the state $|i\rangle$.

In the same "QCD-dominated" regime,
the $\partial_{\vec{k}_{\perp}}$
differentiations of the Colulomb tail and the confining
wave function in (\ref{eq:13.2.2}) lead
to
$\partial_{\vec{k}_{\perp}}\varphi(\vec{k}+{1\over 2}\vec{Q})
\sim \varphi(\vec{k}+{1\over 2}\vec{Q})\vec{k}/\vec{Q}\,^{2}$
and $\partial_{\vec{k}_{\perp}}\phi(\vec{k})
\sim \phi(\vec{k})R^{2}\vec{k}$, so that
we obtain $\left<\rho^{2}\right> \propto 1/Q^{2}$.
The major conclusion [45] from this analysis is that
QCD indeed predicts vanishing strength of FSI
in the quasielastic $(e,e'p)$ scattering, which has its
origin in CT law (\ref{eq:3.4}) and in the QCD Coulomb
interaction in hadrons responsible  for the power-law asymptotics
of electromagnetic FF's.

Making use of the asymptotics (\ref{eq:13.2.7}) in expansion over
the intermediate states (\ref{eq:13.2.3}),
\arr
\Sigma_{ep}(Q) = {1 \over G_{em}(Q)}
\sum_{i}\sigma_{pi}G_{ip}(Q)~~~~~~~~~~~~~~~~~~~~~ \nonumber \\
\propto {1\over G_{em}(Q)}{V(\vec{Q}) \over \vec{Q}^{2}}
\sum_{i}\sigma_{pi}\Psi_{i}^{*}(0)\Psi_{p}(0)
\propto \sum_{i}\sigma_{pi}\Psi_{i}^{*}(0)
\label{eq:13.2.9}
\endarr
we find that the vanishing of $\Sigma_{ep}(Q)$ is related to
CT sum rule (\ref{eq:5.1.7})!
I emphasize that matrix elements $\sigma_{ik}$ of
the diffraction operator are large and do not depend on $Q$.
Eq.(\ref{eq:13.2.9}) demonstrates explicitly
how in QCD the {\sl perturbative} mechanism of hard electromagnetic
scattering, quantified by eqs.(\ref{eq:13.2.6}),(\ref{eq:13.2.7}),
conspires with the ({\sl nonperturbative}) mechanism of
diffractive scattering and produces weak FSI.


\subsection{The effective size of the ejectile [{\sl Ref.6}]}


The statement that
$|\Psi_{E}(\vec{r})|^{2}=|\Psi_{p}(\vec{r})|^{2}$ is an exact
result, if all states $|i\rangle$ are allowed in
expansion (\ref{eq:2.1}). However, in the intranuclear
rescattering problem
only the intermediate states $|i\rangle$ with excitation
energy $\Delta E_{i} \lsim Q^{2}/2m$ can contribute
at finite $Q$, which imposes a kinematical constraint
essentially identical to the constraint (\ref{eq:13.2.8}).
As I have shown in section 6.1, in the relativistic case of the
interest the coherence constraint $m_{i}^{2} \lsim Q^{2}/R_{A}m_{p}$
eq.(\ref{eq:6.3.14}) is much more stringent than the
corresponding kinematical constraint  $m_{i}^{2} \lsim Q^{2}$ .
With allowance  for this
nuclear filtering by the coherence constraint one can confine
himself to a subset of $N_{eff}(Q)$ lowest excitations, for which the
asymptotics (\ref{eq:13.2.7}) is a good approximation.
Projecting the ejectile state onto this subset of lowest
excitations, one finds the effective ejectile state
\beq
\Psi_{E}^{(eff)}(\vec{r}) \propto
 \psi_{p}(0)\sum_{i}^{N_{eff}}\Psi_{i}^{*}(0)\Psi_{i}(\vec{r})
\label{eq:13.3.1}
\endeq
Compare this with an expansion of the delta-function
\beq
\delta(\vec{r})=\sum_{i}\Psi_{i}^{*}(0)\Psi_{i}(\vec{r})
\label{eq:13.3.2}
\endeq

I obtained a seemingly {\sl paradoxial and
remarkable result}: The true ejectile state as probed
at the production vertex has {\sl exactly the same} transverse size
and the same interaction cross section as the free proton.
Asking for the quasielastic scattering, one projects this
ejectile state onto the final state proton. This projection,
in concert with the coherence condition, imposes a constraint
on the subset of $N_{eff}$ frozen out intermediate states
which could contribute coherently
to the intranuclear FSI. The perturbative QCD mechanism
of the hard electromagnetic scattering leads to the effective
ejectile state (\ref{eq:13.3.1}) which on the basis of $N_{eff}$
states is the best approximation to $\delta(\vec{r})$. Because
$N_{eff}(Q)$ rises with $Q$, and since the coherence constraint
guarantees that the form factors of the contributing
states are always in the asymptotic regime (\ref{eq:13.2.7}),
the effective ejectile state has the ever decreasing transverse
size as $Q$ is increased. Similarly, at finite
$Q$ the sum in the second
line of eq.(\ref{eq:13.2.9}) includes only the states $|i\rangle$
which satisfy the coherency constraint (\ref{eq:6.3.14}),
so that $\Sigma_{ep}(Q)$
does not vanish, but decreases gradually with the
increase of the number of frozen out intermediate states,
see section 15.


\subsection{Strength of final state interaction:
the relativistic model [{\sl Ref.45}]}


The above connection between vanishing FSI and QCD origin of
the power-law asymptotics of electromagnetic form factors holds
in the relativistic case too. Here I outline major ideas
of the QCD light cone analysis of form factors,
referring the readers to the extensive literature on the
subject (for the review and references see [83-85]).

In the relativistic case of $Q^{2} \gg m^{2}$ one must use
the Drell-Yan light-cone wave functions (LCWF).
Let $x=k_{+}/p_{+}$ be a fraction of the (light-cone) momentum
$p_{+}=E+p_{z}$ of the meson carried by the struck quark and
let $\vec{Q}$ be along the $y$-axis. The Drell-Yan formula
for the form factor reads [86,87].
\arr
F_{em}(Q) = \int{d^{2}\vec{k}\over (2\pi)^{2}}
{dx \over x(1-x)}
\varphi^{*}(M_{f}^{2})
\varphi(M_{in}^{2}) \nonumber\\
=
\int_{0}^{1} dx \int d^{2}\vec{\rho}
\exp[-i(1-x)\vec{\rho}\vec{Q}]
|\Psi(x,\vec{\rho})|^{2}
\label{eq:13.4.1}
\endarr
where the invariant (radial) variables  of LCWF equal [87]
\arr
M_{in}^{2} = {m_{q}^{2}+\vec{k}^{2}\over x(1-x)}~~~~~~~~~\nonumber\\
M_{f}^{2} =  {m_{q}^{2}+(\vec{k}+(1-x)\vec{Q})^{2}\over x(1-x)}
\label{eq:13.4.2}
\endarr

I am interested in the transverse size of the ejectile state
in the laboratory frame, in which the target proton at rest
receives the purely longitudinal momentum transfer. The Drell-Yan
formula holds only in the
light-cone frame in which the momentum transfer $\vec{Q}$
is purely transversal and in which the so-called pair creation
diagrams do not contribute. Fortunately, the Lorentz transformation
from the light-cone frame to the laboratory frame
leaves intact $\rho_{x}$, so that one can evaluate the strength
of FSI computing $\langle \rho_{x}^{2}\rangle=
\left< \partial_{k_{x}}^{2}\right>$ and making use
of the azymuthal symmetry in the laboratory frame:
$\langle \rho^{2}\rangle = 2\langle \rho_{x}^{2}\rangle$ [45].
Since
$\Psi_{E}(x,\vec{\rho})=\exp[-i(1-x)\vec{\rho}\vec{Q}]
\Psi_{p}(x,\vec{\rho})$, the $\rho_{E}$ retention
property holds in the relativistic case too.
Using  an
explicit form of radial variables eq.(\ref{eq:13.4.2}), we obtain:
\beq
\left<\rho_{x}^{2}\right>= {1\over G_{em}(Q)}
\int { d^{2}\vec{k}\over 2\pi}
{dx \over x(1-x)}
\varphi^{*'}(M_{f}^{2})
\varphi'(M_{in}^{2}) \left[ { 2k_{x} \over x(1-x)} \right]^{2}
\label{eq:13.4.3}
\endeq

With the  Gaussian LCWF
$\varphi(M^{2}) = \varphi_{0}\exp(-{1 \over 2}R^{2}M^{2})$ one
finds from (\ref{eq:13.4.1}) the
form factor $G_{em}(Q) \propto \exp(-RmQ)/Q^{3}$, which
is dominated by the so-called end-point contribution
from
\beq
1-x \sim 2m/Q
\label{eq:13.4.4}
\endeq
and from (\ref{eq:13.4.3}) one finds ([45], see also [81])
\beq
\left<\rho^{2}\right> \sim
\left< { R^{2} \over 1-x } \right> \propto R^{2}{Q \over m_{q}} \, .
\label{eq:13.4.5}
\endeq
This rising with $Q$ strength of FSI is even more striking than $
\left<\rho^{2}\right> = const(Q)$ found with the
nonrelativistic Gaussian wave function.

Introduction of the Coulomb effects in  LCWF is
very much similar to the case of NRQM.
The variable $M^{2}$ plays the
same role as $\vec{k}^{2}$, so that in the relativized version
of eq.(\ref{eq:13.2.4}) one should simply substitute
$\vec{Q}^{2}-2mE$ for $M^{2}-m_{h}^{2}$.  The slight difference
from NRQM is that the nonrelativistic Coulomb scattering amplitude
$V(\vec{Q}) \sim 4m^{2}/Q^{2}$ in which $Q^{2} \ll 4m^{2}$
should be substituted for
by the large angle relativistic Coulomb amplitude
$ V(\vec{Q}) \sim M^{2}/Q^{2}$ with $M^{2} \sim Q^{2}$.
As a result, LCWF of the meson will have the $1/M^{2}$ tail,
the form factor will be dominated by contributions from
regions of $M_{in}^{2} \sim m_{h}^{2}, \, M_{f}^{2} \sim Q^{2}$ (and
vice versa, $M_{f}^{2} \sim m_{h}^{2}, \, M_{in}^{2} \sim Q^{2}$ ),
which leads to the $1/Q^{2}$ FF of mesons ([88], for the
review see [83-85]).
When the transition
form factor $G_{kp}(Q)$ is considered, the above holds if
$Q^{2} \gg M_{k}^{2}$.
In the same asymptotic regime $\varphi'(M_{f}^{2}) \sim
\varphi(M_{f}^{2})/M_{f}^{2}$ and
$\varphi'(M_{in}^{2})\sim
R^{2}\varphi(M_{in}^{2})$, and from eq.(\ref{eq:13.4.3}) I find
$\Sigma_{ep} \propto 1/Q^{2}$.

The most important effect of the one-gluon exchange QCD interaction
in the relativistic case is that it eliminates a dominance of
the end-point, eq.(\ref{eq:13.4.4}), contribution to the FF. The
FF will be dominated by finite, weakly $Q$ dependent,
values of $1-x$. Nevertheless,
a presence of extra $x(1-x)$ in the denominator of
the integrand of (\ref{eq:13.4.3})
enhances a sensitivity of a strength of FSI to the end-point
contribution. Consequently, the onset of
vanishing $\Sigma_{ep}$ is slower than the onset of the power
asymptotics of the FF.


\subsection{Strength of final state interaction:
the model estimates [{\sl Ref.45}]}

Evidently, the onset of CT depends on the relative strength of
the QCD `Coulomb' and confining interactions in hadrons.
To this end, one must be reminded of the observation by
Isgur and Llewellyn Smith on
the very late onset of the QCD power-law
asymptotics of the electromagnetic FF [89].
In fig.17 I present the results from the light-cone
toy model of the scalar pion with the scalar quarks,
which incorporates the principal features of the QCD wave function:
\beq
\varphi(M^{2}) \propto \exp(-{1 \over 2}R^{2}M^{2}) +
\alpha {1 \over (1+R^{2}M^{2}/2)^{n}} \, .
\label{eq:13.5.1}
\endeq
For our purposes, the presence of $\alpha_{S}(M^{2})^{n}$
factor in the `Coulomb' correction is not
essential. To the first order in the `Coulomb' correction the
$\propto 1/Q^{2}$ behaviour of the pion FF
 corresponds
to $n=1$.

Experimentally, the pion FF follows the simple
$\rho^{0}$-dominance monopole formula
$F_{em}(Q)=1/(1+Q^{2}/m_{\rho}^{2})$ up
to $Q^{2}\sim 10\,(GeV/c)^{2}$ [90].
Therefore, in all cases I adjust $R^{2}$ to
$\left<R_{ch}^{2}\right>= 6/m_{\rho}^{2}$.
The Coulomb-admixture
parameter $\alpha$ controls the large-$Q$ normalization
$\Lambda^{2}=Q^{2}F_{\pi}(Q)$, changing from $\Lambda^{2}\rightarrow 0$
for the Gaussian LCWF, to
$\Lambda^{2}\approx m_{\rho}^{2}$ for the
$\rho^{0}$-dominated monopole FF (Since neither the
running QCD coupling, nor the the higher
order QCD effects like the Sudakov FF [91], see below,
are included, my toy model (\ref{eq:13.5.1}) should not
be extended up to very large $Q^{2}$).
As a case in
between I consider LCWF's giving
$\Lambda^{2}\sim 8\pi \alpha_{S}f_{\pi}^{2} \sim 0.1 (GeV/c)^{2}$,
as suggested by the perturbative QCD [89,85].
In the problem of interest it is natural
to use the effective mass of quarks $m_{q}$ somewhere
in between the spectroscopic constituent mass of
$\sim 300\, MeV/c^{2}$ and the current mass $\sim 10\, MeV/c^{2}$.
Here I use $m_{q}=150 \, MeV/c^{2}$.

The principle new findings are:
\begin{enumerate}
\item With the purely Gaussian LCWF,
$\alpha=0$, $R=1.92\, (GeV/c)^{-1}$,
the FF $F_{em}(Q) \propto  \exp(-R^{2}m_{q}Q)/Q^{3}$
follows closely the
monopole FF up to $Q^{2} \sim 5 \,(GeV/c)^{2}$ and
gives the rising strength of FSI.
\item With
$\alpha = 0.1$, $R =\, 2.23 (GeV/c)^{-1}$ one reproduces
the monopole FF. Notice, that this is achieved with the relatively
weak Coulomb correction: $\alpha \ll 1$.
The strength of FSI stays approximately
constant up to $Q^{2} \sim 10-20 (GeV/c)^{2}$, and then decreases rapidly
with $Q^{2}$.
\item The FF with
$\alpha = 0.01$, $R = 1.95\, (GeV/c)^{-1}$ follows the mono\-po\-le FF
up to $Q^{2} \sim 10 \,(GeV/c)^{2}$ and has the large-$Q$
normalization close to the perturbative QCD prediction.
The strength of FSI first
increases  up to a saturation at
$Q^{2} \sim 30-40\, (GeV/c)^{2}$ and
steeply decreases beyond $ \sim 100\, (GeV/c)^{2}$.
\item With the more conservative
$m_{q}=300 MeV/c^{2}$ ( $\alpha=0.01$ and $R=2.72 \,(GeV/c)^{-1}$)
the strength of FSI stays approximately
constant up to $Q^{2} \sim 10 \,(GeV/c)^{2}$ and then starts
decreasing. However, the large-$Q$ normalization of this
FF drops to the perturbative QCD value too rapidly,
in conflict with the experiment [13].
\end{enumerate}

I conclude that the steeper is a decrease of the FF
and the smaller is the large-$Q$ normalization,
the slower is an onset of CT regime. With realistic
LCWF's FSI might remain strong up to a very large
$Q^{2}$. An interesting observation is that a strength of FSI
is very sensitive to the quark structure of the hadron.
The QCD asymptotics $\Sigma_{ep} \propto 1/Q^{2}$ should be even
more elusive in the electron-proton scattering [89,85].


\subsection{Filtering the small size ejectile:
the Sudakov form factor.}

Besides the short distance one-gluon exchange
interaction between (anti)quarks, there is a still another
QCD mechanism which filters the small-size components of
hadrons in $ep$ scattering: the Sudakov form factor [91].
Asking for the elastic scattering one asks for the no-radiation
of gluons. For the quark-antiquark system of size $\vec{\rho}$,
hit with the momentum transfer $\vec{Q}$,
a mean number of the would-be radiated gluons $N_{g}$, which
would have taken a fraction $x_{g}$ of the hadrons momentum
$x_{min} < x_{g} < 1 $, equals [46]
\beq
N_{g}(\rho,Q^{2},x_{min})
\approx
{16 \over 3}\int _{x_{min}}^{1}
{dx \over x }
\int_{0}^{Q^{2}}
{d^{2}\vec{k} \over 2\pi}
{\alpha_{S}(\vec{k}^{2})\over 2\pi}
{1 - \exp(-i\vec{k}\vec{\rho}) \over \vec{k}^{2} }
\, .
\label{eq:13.6.1}
\endeq
Then, a probability of the no-radiation, alias the Sudakov FF,
can be estimated as
\beq
F_{S}(\rho,Q,x_{min}) \approx \exp\left[-N_{g}(\rho,Q^{2},x_{min})\right]
\label{eq:13.6.2}
\endeq
and the Sudakov-modified charge FF (\ref{eq:13.4.1}) takes the form
\beq
F_{em}(Q)=\int_{0}^{1} dx \int d^{2}\vec{\rho}
\exp[-i(1-x)\vec{\rho}\vec{Q}]
|\Psi(x,\vec{\rho})|^{2}F_{S}(\rho,Q,1-x)
\label{eq:13.6.3}
\endeq
(for a more rigorous derivation
and references to the recent work on the Sudakov FF
see [92,93])

The resulting  Sudakov suppression of the
large $\rho$ contribution depends on
the LCWF. With the Gaussian
LCWF we have the end-point eq.(\ref{eq:13.4.4}) dominance,
so that $2\log(1/x_{min})=\log(Q^{2} /m^{2})$ and
the Sudakov FF takes on the standard form
\beq
F_{S}(\rho,Q,x_{min})  \approx
\exp\left\{-{8 \over 27}\log\left({Q^{2} \over m_{q}^{2}}\right)
\log\left[
{\alpha_{S}(1/\rho^{2}) \over \alpha_{S}(Q^{2}) }
\right]\right\}
\label{eq:13.6.4}
\endeq
 Eq.(\ref{eq:13.6.4})
holds before the onset of the Coulomb dominance regime,
when $x_{min}$ will become small but {\sl finite}
and will very weakly depend on $Q$. In this regime the
Sudakov suprression will become a very slow function of
$Q$ and $\rho$ and its effect on the rate of
vanishing of $\Sigma_{ep}$ will be marginal.

Here I must emphasize that
at moderately large $Q^{2}$ the considerations of the last
two sections are of very academic value, though.
Even if eq.(\ref{eq:13.4.3}) with the `good' relativized
wave function will yield $\Sigma_{ep}$ which rapidly decreases
with $Q^{2}$, or even if the Sudakov effects will rapidly
shrink the region of $\rho$ the dominant contribution to
the form factor (\ref{eq:13.6.3})
 comes from [93], it would be premature to
expect a precocious CT regime.
As we have discussed in sections 5,\,6  and 11, the coherency
constraint (\ref{eq:6.3.14}) puts stringent limitations
on the number of intermediate states $N_{eff}(Q)$, which
can contribute to, and control the $Q^{2}$ dependence of,
the QCD observable $\Sigma_{ep}$, see section 15.2.
The principle effect of the Sudakov form factor is that it
enhances a relative contribution from the Coulomb tail of
the wave function and leads to a somewhat sooner onset of
the Sudakov-improved QCD asymptotics (\ref{eq:13.2.7}).




\section{Multiple scattering theory of final state interaction
in $(e,e'p)$ scattering}


\subsection{Final state interaction and the semiexclusive cross
section}

We are interested in the
quasielastic (semiexclusive)
$e+A \rightarrow e+p+A^{*}$ reactions, when one
sums over all the excitations of the target debris $A^{*}$,
excluding   production of secondary particles (mesons).

In the familiar PWIA
the momenta of the recoil electron $e'$ and of the ejected proton
$p$ dofix the Fermi momentum $\vec{k}$ of the quasifree proton.
In the DWIA, with allowance for
 FSI of the struck proton with the target
debris
a direct kinematical determination of the target proton's momentum
is impossible
(for the review and references see, for instance, [94,95], I shall
comment
more on that in section 18).
Summing over all the final states of the target
debris $A^{*}$ amounts to integration over all $(\omega,\vec{k})$, and
the Fermi-momentum uncertainty does not affect the semiexclusve scattering
rate.

For instance, the matrix
element of $(e,e'e)$ scattering on the bound electron of
the hydrogen atom is proportional to $\Phi(\vec{k})/Q^{2}$, where
\beq
\Phi(\vec{k})=\int d^{3}\vec{r}\psi(\vec{r})\exp(-i\vec{k}\vec{r})
\label{eq:14.1.1}
\endeq
and the momentum transfer squared $Q^{2}$ and the
Bjorken varible $x_{Bj}=Q^{2}/2m\nu$ are uniquely fixed by the
electron scattering kinematics.
The semiexclusive cross section, integrated over the ejectile's
momentum, is proportional to
\arr
d\sigma \propto \int{ d^{3}\vec{p}\over (2\pi)^{3}}|M|^{2}
\propto \int d^{3}\vec{r}\,' d^{3}\vec{r}
{ d^{3}\vec{k}\over (2\pi)^{3}}
\psi^{*}(\vec{r}\,')
\psi(\vec{r})\exp[-i\vec{k}(\vec{r}-\vec{r}\,')])\nonumber \\
=
\int d^{3}\vec{r}
|\psi(\vec{r})|^{2}
=\int d^{2}\vec{b}\int dz\,n_{e}(\vec{b},z)=1
\label{eq:14.1.2}
\endarr
and is independent of the detailed form of the `Fermi-motion' of
the bound electron. Here the $\vec{k}_{\perp}$ integration is
straightforward, while $k_{z}$ integration requires scanning
over the Bjorken varibale $x_{Bj}=Q^{2}/2m_{p}\nu$  at fixed $Q^{2}$,
since by the scattering kinematics (we treat the target electron
nonrelativistically)
$Q^{2}=2\nu (m_{p}+k_{z})$ and  $k_{z}=m_{p}(x_{Bj}-1)$.


\subsection{The Glauber model prediction for nuclear trans\-pa\-ren\-cy
[{\sl Refs.5,96}]}

Evidently, the elastic rescattering of the relativistic
ejectile does not absorb the particle and
only broadens the angular distribution of ejectiles
compared to that expected from the transverse Fermi-momentum in the
primary interaction (see section 18.2).
Therefore, we can readily write down a simple
probabilistic formula for the nuclear transparency or the nuclear
transmission coefficient $Tr_{A}^{(Gl)}$, often referred to as
the Glauber model prediction:
\arr
Tr_{A}^{(Gl)}={d\sigma_{A} \over Z d\sigma_{p}}=
{1 \over A}\int d^{2}\vec{b} dz n_{A}(\vec{b},z)
\exp\left[-\sigma_{in}(pN)\int_{z}^{\infty}dz' n_{A}(\vec{b},z')\right]
\nonumber\\
= Tr(\sigma_{in}(pN))={1 \over A\sigma_{in}(pN)}\int d^{2}\vec{b}\left\{1-
\exp\left[-\sigma_{in}(pN)T(\vec{b})\right]\right\}=
{\sigma_{abs}(pA) \over A\sigma_{in}(pN)}
\label{eq:14.2.1}
\endarr
Eq.(\ref{eq:14.2.1}) differs from (\ref{eq:14.1.2}) by presence of the
nuclear attenuation factor and gives the reference value for the
nuclear transparency.

The multiple sacttering theory derivation of (\ref{eq:14.2.1}) goes
as follows:
With allowance for FSI of the struck proton
the matrix element of reaction $eA\rightarrow e'pA_{f}$ reads
\beq
M=\langle f|\exp(-i\vec{k}\vec{r}_{1})
S(\vec{b},\vec{c}_{A},...,\vec{c}_{2})|A\rangle
\label{eq:14.2.2}
\endeq
where
\beq
S(\vec{b},\vec{c}_{A},...,\vec{c}_{2})=
\prod_{i=2}^{A} \left[1-\Gamma(\vec{b}-\vec{c}_{i})\right]
\label{eq:14.2.3}
\endeq
Here I consider the above defined
semiexclusive cross section
$d\sigma =\sum_{f}\int d^{3}\vec{k}|M|^{2}$.
 Applying the closure,
$\sum_{f}|f\rangle \langle f| = 1$, I obtain
\beq
d\sigma_{A}=Zd\sigma_{p}
\langle A|S^{*}(\vec{b},\vec{c}_{A},...,\vec{c}_{2})
S(\vec{b},\vec{c}_{A},...,\vec{c}_{2})|A \rangle
\label{eq:14.2.4}
\endeq
with
\beq
S^{*}(\vec{b},\vec{c}_{A},...,\vec{c}_{2})
S(\vec{b},\vec{c}_{A},...,\vec{c}_{2})
=\prod_{i=2}^{A} \left[1-\Gamma_{in}(\vec{b}-\vec{c}_{i})\right]
\label{eq:14.2.5}
\endeq
where $\Gamma_{in}(\vec{b})=2Re \Gamma(\vec{b})-|\Gamma(\vec{b})|^{2}$
is the profile function of the $pN$ inelastic cross section.
Using the technique of section 6, I find
\arr
Tr_{A}=
{1\over A}\int d^{2}\vec{b} dz_{1}n_{A}(\vec{b},z_{1})
\left[1-{1\over A}\sigma_{in}(pN)
t(b,\infty,z_{1})\right]^{A-1}
\label{eq:14.2.6}
\endarr
where $t(b,z_{2},z_{1})$ is a partial optical thickness
\beq
t(b,z_{2},z_{1})=
\int_{z_{1}}^{z_{2}}dz n_{A}(\vec{b},z) \, .
\label{eq:14.2.7}
\endeq
The exponentiation in the {\sl r.h.s.} of eq.(\ref{eq:14.2.6})
gives eq.(\ref{eq:14.2.1}) with the attenuation given by
the {\sl inelastic}, not the {\sl total},
free-nucleon cross section [5,96].
Above I have followed Glauber and Matthiae [97], who gave
MST description of the quasielastic scattering $pA\rightarrow
pA^{*}$ and have derived in that process an
attenuation with the inelastic cross section.
With certain caution
eq.(\ref{eq:14.2.1}) for the nuclear transparency can be
applied to $\vec{k}_{\perp}$ integrated
differential cross section too, not far off the
quasileastic peak at
$k_{z}=0$, $x_{Bj}\approx 1$ (see section 18 below).


\subsection{Gribov's inelastic shadowing and
nuclear transpa\-ren\-cy in $(e,e'p)$ scattering [{\sl Ref.5}]}

Incorporation of Gribov's  inelastic shadowing using the coupled-channel
technique [14] is straightforward in the
high-energy limit.
Consider the reaction
$eA\rightarrow e'kA_{f}^{*}$, where $|k\rangle$
stands for the proton or its
excitations: $|k\rangle=|p\rangle$, $|p^{*}\rangle$, $|N\pi\rangle,....$.
Introduce
the diffraction-eigenstate expansions
of the observed baryonic state
$|k\rangle=\sum_{\alpha}a_{\alpha}|\alpha \rangle$
and of the ejectile state
$ |E\rangle = \sum_{\alpha}c_{\alpha}|\alpha \rangle $.
The diffraction eigenstates diagonalize the nuclear profile function, so
that
\beq
M=\sum_{\alpha}
a^{*}_{\alpha}c_{\alpha}
\langle f|\exp(-i\vec{k}\vec{r}_{1})
S_{\alpha}(\vec{b},\vec{c}_{A},...,\vec{c}_{2})
|A \rangle \, .
\label{eq:14.3.1}
\endeq
Repeating the derivation of eq.(\ref{eq:14.2.4}) I obtain
\beq
d\sigma_{A} \propto \sum_{\alpha,\beta}
a^{*}_{\alpha}c_{\alpha}c^{*}_{\beta}a_{\beta}
\langle A|S^{*}_{\alpha}(\vec{b},\vec{c}_{A},...,\vec{c}_{2})
S_{\beta}(\vec{b},\vec{c}_{A},...,\vec{c}_{2})|A \rangle   \, ,
\label{eq:14.3.2}
\endeq
\beq
S^{*}_{\alpha}(\vec{b},\vec{c}_{A},...,\vec{c}_{2})
S_{\beta}(\vec{b},\vec{c}_{A},...,\vec{c}_{2})
=\prod_{i=2}^{A} \left\{
\left[1-\Gamma^{*}_{\alpha}(\vec{b}-\vec{c}_{i})\right]
\left[1-\Gamma_{\beta}(\vec{b}-\vec{c}_{i})\right]\right\}\,\, .
\label{eq:14.3.3}
\endeq
Calculation of the nuclear matrix elements gives the nuclear
attenuation factor of the form
\beq
\langle A|S^{*}_{\alpha}(\vec{b},\vec{c}_{A},...,\vec{c}_{2})
S_{\beta}(\vec{b},\vec{c}_{A},...,\vec{c}_{2})|A \rangle
=
\exp\left[-\Xi_{\alpha\beta}
t(b,\infty,z)\right]
\label{eq:14.3.4}
\endeq
where
\beq
\Xi_{\alpha\beta}=
{1 \over 2}(\sigma_{\alpha} + \sigma_{\beta}) -
\int d^{2}\vec{b}\Gamma_{\alpha}^{*}(\vec{b})\Gamma_{\beta}(\vec{b})
\label{eq:14.3.5}
\endeq
The resulting nuclear transparency equals [5]
\arr
Tr_{A} = {d\sigma_{A} \over Z d\sigma_{p} }=
{1 \over A|\sum_{\alpha}a_{\alpha}^{*}c_{\alpha}|^{2} }
\sum_{\alpha,\beta}
a_{\alpha}^{*}c_{\alpha}
c_{\beta}^{*}a_{\beta}\int d^{2}\vec{b}dz n_{A}(\vec{b},z)
exp[-\Xi_{\alpha \beta}t(b,\infty,z)]\nonumber\\
=
{1 \over |\sum_{\alpha}a_{\alpha}^{*}c_{\alpha}|^{2} }
\sum_{\alpha,\beta}
a_{\alpha}^{*}c_{\alpha}
c_{\beta}^{*}a_{\beta}Tr(\Xi_{\alpha\beta})       \,  ,~~~~~~~~~~~~~
\label{eq:14.3.6}
\endarr
where $Tr(\sigma)$ is defined by eq.(\ref{eq:14.2.1}).
The emergence in (\ref{eq:14.3.4},\ref{eq:14.3.6}) of
the non-trivial quantity $\Xi_{\alpha\beta}$, which contains
the convolution
$\int d^{2}\vec{b}\,\Gamma_{\alpha}^{*}(\vec{b})\Gamma_{\beta}(\vec{b})$
of elastic scattering amplitudes
of the two different eigenstates $|\alpha\rangle$ and $|\beta\rangle$,
demonstrates the importance of the quantal
interference effects in the quasielastic
multiple scattering problem [14].


\section{The coherency constraint, the observable $\Sigma_{ep}$
and the signal of co\-lour trans\-pa\-ren\-cy  in $(e,e'p)$
scattering [{\sl Refs.5,6}]}


\subsection{The coherency condition and the effective ejectile state.}

Eq.(\ref{eq:14.3.6}) gives the transparency in the high-energy limit.
At moderate energies, one has to repeat considerations
of section 6.3 which have
lead to the effective diffraction operator (\ref{eq:6.3.13}).
In this case the typical $\nu$-fold scattering matrix element
will be of the form $\langle p|\hat{\sigma}^{\nu-1}J_{em}|p\rangle$
which must be expanded over the intermediate states precisely
like in eq.(\ref{eq:6.3.2}). From the kinematics
of deep inelastic scattering
\beq
m_{i}^{2}-m_{p}^{2}=2m_{p}\nu - Q^{2}-2\nu k_{z}\, \, ,
\label{eq:15.1.1}
\endeq
and  different components of the ejectile wave packet are produced
on nucleons having different Fermi momenta, with the difference
given by precisely eq.(\ref{eq:6.3.1}). The only difference is
that in the primary interaction vertex $\sigma_{kp}$ is
substituted for by the electromagnetic form factor $G_{kp}(Q)$,
so that the
rule (\ref{eq:6.3.13}) must be applied to the  electromagnetic
vertex too [5]
\beq
G_{kp}(Q) \rightarrow
G_{kp}(Q)G_{A}(\kappa_{kp}) \, .
\label{eq:15.1.2}
\endeq
Consequently, the attenuation factor in eq.(\ref{eq:6.3.14}) must
be computed with the effective diffraction operator
$\hat{\sigma}^{eff}$, whereas eq (\ref{eq:15.1.2}) leads to an
effective ejectile state
\beq
|E_{eff}\rangle =\sum_{i}G_{ip}(Q)G_{A}(\kappa_{ip})|i\rangle
\label{eq:15.1.3}
\endeq
Eq.(\ref{eq:15.1.3}) describes quantitatively
projection of the full ejectile
state $|E\rangle$
onto the subset of $N_{eff}(Q)$ frozen out states, which
satisfy the
coherence condition (\ref{eq:6.3.14}).

One can call this mechanism of formation of an wave
packet $|E_{eff}\rangle$ of size wich decreases with $Q$
from the wave packet $|E\rangle$ of the full proton size
the `nuclear filtering'. Nuclear filtering
 by the coherence
condition (\ref{eq:6.3.14}) is different from the nuclear filtering
by attenuation strength, discussed by Zamolodchikov et al [1]
(see section 4.2).


\subsection{Final state interaction at asymptotic $Q^{2}$ and
the observable $\Sigma_{ep}$ [{\sl Refs.6,49}]}

In the limit of asymptotically large $Q^{2}$ in
the diffraction matrix all $F(\kappa_{ik})=1$ and $\sigma_{min}=0$.
In this limit the nuclear transparency will be dominated by
small $\sigma_{\alpha}$, and one can expand the integrand of
$Tr_{A}$ in powers  of a small parameter
$\Xi_{\alpha\beta}T(b)$. As far as the leading term of
FSI is concerned, $\Xi_{\alpha\beta}\approx
{1\over 2}[\sigma_{\alpha}+\sigma_{\beta}]$, and
\arr
Tr_{A} =
{1 \over A}
\int d^{2}\vec{b}dz n_{A}(b,z)
\left|
{
\langle p|
\exp
\left[
-{1\over 2}\hat{\sigma}t(b,\infty,z)
\right]|E\rangle
\over
\langle p|E\rangle
}
\right|^{2}
\nonumber \\
= {1 \over A}
\int d^{2}\vec{b}dz n_{A}(b,z)
\left\{
1-
{
\langle p|\hat{\sigma}|E\rangle
\over
\langle p|E\rangle
}
t(b,\infty,z)
\right\}
=
1-\Sigma_{ep}(Q){1\over 2A} \int d^{2}\vec{b}\,T(b)^{2}\,  .
\label{eq:15.2.1}
\endarr
The observable $\Sigma_{ep}$ has emerged as
a measure of a strength of FSI.

Hitherto I have treated the operator $\hat{\sigma}J_{em}$ in
the definition of $\Sigma_{ep}$ as a local operator. By the
nature of CT experiments the ejectile state produced on one
nucleon is probed by its interaction with other nucleons
a distance $\Delta z \sim R_{A}$ apart.
In application to the expansion
of the matrix element $\langle p|\hat{\sigma}J_{em}|p\rangle$
over the intermediate states
this amounts to  a sequense of substitutions
\arr
\langle p|\hat{\sigma}J_{em}|p\rangle =
\sum_{i}\langle p|\hat{\sigma}|i\rangle\langle i|J_{em}|p\rangle
{}~~~~~~~~~~~~~~~~~~~~~~~~~~~
\nonumber\\
\Longrightarrow
\sum_{i}\langle p|\hat{\sigma}|i\rangle\langle i|J_{em}|p\rangle
\exp[i\kappa_{ip}(z_{2}-z_{1})]
\Longrightarrow
\sum_{i}\langle p|\hat{\sigma}|i\rangle\langle i|J_{em}|p\rangle
G_{A}(\kappa_{ip})^{2}\,\, ,
\label{eq:15.2.2}
\endarr
which leads to a finite-energy expansion
\beq
\Sigma_{ep}(Q)=\sigma_{tot}(pN)+
\sum_{i\neq p}{G_{ip}(Q) \over G_{em}(Q)}\sigma_{ip}G_{A}(\kappa_{ip})^{2}
\label{eq:15.2.3}
\endeq
In the low-energy limit of $l_{f} \ll R_{A}$ all inelastic
channels decouple, FSI reduces to the conventional
single-channel attenuation
with  $\sigma_{\alpha}=\sigma_{tot}(pN)$,
$\Xi_{\alpha\alpha}= \sigma_{in}(pN)$ and
$\Sigma_{ep}=\sigma_{tot}(pN)$, and FSI measures the free-nucleon
cross section. The observable $\Sigma_{ep}(Q)$ is simlar, although
not identical, to $\sigma_{min}(Q)$. Since
\beq
\int d^{2}\vec{b}T^{2}(b) \sim {A^{2} \over \pi R_{A}^{2} }\, ,
\label{eq:15.2.4}
\endeq
the expansion (\ref{eq:15.2.1}) holds at $\sigma_{min} \ll 30A^{-1/3}\,mb$.


\subsection{The observable $\Sigma_{ep}$ and
the onset of colour transparency at moderate $Q^{2}$ [{\sl Ref.6}]}

The reference value of nuclear transparency is given by
the Glauber formula, which can be referred to as the
strong-attenuation limit formula.
In this section I shall derive a strength of FSI and
CT signal in the strong attenuation limit.
I start with rewriting the Glauber formula (\ref{eq:14.2.1}) in
the form which emphasizes the departure from
$Tr_{A}^{(IA)}=1$ in the impulse
approximation:
\arr
Tr_{A}^{(GL)}= 1-
{1 \over A}\int d^{2}\vec{b} dz n_{A}(\vec{b},z)
\left\{
1-\exp\left[-\sigma_{in}(pN)t(b,\infty,z)\right]
\right\}
\nonumber\\
=1-
{1 \over A}\int d^{2}\vec{b} dz n_{A}(\vec{b},z)
\sigma_{in}(pN)\int_{z}^{\infty}dz_{1}n_{A}(\vec{b},z_{1})
\exp\left[-\sigma_{in}(pN)t(b,\infty,z_{1})\right]
\label{eq:15.3.1}
\endarr
In the latter form eq.(\ref{eq:15.3.1}) shows that a strength of
FSI is measured by $\sigma_{in}(pN)$. I shall show that, within
certain plausible approximations,
the observable $\Sigma_{ep}$ emerges as a measure
of strength of FSI in the strong attenuation limit too.

The onset of CT is closely related to a mixing of elastic and
inelastic intermediate states via Gribov's off-diagonal
diffractive transitions.
The off-diagonal matrix elements $\sigma_{ik}^{(DE)}$ are small
compared to $\sigma_{tot}(iN)$ (here $DE$ stands for the
diffraction excitation).
In the same approximation we can take all diagonal
elements of $\hat{\sigma}^{eff}$ equal to each other:
$\sigma_{ii}=\sigma_{tot}(iN)\approx \sigma_{tot}(pN)=\sigma_{N}$,
so that
\beq
\sigma_{ik}=\sigma_{N}\delta_{ik}+\sigma_{ik}^{(DE)}
\label{eq:15.3.2}
\endeq
Since $\sigma_{el}/\sigma_{tot} \ll 1$,
the analysis is greatly simplified by a convenient approximation
$\Xi_{\alpha\beta}\approx {1 \over 2}(\sigma_{\alpha}+\sigma_{\beta})$
(to the same accuracy  $\sigma_{in}(pN) \approx \sigma_{tot}(pN)$),
what leads to eq.(\ref{eq:15.2.1}) in which now we expand in
$\hat{\sigma}^{(DE)}$ keeping to all orders in $\sigma_{N}$:
\beq
\exp\left[-{1\over 2}\hat{\sigma}t\right]=
\exp\left[-{1\over 2}\sigma_{N}t\right]
\left(1-{1\over 2}\hat{\sigma}^{(DE)}t\right)
\label{eq:15.3.3}
\endeq
Making use of $\langle p|\hat{\sigma}^{(DE)}|E\rangle =
\langle p|\hat{\sigma}|E\rangle-
\langle p|\hat{\sigma}|p\rangle \langle p|E\rangle=
\langle p|\hat{\sigma}|E\rangle-
\sigma_{N}\langle p|E\rangle$.
and separating the impulse approximation term following (\ref{eq:15.3.1}),
I find
\arr
1-Tr_{A}
={\langle p |\hat{\sigma}|E\rangle \over
 \langle p |E\rangle }
{1 \over A}\int d^{2}\vec{b}dz n_{A}(\vec{b},z)
t(b,\infty,z)
\exp\left[-\sigma_{N}t(b,\infty,z)\right]\nonumber\\
+{1 \over A}\int d^{2}\vec{b}dz n_{A}(\vec{b},z)
 \left\{
1-\exp\left[-\sigma_{N}t(b,\infty,z)\right]-
\sigma_{N}t(b,\infty,z)\exp\left[-\sigma_{N}t(b,\infty,z)\right]
\right\}\nonumber\\
=-{\Sigma_{ep}(Q) \over \sigma_{N}} \sigma_{N}
Tr^{(1)}(\sigma_{N})
+\left[1-Tr(\sigma_{N})+\sigma_{N}Tr^{(1)}(\sigma_{N})\right]
\,\, , ~~~~~~~~~~~~~
\label{eq:15.3.4}
\endarr
where $Tr^{(\nu)}(\sigma)=d^{\nu}Tr(\sigma)/d\sigma^{\nu}$.
 At $\sigma_{N} \sim \sigma_{tot}(pN)$ the coefficient
$\sigma_{N}Tr(\sigma_{N})$ is numerically small,
which makes CT signal in the transmission coefficient
numerically much smaller than
$\Delta \Sigma_{ep}(Q)/\sigma_{tot}(pN) \sim
\Delta \sigma_{min}(Q)/\sigma_{tot}(pN)$.

A transition to the weak FSI  with increasing $Q$
qualitatively proceeds as follows: Firstly,
$\Sigma_{ep}(Q) \rightarrow 0$ and, secondly, to higher
orders in the off-diagonal diffractive transitions the
central value  $\sigma_{N}$ will effectively be
substituted for by $\sigma_{min}(Q) \sim \Sigma_{ep}(Q)$
(see the next section 15.4).
In this limit the attenuation factor
$\exp(-\sigma_{N}t)\rightarrow 1$, so that the first term in the
bottom line of eq.(\ref{eq:15.3.4}) leads to expansion
($\ref{eq:15.2.1}$). The multiple scattering expansion
of the second term
contains only the second and higher orders in $\sigma_{N}$,
and at $\sigma_{N} \sim \sigma_{min} \rightarrow 0$ it vanishes
more rapidly than the leading first term $\propto \Sigma_{ep}$.


\subsection{Realistic estimate of the signal of colour
trans\-pa\-ren\-cy
[{\sl Refs.5,6}]}

Eqs.(\ref{eq:6.3.13}),(\ref{eq:14.3.6}) and (\ref{eq:15.1.3}) give a
closed solution of the nuclear transparency problem, but
neither the full set of electromagnetic form factors (for
the review see [98]), nor the full diffraction operator are
known.
Our estimates of CT signal derive from considerations
of sections 15.2 and 13.3. Very important hint from section 13.3
is that in QCD the effective ejectile state is a superposition of
the $N_{eff}(Q)$ states $|i\rangle$ satisfying the coherence constraint
and making the best approximation to the vanishing size
state in this limited basis $|i\rangle$. The major loophole
in this argument is that it is based on the asymptotic form
of the perturbative form factor (\ref{eq:13.2.7}), whereas an
applicability of asymptotic formulas to  baryon
formfactors at momentum transfer $Q$ of
the practical interset is as yet controversial [89,85].
To this end, a significance of the Sudakov form factor must
be emphasized, as it lowers $Q^{2}$ needed for the onset of
the QCD asymptotics of form factors.

Solving for the eigenvalues of $\hat{\sigma}^{eff}$, one can
find the eigenstate $|E_{min}\rangle$ with the minimal eigenvalue
$\sigma_{min}(Q)$. This eigenstate $|E_{min}\rangle$ is an
wave packet which
minimizes the interaction cross section and, to the crude
approximation, can be taken for the effective ejectile state
$|E_{eff}\rangle$. This will give a realistic estimate of
CT signal (I also dubb it the `optimized' nuclear transparency).
In this case the nuclear transarency
will be given by $Tr(\sigma)$ with
\beq
\sigma \approx \sigma_{min}\left(1-
{\sigma_{min} \over \sigma_{tot}(pN) }
{\sigma_{el}(pN) \over \sigma_{tot}(pN) }\right)     \,  .
\label{eq:15.4.1}
\endeq
(When evaluating the second term in the brackets, which
is numerically small, I neglect
small variation of the diffraction slope with the
eigenvalue size.)
Hence the primary task is a construction of the realistic
diffraction operator $\hat{\sigma}$ and study of the
energy (Q) dependence of the minimal eigenvalue $\sigma_{min}(Q)$
of $\hat{\sigma}^{eff}$.


\subsection{Conspiration of the
two states does not produce colour transparency}

As a prelude to the further construction of the
diffraction operator $\hat{\sigma}$ I comment
on significance of having many interfering states,
$N_{eff} \gg 1$, for large CT signal, i.e., for
having $\sigma_{min}$ and
$\Sigma_{ep}$ significantly smaller than $\sigma_{tot}(pN)$.

The case of $N_{eff}=1$ corresponds to
$\sigma_{min} = \sigma_{11}=\sigma_{tot}(NN)$ and
the conventional nuclear attenuation. The $N_{eff}=2$ model gives
the eigenvalues
\beq
\sigma_{\alpha}= {1 \over 2}(\sigma_{11}+\sigma_{22}) \pm
{1 \over 2}\sqrt{ (\sigma_{11}-\sigma_{22})^{2}+4\sigma_{12}^{2}}
\label{eq:15.5.1}
\endeq
so that
\beq
\sigma _{min} \geq \sigma_{tot}(pN)-|\sigma_{12}|\,\,.
\label{eq:15.5.2}
\endeq
Similarly, in this case
\beq
\Sigma_{ep} =\sigma_{tot}(pN)+
\sigma_{12}G_{12}(Q)/G_{em}(Q)
\label{eq:15.5.3}
\endeq

In order to set up the scale for $\sigma_{12}$, one can take
the prominent diffraction dissociation channel
$pN\rightarrow N^{*}(1680)p$ which has the cross section
$\sigma(NN\rightarrow
N^{*}(1680) \approx 0.17 mb$ [99,100]
compared to $\sigma_{el}(NN\rightarrow NN) \approx 7mb$.
Evidently, one
can not gain much in transmission when the number of interfering
states is small: the two interfering
states are simply not enough to
form a wave packet of vanishing size, vanishing $\sigma_{min}$
and vanishing strength of FSI.

The two-channel approximation, in which all the
excited states are substituted for by a single resonance
of mass $m^{*} \sim 1.5-1.6 \,GeV/c^{2}$
is a convenient starting point for the qualitative discussions,
but it does not properly account for the
coherenecy-controlled late freezing out
of the higher mass components of the diffractive excitation.
 The two-state model
will evidently overestimate CT signal.

The repulsion of eigenstates, eq.(\ref{eq:15.5.1}), shows that
$\sigma_{min}$ is separated from other eigenstates by a gap.
A presence of this gap adds to a credibility of
$Tr(\sigma_{min})$ as a realistic
estimate of the signal of CT.



\section{QCD motivated diffraction operator.}


\subsection{Overview of diffraction dissociation of protons.}

The available experimental information on diffractive reactions
is not sufficient for reconstruction of the diffraction matrix
$\hat{\sigma}$ directly from the experimental data
(for the extensive review and references see [100]).
Experimentally, one can measure the single diagonal
element $\sigma_{tot}(pN)$ and one row of the off-diagonal
amplitudes $\sigma_{ip}=\sigma_{pi}$ of the
diffraction dissociation of protons
$pN\rightarrow iN$.

Therefore, my strategy is to construct $\hat{\sigma}$ in the
QCD motivated framework, so that by construction my diffraction
operator will satisfy the CT sum rule.
We shall work in
the realistic model of the nucleon and its excitations and
require that the so obtained
diffraction operator reproduces the gross features of the
diffraction interactions of hadrons.
Starting with the QCD diffraction operator in
the basis of the quark Fock states, I construct the diffraction
operator $\hat{\sigma}$ in the hadronic basis. In fact, doing so
I shall
exploit the {\sl quark-hadron duality}. Making use of the
hadronic basis is quite essential, since the
coherency condition can be formulated, and the effective diffraction
operator $\hat{\sigma}^{eff}$ can be defined,
only in the hadronic basis.

In high-energy QCD, the quark-spin changing diffractive
transitions are weak.  Therefore, the
candidate states which could conspire to produce the small
{\sl transverse} size ejectile are
$N(938,{1\over 2}^{+}),~N^{*}(1680,{5\over 2}^{+}),
{}~N^{*}(1710,{1\over 2}^{+}),
{}~N^{*}(2220,{9\over 2}^{+})$ and higher excitations.
Experimentally, excitation of $N^{*}(1680,{5\over 2}^{+})$
is one of the prominent channels of the diffraction dissociation
of nucleons with $\sigma(NN\rightarrow
N^{*}(1680,{5\over 2}^{+}) \approx 0.17 mb$ [99,100]
compared to $\sigma_{el}(NN\rightarrow NN) \approx 7mb$.
The inclusive cross section of diffraction excitation sums up
to $ \sim 2mb$ (the so-called triple-pomeron mechanism
of production of very heavy states adds extra
$\sim 1 mb$ to the inclusive diffraction dissociation cross section.

By now it is well established that nucleons are best described
by the hybrid pion--nucleon model (or still better, by the
hybrid pion--quark core model).  Here, the $\pi (3q)$ Fock state
is of obviously non-perturbative origin and has a large size.
To the first approximation, the $3q$ Fock state can be taken
for the bare nucleon, and excitation
of the $\pi (3q)$ Fock state describes the diffractive
dissociation of protons into the $\pi N$ system via the
so-called Drell-Hiida-Deck mechanism [100]. The $\pi N$
mass spectrum is peaked at $m_{\pi N} \sim 1.4 \,GeV/c^{2}$
and extends little beyond $m_{\pi N} \gsim 2 \,GeV/c^{2}$.
The diffraction excitation of the Roper resonance
$N(1440,{1\over 2}^{+})$ is experimentally invisible on the $\pi N$
continuum background.
Besides, the form factor of electroproduction
of the Roper resonance vanishes very rapidly with $Q$ [98], and
expansion (\ref{eq:15.2.3}) suggests that
the Roper resonance is of little relevance to CT problem.
Evidently, stripping of pions off the $3q$ core contributes
to CT signal, since that lowers $\sigma_{\min}$ from
$\sigma_{tot}(pN)$ down to $\sigma_{tot}((3q)N)$ (see below,
section 16.4).

Diffractive production of the higher mass states comes
from the diffraction excitation of the $3q$ core.
Further reduction of $\sigma_{min}(Q)$ down from
$\sigma_{tot}((3q)N)$
is connected with
the small size $3q$ Fock states, interactions of which
can be described in terms of the interquark separation
cross section $\sigma(\rho)$, which is to a large extent
understood in the perturbative QCD framework.


\subsection{QCD diffraction operator in the quark basis [{\sl Refs.5,6}]}

Of major interest for CT signal is the diffraction matrix
$\hat{\sigma}(3q)$ constructed on the basis of the $3q$ core states.
A useful property of the higher excitations of the $3q$ state
is a strong quark-diquark clustering
(for the recent review on diquarks see [101]), and
we shall use the quark-diquark model with the
three-dimensional harmonic oscillator (HO)
wave functions.

The crucial ingredient is the $\rho$-dependence of
the interaction cross section studied in [39] and shown
in Fig.2. Its gross features are well reproduced by
a simple parametrization
\beq
\sigma(\rho)=\sigma_{0}
\left[1-\exp\left(-{\rho^{2}\over R_{0}^{2}}\right)\right]    \, ,
\label{eq:16.2.1}
\endeq
which has the $\propto \rho^{2}$
behaviour in the perturbative region of small $\rho$,
and saturates in the nonperturbative region of large
$\rho$ in agreement with the confinement.

The diffraction operator (\ref{eq:16.2.1}) is defined in the
quark basis of the fixed-$\vec{\rho}$ states.
In the hadronic basis it is given by
its matrix elements (\ref{eq:5.2.1}).
CT property of the cross section (\ref{eq:16.2.1}) ensures that
the so-constructed diffraction operator $\hat{\sigma}^{eff}$
defined
on the hadronic basis has the vanishing eigenvalue $\sigma_{min}=0$
and satisfies CT sum rule. Here I make a full use of the
quark-hadron duality and  combine our
understanding of QCD mechanism of the diffractive scattering
with the hadronic spectroscopy and, furthermore, can test the
resulting diffraction matrix against the experimental data on
the diffraction dissociation of protons. Good agreement with
the latter data, section 17.6,
adds to a credibility of such a caluclation of
$\sigma_{min}(Q)$ and predictions of the nuclear transparency
$Tr_{A}(Q)$.

The diffraction operator $(\ref{eq:16.2.1})$ does not depend on the
quark spin variables and on the azymuthal angle $\phi$.
Consequently, in the model of the scalar diquarks we have
the property of spin retention and suppression of
the spin-flip transitions. The diffraction operator (\ref{eq:16.2.1})
is an even function of the polar angle $\theta$, and we have the dominant
diffraction transitions from the proton ($S$-wave state)
to the even angular momentum and positive parity
quark-diquark states with the angular momentum projection $L_{z}=0$.
Production of $N^{*}(1680,{5\over 2}^{+})$ is a
prominent feature of the forward
diffraction dissociation of nucleons, which suggests
$\Delta m =2\hbar \omega \approx 700 MeV/c^{2}$.
Following calculations in Ref.39, we take
$\sigma_{0} \approx 1.6\sigma_{tot}(pN)$ and
$R_{0}^{2} \approx 0.55\, fm^{2}$, so that the $3q$-core interacts
with the nucleon with the cross section $\approx 0.8 \sigma_{tot}(pN)$
(see section 16.3 below).


\subsection{The hybrid pion-nucleon model: size of the 3-quark core
[{\sl Refs.5,6}]}

The hybrid model of the nucleon
with the 3-quark core surrounded by the pionic
cloud [102-104] yields the
wave function
\beq
 |N\rangle = \cos\theta|3q\rangle + \sin \theta|\pi(3q)\rangle \,  .
\label{eq:16.3.1}
\endeq
The mixing angle is related to the number of pions $n_{\pi}$ in
the pionic cloud by $\sin^{2}\theta \approx n_{\pi}/(1+n_{\pi})$.
The diffractive excitation of the $N+\pi$ Fock component of the
nucleon is similar to the diffraction dissociation of the deuteron
on nuclei and its salient feature is production of the
continuum $\pi N$ states with the relatively small mass.
The $\pi NN$ vertex parameters, used in evaluations of $n_{\pi}$ in
[102-104], are close to the ones used in the successful
Drell-Hiida-Deck description of dissociation into $N\pi$ pairs [100].
The resulting mass spectrum is strongly peaked at
$m \approx 1.4 GeV/c^{2}$ and extends very little
beyond $m \sim 2 GeV/c^{2}$
(for the review see [100]).

In the hybrid model, the nucleon-nucleon total cross section
can be estimated as
\arr
\sigma_{tot}(NN)=\cos^{2}\theta \sigma_{tot}((3q)N)+
\sin^{2}\theta\left[\sigma_{tot}((3q)N)+\sigma_{tot}(\pi N)\right]
\nonumber\\
 \approx
\sigma_{tot}((3q)N) +
\sin^{2}\theta\sigma_{tot}(\pi N) \, .
\label{eq:16.3.2}
\endarr
Here I have neglected the Glauber shadowing in the $3q+\pi$ system,
so that
\beq
\sigma_{tot}((3q)N)= \sigma_{tot}(pN)-\sin^{2}\theta\sigma_{tot}(\pi N)
\label{eq:16.3.3}
\endeq
gives a lower bound for $\sigma_{tot})(3q)N)$. Evaluation
of the number of pions in a nucleon
with the realistic $\pi N$ vertex function
gives $n_{\pi} \sim 0.4$ ($\pi \Delta$ component included)[102-104].
This estimate of $n_{\pi}$
suggests rather weak,
$\sim 10\%$, reduction of the $3q$-core radius compared to the
proton radius.


\subsection{How stripping of pions off nucleons
contributes to the signal of colour transparency [{\sl Refs.5,6}]}

In terms of the $|3q\rangle$, $|\pi (3q)\rangle$ Fock states the
$|\pi N\rangle$ state orthogonal to $|N\rangle$ equals
\beq
 |\pi N\rangle = \sin\theta|3q\rangle -\cos \theta|\pi(3q)\rangle \,  .
\label{eq:16.4.1}
\endeq
Making use of eq.(\ref{eq:16.3.3}), I find
\arr
\sigma_{tot}((\pi N)N)=\sigma_{tot}(3qN)+\cos^{2}\theta\sigma_{tot}(\pi N)
\nonumber\\
=\sigma_{tot}(NN)+(\cos^{2}\theta -\sin^{2}\theta)\sigma_{tot}(\pi N)
\label{eq:16.4.2}
\endarr
\beq
\sigma(N\rightarrow \pi N)=-\cos\theta \sin\theta \sigma_{tot}(\pi N)
\label{eq:16.4.3}
\endeq

At large $Q$ the electromagnetic form factor of the $\pi(3q)$
Fock state vanishes more rapidly than the form factor of the
$3q$ Fock state, so that asymptotically
\beq
G_{em}(Q)=G_{pp}(Q)\approx \cos^{2}\theta G_{3q}(Q)
\label{eq:16.4.4}
\endeq
and
\beq
G_{(\pi N)p}(Q)\approx \cos\theta \sin\theta G_{3q}(Q)
\label{eq:16.4.5}
\endeq
Evidently, the stripping of pions off nucleons leads to
$\sigma_{min} \approx \sigma_{tot}((3q)N)$. It is interesting
to see how the weaker FSI shows itself up via $\Sigma_{ep}$.
Making use of
eqs.(\ref{eq:16.3.3}), (\ref{eq:16.4.4}), (\ref{eq:16.4.5})
in eq.(\ref{eq:15.2.3}), and saturating $\Sigma_{ep}$ by the $N$ and
$\pi N$ intermediate states, I find
\beq
\Sigma_{ep}  = \sigma_{tot}(pN)-\sin^{2}\theta \sigma_{tot}(\pi N)\, .
\label{eq:16.4.6}
\endeq
This simple model shows that signs of the diffractive transition
amplitudes and of the electromagnetic form factors {\sl conspire} so as
to {\sl reduce} a strength of FSI. Incidentally, $\Sigma_{ep}$ given by
eq.(\ref{eq:16.4.6}) precisely equals $\sigma_{tot}((3q)N)$, which
also equals $\sigma_{min}$. This shows that FSI of the $\sigma_{min}$
eigenstate gives a realistic estimate for the CT signal.
It also shows that at sufficiently high $Q^{2}$ one can concentrate
on excitations of the $3q$ core.

%

\section{Predictions for nuclear transparency $Tr_{A}$
in $(e,e'p)$ scattering [{\sl Refs.5,6}]}


\subsection{It is not easy to shrink the proton}

In Table 1 I show the structure of the diffraction matrix
$\hat{\sigma}$ [5]. Its off-diagonal matrix elements give
realistic estimates for the diffraction dissociation cross
sections
$\sigma_{D}(pN\rightarrow ip) \sim
(\sigma_{ip}/\sigma_{pp}^{tot})^{2} \sigma_{pp}^{el}$
(see section 17.6). In Table 2 I show the minimal eigenvalue
of the truncated diffraction matrix $\hat{\sigma}$
{\sl vs.} the number $K$ of the interfering shells.
The departure of the minimal eigenvalue $\sigma_{min}$ from
zero measures the rate of saturation of the CT sum rule in
the limited basis of $N_{eff}$ states. The minimal
eigenvalue $\sigma_{min}$ is also a measure of how small
is size  $\rho_{E}$ of the effective ejectile wave packet.
To a crude approximation
\beq
\rho_{E} \sim R_{(3q)}
\sqrt{  {\sigma_{min} \over \sigma_{(3q)N} } }
\label{eq:17.1.1}
\endeq
The minimal eigenvalue $\sigma_{min}$ decreases at large $N_{eff}$,
but this decrease is very slow. The states of at least $K \gsim 6$
shells must conspire in order to produce the wave packet of the
transverse size half of that of the free proton.

The origin
of this slow shrinkage is quite obvious: In order to produce
the wave packet of small size, one has to superimpose with the
proton the higher excited states, which individually have a size
larger than the size of the proton. Mixing with the proton and
biting into the proton's periphery, such excited states bring in
their large-$\rho$ tale, cancellation of which requires mixing
with still higher excitations. For instance, take the
extreme case of the wave packet (\ref{eq:13.3.2}) of vanishing size.
Here the expansion coefficients equal $\Psi_{i}(0)$, and
in  any potential model the convergence to $\delta(\vec{r})$ is
very slow (for instance, see
the textbook [105] and the review on the potential models in [106]).

The $Q$ dependence of $\sigma_{min}(Q)$ is controlled by $N_{eff}(Q)$.
There are three limiting factors on $N_{eff}(Q)$:
In the deep inelastic $ ep$ scattering at $x_{Bj} \sim 1$ and
fixed $Q$ there is a kinematical upper bound on
the mass $m$ of the produced final state:
\beq
 m ^{2} \lsim Q^{2}
\label{eq:17.1.2}
\endeq

In the quasielastic electron-nucleus scattering one rather
studies the fixed-$\nu$, fixed-$Q^{2}$ scattering at $x_{Bj}\approx 1$.
In this case
one can gain in the mass $m$ of the intermediate states
only making use of the Fermi motion of the target nucleon,
see eq.(\ref{eq:15.1.1}):
\beq
m^{2}-m_{p}^{2} \lsim Q^{2} {k_{F} \over m_{p}}
\label{eq:17.1.3}
\endeq
The Fermi-motion constraint is already more stringent than
the kinematical constraint (\ref{eq:17.1.2}). However, since
\beq
{1 \over R_{A}m_{p}} \ll {k_{F} \over m_{p}} \, ,
\label{eq:17.1.4}
\endeq
the real limiting factor on $N_{eff}(Q)$ is the coherence
constraint (\ref{eq:6.3.14}).

In fig.18 I show the lowest eigenvalue
of $\hat{\sigma}^{eff}$ as a function of $Q^{2}$ for a few nuclei.
The corresponding effective number of coherent shells $K$ can
easily be deduced from Table 2.
For instance, at $E \sim 30\, GeV$ only the ground
state (proton, $K=1$) and the two excited shells, $K=2,3$,
can interfere coherently
in $(e,e'p)$ reaction on the carbon nucleus.
Notice, that excitation of the same number of shells
on  different nuclei
requires an energy $\nu \propto R_{ch}$, so that
$\sigma_{min}(Q)$ scales with $Q^{2}/R_{ch}$ and we predict a
very slow
onset of reduced nuclear attenuation for heavy nuclei.
In fact, in the limited  range
$15\,(GeV/c)^{2} \lsim Q^{2} \lsim 200\, (GeV/c)^{2}$
which is of the practical interest, our calculations correspond to
\beq
\sigma_{min} \propto \left(R_{ch}\over {Q^{2}  }\right)^{\eta}
\label{eq:17.1.5}
\endeq
with the (model-dependent) exponent $\eta \approx 0.2$. This
scaling behavior of $\sigma_{min}(Q)$ is shown in fig.18.

The expansion (\ref{eq:6.1.3}) for the effective ejectile
state shows, that even if
the admixture of states of higher shells, $K \gsim 4$, is
present in the initial ejectile state $|E\rangle$,
because of the nuclear filtering by $G_{A}(\kappa)$,
eq.(\ref{eq:15.1.3}), they
can affect little the nuclear trasnparency $Tr_{A}$ in
the energy range of SLAC-EEF.

The coherence constraint invalidates the much discussed and
often used [38,39,70,78-80]
Brodsky-Mueller dimensional estimates [3,4]
$\rho_{E} \sim 1/Q$ and $\sigma_{min} \sim 1/Q^{2}$ (see below,
section 17.5) -
we find much slower shrinkage (\ref{eq:17.1.5})
of the effective ejectile state, fig.19. Jennings and
Kopeliovich [53] have suggested that because
of the Fermi-motion constraint $\rho_{E}$ is larger than
the Brodsky-Mueller estimate,
\beq
\rho_{E}^{2} \sim {m_{p}\over k_{F}} {1 \over Q^{2}}\,\, ,
\label{eq:17.1.6}
\endeq
but the real limiting factor is the coherency constraint
 rather than the
Fermi-motion constraint, and (\ref{eq:17.1.6}) is not born out
by our more accurate analysis.


\subsection{Quantitative predictions for the transmission $(e,e'p)$
expe\-ri\-ment.}

In fig.20 I show the energy dependence of the nuclear
transparency evaluated [5] for the effective ejectile state
$|E_{eff}\rangle$ identified
with the $\sigma_{min}$-eigenvalue $|E_{min}\rangle$
of the diffraction
matrix $\hat{\sigma}^{eff}$. For the reference
point one should take the Glauber model predictions from formula
(\ref{eq:14.2.1}) with $\sigma=\sigma_{in}(pN)\approx 30 mb$: ~
$Tr_{C}=0.61$, ~~$Tr_{Al}=0.49$,~~$Tr_{Cu}=0.40$,~~$Tr_{Pb}=0.26$,
which are also shown in fig.20. As I have discussed above, the
stripping of pions off nucleons already contributes
to CT as reduces the attenuation cross section from the free-nucleon
cross section $\sigma_{tot}(pN)$ down to
$\sigma_{tot}((3q)N) \approx 0.8\sigma_{tot}(pN)$. The further
reduction of $\sigma_{min}$ at higher $Q$ is an effect of
shrinking the $3q$ core.
At $\nu \sim 14\, GeV$ and $Q^{2}\sim 25\, (GeV/c)^{2}$
we find $ \sim 35\%$ reduction of attenuation for carbon,
and $ \sim 25\%$ effect for aluminum.

The above choice of the diffraction eigenstate $|E_{min}\rangle$
with the minimal eigenvalue $\sigma_{min}$ for the initial
ejectile state is consistent with the experimental data on
the elctroproduction of the nucleon resonances.
For instance, when $K=1,2,3$ shells are included,
the eigenstate $|E_{min}\rangle$
with $\sigma_{min}=17.5 mb$, appropriate for an
energy $\nu \sim 30 GeV$,
corresponds to the superposition of the harmonic oscillator
$|n_{r},L\rangle$ states
\arr
|1,0\rangle\,:\,
|2,0\rangle\,:\,
|1,2\rangle\,:\,
|3,0\rangle\,:\,
|2,2\rangle\,:\,
|1,4\rangle  =   \nonumber \\
0.77\,:\,0.44\,:\,0.31\,:\,0.26\,:\,0.22\,:\,0.11~~~~~~ \,  .
\label{eq:17.2.1}
\endarr
It corresponds to the electromagnetic transition form factors
which are of the same order of magnitude as the elastic scattering
form factor, which is consistent with the QCD predictions,
see eq.(\ref{eq:13.2.7}) and with Stoler's
analysis [98] of the SLAC electroproduction data.

The admixture of the lowest states in $|E_{min}\rangle$ changes little
with the number of the conspiring shells:
With the two shells ($\sigma_{min}=22.4 mb$) and  four shells
($\sigma_{min}=13.8 mb$) the corresponding eigenstates $|E_{min}\rangle$
are:
\beq
|1,0\rangle\,:\,
|2,0\rangle\,:\,
|1,2\rangle
 =
0.85\,:\,0.43\,:\,0.30       \,  ,
\label{eq:17.2.2}
\endeq
\arr
|1,0\rangle\,:\,
|2,0\rangle\,:\,
|1,2\rangle\,:\,
|3,0\rangle\,:\,
|2,2\rangle\,:\,
|1,4\rangle\,:\,
|4,0\rangle\,:\,
|3,2\rangle\,:\,
|2,4\rangle\,:\,
|1,6\rangle  =   \nonumber \\
0.70\,:\,0.43\,:\,0.31\,:\,0.28\,:\,0.24\,:\,0.18\,:\,0.17\,:\,
0.10\,:\,0.04~~~~~~
\label{eq:17.2.3}
\endarr

The above multiple-scattering formalism and all conclusions
are fully applicable to nuclear transparency in
the pion electroproduction reactions $eA \rightarrow e'\pi A_{f}$  too.
In this case the first excited shell of the $q\bar{q}$ system
corresponds to $\pi(1300,0^{-})$,\,$A_{1}(1260,1^{+})$,\,
$\pi_{2}(1670,2^{-})$ resonances, and the appropriate energy
scale will be set by $\Delta m \sim 1.2 GeV/c^{2}$. Therefore,
the onset of CT in the
electroproduction of pions and protons should be
similar.


\subsection{Production of resonances and conti\-nu\-um
 states [{\sl Refs.6,49}]}

Electroproduction of the nucleon resonances $(e,e'p^{*})$ is of
particular interest. Generalization of derivation of
eq.(\ref{eq:14.2.1}) shows that in the low-energy limit nuclear
transparency measures $\sigma_{in}(p^{*}N)$:
\beq
Tr_{A}(e,e'p^{*})=Tr(\sigma_{in}(p^{*}N))
\label{eq:17.3.1}
\endeq

The QCD mechanism of electroproduction predicts simultaneous
onset of CT in $(e,e'p^{*})$ and $(e,e'p)$ scattering. This
is obvious from a comparison of observables $\Sigma_{ep}$
and $\Sigma_{ep^{*}}=\langle p^{*}|\hat{\sigma}|E\rangle /
\langle p^{*}|E\rangle$. Repeating an analysis of section 3
one readily finds that with the QCD asymptotics of
electromagnetic FF's for any produced state $|i\rangle $
the observable $\Sigma_{ei}$ vanishes asymptotically by
virtue of CT sum rule (\ref{eq:5.1.7}).
To the extent that $|E_{eff}\rangle \approx |E_{min}\rangle$,
the nuclear transparency in $(e,e'p^{*})$ scattering
will be similar to that in $(e,e'p)$ scattering:
\beq
Tr_{A}(e,e'p^{*})\approx Tr_{A}(e,e'p)\approx Tr(\sigma_{min})
\label{eq:17.3.2}
\endeq

One has to compare
\beq
\Sigma_{ep^{*}}(Q)=\sigma_{tot}(p^{*}N)+
\sum_{i\neq p^{*}}{G_{ip}(Q) \over G_{p^{*}p}(Q)}\sigma_{ip^{*}}
G_{A}(\kappa_{ip^{*}})^{2}
\label{eq:17.3.3}
\endeq
with $\Sigma_{ep}$ eq.(\ref{eq:6.2.2}). CT sum rule
guarantees that asymptotically the off-diagonal contributions
to  $\Sigma_{ep^{*}}$ and $\Sigma_{ep}$ cancel the
corresponding total cross sections $\sigma_{tot}(p^{*}N)$
and $\sigma_{tot}(pN)$. However, at moderate energies, when
only few intermediate states are frozen out, the
magnitude and even the sign of the off-diagonal contribution are
model-dependent. For instance, if the electroproduction
form factor $G_{p^{*}p}(Q)$ is numerically small compared to
the elastic form factor $G_{em}(Q)=G_{pp}(Q)$ (the limited
data [98] on electroproduction of resonances allow such a
possibility), then a
contribution $\propto \sigma_{p^{*}p}$ to
$\Sigma_{ep}$ is proportional to the small factor
$G_{p^{*}p}(Q)/G_{el}(Q)$, whereas the corresponding
contribution to $\Sigma_{ep^{*}}$ is enhanced by a
large factor $G_{el}(Q)/G_{p^{*}p}(Q)$. One can not
exclude a possibility of much stronger cancellations in
$\Sigma_{ep^{*}}$ resulting in
$\Sigma_{ep^{*}} \ll \Sigma_{ep}$ and stronger CT signal
in $(e,e'p^{*})$ scattering  in spite  of
$\sigma_{tot}(p^{*}N) > \sigma_{tot}(pN)$.
One can think of even an interesting possibility of
antishadowing
$\Sigma_{ep^{*}} <0$ and $Tr_{A}(e,e'p^{*}) > 1$.

Still another possibility is a similar enchancement of the
diffractive transition with the off-diagonal
amplitude $\sigma_{ik} > 0$, in which case
 the onset of CT  is preceeded at moderate
$Q$ by enhancement of the
nuclear attenuation compared to the Glauber model prediction.
To this end, electroproduction of nucleon resonances can
be as rich in the anomalous shadowing and antishadowing
phenomena as electroproduction of vector mesons
discussed in sections 10,11,
 which makes experiments on electroproduction
of resonances  particularly interesting. The present
poor knowledge of electroproduction of resonances
[98] does not allow us to list the
candidate reactions for anomalous nuclear attenuation.

One specific soluble model is the hybrid pion-nucleon
model of sections 16.3,\,16.4 at moderate $Q^{2}$, when
the coherency constraint suppresses contribution from
the higher-lying excitations of the $3q$-core.
In this case, repeating the analysis
of section 16.4 one finds
\beq
\Sigma_{e(\pi N)}=\Sigma_{ep}=\sigma_{min}=\sigma_{tot}((3q)N)\,\, .
\label{eq:17.3.4}
\endeq
and nuclear transparency in $(e,e'(\pi N))$ production must be
close to that in $(e,e'p)$ scattering.




\subsection{Scaling properties of nuclear transparency
in $(e,e'p)$ scattering [{\sl Refs.6,49}]}


It is interesting to compare the onset of nuclear transparency for
different nuclei. We have to study  $Tr_{A} = Tr(\sigma_{min})$
as a function of $Q$ and atomic number $A$.

The starting point is that $Tr(\sigma)$ scales with
\beq
x={3A\sigma \over 10\pi R_{ch}^{2}}\,\,.
\label{eq:17.4.1}
\endeq
A variable of this form emerges, for instance,
with the Gaussian parametrization of the nuclear density
(for instance, see [16] and references therein). For the
realistic Woods-Saxon nuclear density $n(r) \propto
\left\{1+\exp\left[-(R_{A}-r)/c\right]\right\}^{-1}$
the convenient relationship is
\beq
R_{ch}^{2}=\langle r_{A}^{2}\rangle_{ch}=
{3\over 5}(R_{A}^{2}+{7\over 3}\pi^{2}c^{2})
\label{eq:17.4.2}
\endeq
The accuracy of this scaling is illustrated in fig.21,
where we plot $Tr(\sigma)$ for different nuclei
as a function of the scaling variable $x$.

With the scaling $\sigma_{min}(Q)$
eq.(\ref{eq:17.1.5}) we predict the
nuclear transparency $Tr_{A}$ (in the limited range of
applicability of parametrization (\ref{eq:17.1.5}) )
 to be a scaling function of the
variable
\beq
t={A\sigma_{min} \over R_{ch}^{2}} \propto
{ A \over R_{ch}^{2} } \left(R_{ch}\over {Q^{2}  }\right)^{\eta}
\label{eq:17.4.3}
\endeq

Based on the dimensional
considerations, Raston and Pire [107] have suggested that
nuclear transparency scales with
\beq
t_{RP} = {A \over R_{ch}^{2} Q^{2}} \propto {A^{1/3} \over Q^{2}}\,  ,
\label{eq:17.4.4}
\endeq
The Ralston-Pire variable corresponds to the assumption
$\sigma_{min} \propto 1/Q^{2}$, which does not account for the
coherency constraint, see above section 17.1. For this reason
the Ralston-Pire scaling law has no region of applicability.
For instance, at $\eta = 1$ the both variables
(\ref{eq:17.4.3}) and (\ref{eq:17.4.4}) have similar
$ \propto 1/Q^{2}$ behaviour, but our eq.(\ref{eq:17.4.3}) will
give
$t \propto A/R_{ch}Q^{2} \propto A^{2/3}/Q^{2}$ which differs
from (\ref{eq:17.4.4}).


\subsection{Scaling properties of nuclear transparency in virtual
photoproduction of vector mesons [{\sl Ref.48}]}

It is instructive to compare nuclear transparency in $(e,e'p)$
scattering and in the virtual photoproduction of vector mesons.
In the former case the energy $\nu$ and virtuality $Q^{2}$
are kinematically related and the coherency constraint does
not allow to make full use of the potentially rapid decrease
of the effective ejectile's size with $Q^{2}$. In the latter case
$Q^{2}$ and the energy $\nu$ are the two independent variables,
and high energy $\nu$ allows to make full use of the small
initial size $\rho_{Q}$ of the ejectile state. Furthermore,
in the vector meson photoproduction the ejectile wave function
is directly calculable.

According to eqs.(\ref{eq:10.1.6}),\,(\ref{eq:10.1.7}),\,
(\ref{eq:10.2.1}), at asymptotic energy $\nu$ FSI effect
scales with
\beq
t_{V}={A \over R_{ch}^{2}}{1 \over m_{V}^{2}+Q^{2} }
\label{eq:17.41.1}
\endeq
Of course, the above dicussed coherency constraint is
applicable here too. Remarkably, the coherency constraint is
relatively less stringent for the heavy quarkonia, in which
the level splitting $\Delta m \ll m_{V}$, and is more
noticable for the light vector mesons, as here the
$m^{2}$-splitting rises rapidly with the mass. In fig.22
I present our results [48] for the $Q^{2}$ dependence
of $1-Tr_{A}(\rho^{0})$ at fixed energy $\nu$ - it is
much weaker than $\propto 1/(m_{\rho}^{2}+Q^{2})$. At
asymptotic energies the scaling law (\ref{eq:10.2.1})
works for all the vector mesons. A somewhat later onset
of the scaling law (\ref{eq:10.2.1}) for the $\rho^{o}$
can be understood in terms of the stronger initial
shadowing and the stronger node effect in the light vector mesons.


\subsection{Testing QCD diffraction operator against the data
on the diffraction dissociation of protons [{\sl Refs.6,49}]}

Here I evaluate the shape of the mass spectrum and inclusive cross
section in diffraction dissociation of protons predicted by the
diffraction matrix of our hybrid pion-quark core model of the
proton. The starting point is eq.(\ref{eq:5.1.3}).

In our hybrid pion-quark  model the spectrum of the final states consists
of the $\pi N$ continuum and excitations of the $3q$ core.
The transition amplitude (\ref{eq:16.4.3}) gives
\beq
{d \sigma_{D}(p\rightarrow \pi N) \over dt}\biggr\vert_{t=0}=
{n_{\pi}(1-n_{\pi}) \over (1+n_{pi}^{2})^{2} }
\left( { \sigma_{tot}(\pi N) \over \sigma_{tot}(pN) } \right)^{2}
{d \sigma_{el}(pN) \over dt}\biggr\vert_{t=0}
\label{eq:17.5.1}
\endeq
Experimentally, the diffraction slope $B_{D}(N\rightarrow \pi N)$
of the diffractive excitation of the low-mass
$N\pi$ system is significantly ($\sim 50\%$) larger
than $B_{el}(pN)$ [100], which correlates with the large radius of the
$\pi N$ Fock state. With $n_{\pi} \sim 0.2-0.3$
[102-104] we find
\arr
\sigma_{D}(p \rightarrow \pi N) \approx  ~~~~~~~~~~~~~~~~~~~~
{}~~~~\nonumber\\
{n_{\pi}(1-n_{\pi}) \over (1+n_{pi})^{2} }
{B_{el}(pN) \over B_{D}(N\rightarrow \pi N)}
\left( { \sigma_{tot}(\pi N) \over \sigma_{tot}(pN) } \right)^{2}
{\sigma_{el}(pN) \over dt}\biggr\vert_{t=0} \approx
0.04 \sigma_{el}(pN)
\label{eq:17.5.2}
\endarr
which
is consistent with the experimental data [99,100].

The inclusive cross section of excitation of the $3q$-core
summed over all excited states is $\propto
\cos^{2}\theta \left[\langle 3q|\hat{\sigma}^{2}|3q\rangle -
\langle 3q|\hat{\sigma}|3q\rangle^{2}\right]$ and can most
easily be estimated in the $\vec{\rho}$-representation using
$\sigma(\rho)$ given by (\ref{eq:16.2.1}).
 With the Gaussian
wave function of the quark-diquark Fock state
\beq
|\Psi_{3q}(\vec{\rho})^{2}| = {1 \over \pi\rho_{0}^{2}}
\exp\left(-{\rho^{2} \over \rho_{0}^{2} }\right)
\label{eq:17.5.3}
\endeq
one easily finds $ \langle 3q |\hat{\sigma}| 3q \rangle =
\sigma_{0}/( 1+\chi)$, where
$\chi = R_{0}^{2}/ \rho_{0}^{2} \approx 1$.
Similar calculation of the second moment
$\langle 3q|\hat{\sigma}^{2} |3q\rangle$
gives
\arr
{d\sigma_{D} (3q \rightarrow X)\over dt}\biggr\vert_{t=0}=
{}~~~~~~~~~~~~~~~~~~~~~~~~~\nonumber \\
\cos^{2}\theta{\chi \over 2+\chi}
\left({\sigma_{tot}((3q)N)\over \sigma_{tot}(pN) }\right)^{2}
{d\sigma_{el}(pN) \over dt}\biggr\vert_{t=0} \approx 0.12
{d\sigma_{el}(pN) \over dt}\biggr\vert_{t=0}
\label{eq:17.5.4}
\endarr
Our estimates (\ref{eq:17.5.1}),(\ref{eq:17.5.2}) and
(\ref{eq:17.5.4}) nearly saturate the experimentally
observed forward diffraction dissociation cross section
$d\sigma_{D}(p\rightarrow X) /d\sigma^{el}(pN) \approx 0.3-0.35$
(see, for instance, estimates for $d\sigma_{D} /d\sigma^{el}$ in [16]).

We can subject our diffraction matrix to the further
test against the experimentally observed  mass spectrum
of the diffractively produced states. The model mass spectrum is a
superposition of many resonance lines. The harmonic
oscillator model gives too
high a degeneracy of levels. To the crude approximation, we can model
the lifting of the degeneracy
describing a contribution of each shell of mass $m_{K}$ by the
Breit-Wigner formula
\beq
BW(m,m_{K})= {1 \over \pi } { m_{K}\Gamma
 \over (m^{2}-m_{K}^{2})^{2}+m_{K}^{2}\Gamma^{2} } \, .
\label{eq:17.5.5}
\endeq
With the width $\Gamma \approx \Delta m /2$ the resulting
inclusive mass spectrum is a smooth function of mass:
\beq
{d\sigma_{D} \over dt dm^{2}}\biggr\vert_{t=0}\approx
\sum_{k}{\sigma_{kp}^{2} \over 16\pi}BW(m,m_{k})
\label{eq:17.5.6}
\endeq
and is  in a reasonable agreement
with the experimental data on the inclusive diffraction
dissociation of protons compiled in [100], fig.23.
Because we confine ourselves to the lowest Fock state of
the nucleon,
our diffraction matrix does not include the triple-pomeron
mechanism of excitation of very heavy states, and
underestimates $d\sigma_{D}/dm^{2}$ beyond
$m^{2} \gsim 10\,(GeV/c^{2})^{2}$. Because of the coherence
constraint such heavy states do not contribute to CT signal
at values of $Q$ of the interest at SLAC or EEF.
Therefore, we have good reasons to believe that our QCD
motivated diffraction operator gives a reasonable estimate
of the $Q$ dependence of $\sigma_{min}(Q)$.


\section{Effects of the Fermi motion}


\subsection{The Fermi-motion bias in nuclear transparency
[{\sl Refs.50,6}]}

Above I have discussed $Tr_{A}$ either integrated over the Fermi
motion, or at the quasielastic peak $x_{Bj}\approx 1$.
In PWIA quasielastic
scattering of electrons measures the Fermi momentum
distribution ([108-110,95],
\beq
{ dn_{F} \over d^{3}\vec{k}}={1 \over (2\pi)^{3} }
\int d^{3}\vec{r}  d^{3}\vec{r}\,'
\rho(\vec{r}\,',\vec{r})
\exp(i\vec{k}\vec{r}\,'-i\vec{k}\vec{r})
\label{eq:18.1.1}
\endeq
where $\rho(\vec{r}\,',\vec{r})$ is the nuclear density
matrix
\beq
\rho(\vec{r}\,',\vec{r}) =\int \prod_{i=2}^{A} d^{3}\vec{r}_{i}
\Psi^{*}(\vec{r}_{A},....,\vec{r}_{2},\vec{r}\,')
\Psi(\vec{r}_{A},....,\vec{r}_{2},\vec{r})
\label{eq:18.1.2}
\endeq

The experimentally measured $x_{Bj}$ distribution $d\sigma/d x_{Bj}$
is proporional to (this is essentially West's $y$-scaling function
[109] at $y\approx k_{z}$)
\beq
F(x_{Bj})=\int dn_{F} \delta(x_{Bj}-1-k_{z}/m_{p}) \, .
\label{eq:18.1.3}
\endeq
Changing the Bjorken variable one changes the central value of
the Fermi momentum $k_{z}$. As Jennings and Kopeliovich have
noticed, the Fermi-momenum difference $\kappa_{ik}$ in production
of different components of the ejectile, eq.(\ref{eq:15.1.1}),
 gives an interesting
handle on the composition of the ejectile wave packet [50].

FSI introduces the extra factor $S^{*}(\vec{r}\,')S(\vec{r})$,
eqs.(\ref{eq:14.2.5}),\,(\ref{eq:14.3.3}),
in the integrand of (\ref{eq:18.1.1}).
Besides the overall attenuation, it changes the effective
Fermi momentum distribution too (for the
review and references to the early work on DWIA and
FSI at intermediate
energies see [94]). Apart from the tail of
the distribution, the effect is not dramatic, though:
Since $k_{F} \gg 1/R_{A}$, in a crude approximation
\beq
\rho(\vec{r}\,',\vec{r})=W(\vec{\Delta})n(\vec{R}) \,  ,
\label{eq:18.1.4}
\endeq
where
$\vec{\Delta}= \vec{r}\,'-\vec{r}$ and
$\vec{R}={1\over 2}(\vec{r}\,'+\vec{r})$. Here
$W(\vec{\Delta})$ is a steep function changing rapidly on a scale
$1/k_{F}$, the nuclear density $n(\vec{r})$ can be
considered nearly uniform, and the attenuation factor
$S^{*}(\vec{r}\,')S(\vec{r})$ changes on the interaction length
$l=2/n(0)\sigma_{tot}(pN)$ which satisfies a weak
inequality $ l \gsim 1/k_{F}$. For this reason, the
order of magnitude of the Fermi-motion bias can be evaluated in PWIA.
Here I confine myself to the $\vec{k}_{\perp}$ integrated
cross section.

A contribution to the quasileatsic cross section from the
interference of amplitudes of the elastic
$$
ep  \stackrel{(em)}{\rightarrow} p \stackrel{(diff)}{\rightarrow} p
$$
and inelastic
$$
ep  \stackrel{(em)}{\rightarrow} p^{*} \stackrel{(diff)}{\rightarrow} p
$$
rescatterings will be proportional to
\arr
\int d^{3}\vec{r}  d^{3}\vec{r}\,'
\rho(\vec{r}\,',\vec{r})
\exp(i\vec{k_{2}}\vec{r}\,'-i\vec{k}_{1}\vec{r})=\nonumber\\
\int d^{3}\vec{\Delta}\,\,
W(\vec{\Delta})
\exp\left[{i\over 2}(\vec{k}_{2}+\vec{k}_{1})\vec{\Delta}\right]
\int d^{3}\vec{R} \,\, n(\vec{R})
\exp\left[i(\vec{k_{2}}-\vec{k}_{1})\vec{R}\right]
\label{eq:18.1.5}
\endarr
The latter integral produces the familar longitudinal form
factor $G_{A}(\kappa_{p^{*}p})$, see eqs.(\ref{eq:15.1.3}),\,
(\ref{eq:15.2.3}), whereas the former shows that
the interference term in the cross section enters at the
shifted value of the Bjorken variable
$x_{Bj}\rightarrow x_{Bj}+\Delta x_{Bj}$ with
\beq
\Delta x_{Bj} ={\kappa_{p^{*}p} \over 2m_{p}}
\label{eq:18.1.6}
\endeq
This leads to $x_{Bj}$
dependence of the effective ejectile state and
observable $\Sigma_{ep}$:
\beq
|E_{eff}\rangle =\sum_{i}G_{ip}(Q)G_{A}(\kappa_{ip})
{F(x_{Bj}+\Delta x_{Bj}) \over F(x_{Bj})}|i\rangle \,\, ,
\label{eq:18.1.7}
\endeq
\beq
\Sigma_{ep}(Q)=\sigma_{tot}(pN)+
\sum_{i\neq p}{G_{ip}(Q) \over G_{em}(Q)}
{F(x_{Bj}+\Delta x_{Bj}) \over F(x_{Bj})}
\sigma_{ip}G_{A}(\kappa_{ip})^{2} \,\,  .
\label{eq:18.1.8}
\endeq

Varying $x_{Bj}$ one can suppress (at $x_{Bj} > 1$) or
enhance (at $x_{Bj} < 1$) the resonance contribution in
(\ref{eq:18.1.8}) and, correspondingly, CT signal too [50].
For the presence of $G_{A}(\kappa_{ip})^{2}$ in (\ref{eq:18.1.8})
we have the coherence constraint on the shift of $x_{Bj}$:
\beq
\delta x_{Bj} \lsim {1\over R_{A}m_{p}} \sim 0.15 A^{-1/3}
\label{eq:18.1.9}
\endeq
Because $F(x_{Bj})$ changes on a scale $x_{F}= k_{F}/m_{p} \sim 0.25$,
the Fermi-motion bias can be significant only at the tail of the
Fermi-momentum distribution. Although distortion of $F(x_{Bj})$ by
FSI can be quite significant in this region (one should
evaluate $Ad\sigma_{N}$ in denominator of $Tr_{A}$ using this
distorted Fermi distribution!), one can use (\ref{eq:18.1.7}),
(\ref{eq:18.1.8})
for the estimation purposes. Jennings and Kopeliovich [50]
estimated the effect in a crude two-channel approximation
criticised in section 15.5 assuming that the
ejectile state coincides with the diffraction eigenvalue with the
vanishing cross section (for a criticism see section 13.1),
and found very large enhancement of nuclear transparency already at
$x_{Bj} \approx 0.8$. The  more
realistic estimates will be presented elsewhere [111].

Distortion of the apparent longitudinal Fermi momentum
distribution by FSI is a very interesting CT observable. The
realistic estimate of the strength of FSI is given by $\sigma_{min}$,
which decreases with $Q^{2}$, so that the distortion effect
should vanish at $Q^{2}\rightarrow \infty$. According to
Benhar et al. [95,112] the CT effect on the observed Bjorken
distribution $F(x_{Bj})$ could be quite large, but their
numerical estimates are based on the erroneous
Farrar-Frankfurt-Liu-Strikman model [70] for FSI, and they
most likely
overestimate the CT signal in $F(x_{Bj})$ at the moderately
large
values of $Q^{2}$ consireded in [95,112].


\subsection{Rescattering broadening of angular distribution [{\sl Ref.6}]}

In PWIA the $\vec{k}_{\perp}$ distribution of ejected protons
measures the Fermi momentum distribution. Evidently, elastic
rescatterings of the ejected proton broaden the $\vec{k}_{\perp}$
distribution [97]. This broadening must vanish in the limit of
vanishing FSI, which can be used as a novel test of CT mechanism.
I present here major ideas of this test, the detailed results
will be presented elsewhere [6,111].

Following Glauber and Matthiae [97] and Nikolaev [14],
one can easily develop the
multiple scattering expansion for the angular distribution
of ejected protons making use of the strong inequality
$B \ll R_{A}^{2}$. The starting point is a generalization of
approximation (\ref{eq:18.1.3}) for the density matrix:
\beq
\Psi^{*}(\vec{r}_{A},....,\vec{r}_{2},\vec{r}\,')
\Psi(\vec{r}_{A},....,\vec{r}_{2},\vec{r}) =
W(\vec{\Delta})n(\vec{R})\prod_{j=2}^{A} n_{j}(\vec{r}_{j}) \,  .
\label{eq:18.2.1}
\endeq
For the sake of simplicity let us consider the
$k_{z}$ integrated cross section, so that $\vec{\Delta}$ is
the two-dimensional vector in a plane normal to ejectile's
momentum.
$W(\vec{\Delta})$ and the nucleon scattering profile
function $\Gamma(\vec{b})$ are much steeper functions than
$n(\vec{r})$, so that in calculation of the nuclear matrix
element of the attenuation factor
$S^{*}(\vec{r}\,',...)S(\vec{r},...)$ one can put
$\vec{r}\,'=\vec{r}=\vec{R}=(\vec{b},z)$ whenever it is
legitimate. The typical single-nucleon matrix element
one encounters is
of the form
\arr
\int_{z}^{\infty}dz_{1}d^{2}\vec{c} n(\vec{b},z)
[\Gamma^{*}(\vec{b}+{1\over 2}\vec{\Delta} -\vec{c})
+\Gamma(\vec{b}-{1\over 2}\vec{\Delta} -\vec{c})
-\Gamma(\vec{b}+{1\over 2}\vec{\Delta} -\vec{c})^{*}
\Gamma(\vec{b}-{1\over 2}\vec{\Delta} -\vec{c})]\nonumber \\
={1\over A}t(b,\infty,z)\left[\sigma_{tot}(pN)-
\int d^{2}\vec{c}\Gamma^{*}(\vec{\Delta}+\vec{c})\Gamma(\vec{c})
\right]~~
{}~~~~~~~~
\label{eq:18.2.2}
\endarr
The convolution term in the square brackets can conveniently
be represented as
\beq
\xi(\vec{c})=
\int d^{2}\vec{c}\Gamma^{*}(\vec{\Delta}+\vec{c})\Gamma(\vec{c})]
={1 \over \pi}
\int d^{2}\vec{q}{d\sigma_{el} \over d\vec{q}^{2}}
\exp(-i\vec{\Delta}\vec{q})   \, .
\label{eq:18.2.3}
\endeq
It is a steep function of $\vec{\Delta}$, whereas one can neglect
$\vec{\Delta}$ in $\Gamma^{*}(\vec{b}+{1\over 2}\vec{\Delta} -\vec{c})
+\Gamma(\vec{b}-{1\over 2}\vec{\Delta} -\vec{c}) \approx
2\Gamma^{*}(\vec{b} -\vec{c})$, which I have already
used in (\ref{eq:18.2.2}).
Since $\xi(\vec{c}) \leq \sigma_{el}$, the quantity
$ \xi(\vec{c})t(b,\infty,z)$ in the exponent of the attenuation
factor is a good expansion
parameter. The corresponding {\sl multiple elastic rescattering
expansion} for the nuclear cross section will be of the form
\arr
{d\sigma_{A} \over d\sigma_{N} d\vec{k}^{2} }=
\int dz d^{2}\vec{b}n_{A}(\vec{b},z)
\exp[-\sigma_{tot}(pN)t(b,\infty,z)]~~~~~~~~~~~~~~\nonumber\\
+\sum_{\nu \geq 1}{t(b,\infty,z)^{\nu}\over \pi^{\nu}\nu!}
\int d^{2}\vec{k}\exp(i\vec{k}\vec{\Delta})W(\vec{\Delta})
\left[
\int d^{2}\vec{q}{d\sigma_{el} \over d\vec{q}^{2}}
\exp(-i\vec{\Delta}\vec{q})\right]^{\nu}\nonumber \\
=Tr(\sigma_{tot}(pN)){dn_{F} \over d^{2}\vec{k}}+
 \sum_{\nu\geq 1} {1\over  \nu!}Tr^{(\nu)}(\sigma_{tot}(pN))
\sigma_{el}(pN)^{\nu}{B \over \nu}
\left<\exp\left[-{B\over \nu}\vec{q}\,^{2}\right]\right>
\label{eq:18.2.4}
\endarr
Here $\left<\exp\left[-{B\over \nu}\vec{q}\,^{2}\right]\right>$
stands for the $\vec{q}\,^{2}$ dependence of the $\nu$-fold
elastic rescattering, folded with the transverse Fermi-momentum
distribution:
\beq
\left<\exp\left[-{B\over \nu}\vec{q}^{2}\right]\right>=
\int d^{2}\vec{k} {dn_{F} \over d^{2}\vec{k}}
\exp\left[-{B\over \nu}(\vec{q}-\vec{k})^{2}\right]
\label{eq:18.2.5}
\endeq
For the sake of simplicity, here I use the conventional
parametrization of the diffraction peak of elastic scattering
$d\sigma_{el}/dq^{2} \propto \exp(-Bq^{2})$. Obviously, the
$\vec{k}$ integrated cross section (\ref{eq:18.2.4}) gives
the Glauber model formula (\ref{eq:14.2.1}).

The $\nu=0$ term gives the `unperturbed' Fermi-momentum
distribution. As here
we ask for the no-elastic-rescatterings, the Fermi-distribution
peak is attenuated with the {\sl total} cross section
$\sigma_{tot}(pN)$.

Expansion (\ref{eq:18.2.5}) admits
a simple probabilistic interpretation,
but its coupled-channel generalization, which
can easily be derived following [14], defies this
probabilistic interpretation. It takes a particlularily
simple form if the diffraction slope $B$ is taken universal
for all eigenstates of the diffraction matrix:
\arr
{d\sigma_{A} \over d\sigma_{N} d\vec{k}^{2} }=
{dn_{F} \over d\vec{k}^{2} }
\int dz d^{2}\vec{b}n_{A}(\vec{b},z)
\left|
{ \langle p|
\exp\left[-{1\over 2}\hat{\sigma}t(b,\infty,z)\right]|E\rangle
\over \langle p|E\rangle }\right|^{2}~~~~~~~~~
\\
+\sum_{\nu \geq 1}
\int d^{2} \vec{b}dz {t(b,\infty,z)^{\nu}\over  \,\nu!}
\sigma_{el}(pN)^{\nu}
\left|
{ \langle p|
\hat{\sigma}^{\nu}
\exp\left[-{1\over 2}\hat{\sigma}t(b,\infty,z)\right]|E\rangle
\over \langle p|E\rangle \sigma_{tot}(pN)^{\nu} }\right|^{2}
{B\over \nu}
\left< \exp\left[-{B\over \nu}\vec{q}^{2}\right]\right>
\nonumber
\label{eq:18.2.6}
\endarr
A presence of the Fermi-folded elastic rescattering tail
in the transverse momentum dustribution is a direct evidence
for FSI. A strength of the $\nu$-fold rescattering in
(\ref{eq:18.2.6}) is more senistive to departure of
$\Sigma_{ep}(Q)$ from $\sigma_{tot}(pN)$ than a strength
of the $\nu=0$
Fermi-distribution peak, which measures the
transmission coefficient. For instance, if the
coherence constraint is lifted, then asymptotically
$\langle p |\hat{\sigma}^{n}J_{em}(Q)|p\rangle
\propto G_{em}(Q) /(Q^{2})^{n}$. At moderate energy
we can use $|E_{eff}\rangle = |E_{min}\rangle$, and
expansion (\ref{eq:18.2.6}) takes a particularly
simple form
\arr
{d\sigma_{A} \over d\sigma_{N} d\vec{k}^{2} }=
Tr(\sigma_{min}){dn_{F}\over d\vec{k}^{2}}~~~~~~~~~~~~~~~~~\\
+\sum_{\nu \geq 1}{1 \over  \,\nu!}Tr^{(\nu)}(\sigma_{min})
\sigma_{el}(pN)^{\nu}
\left[ {\sigma_{min}
\over \sigma_{tot}(pN) }\right]^{2\nu}{ B \over \nu}
\left< \exp\left[-{B\over \nu}\vec{q}^{2}\right]\right> \nonumber
\label{eq:18.2.7}
\endarr
This makes the experimental study
of $\vec{k}_{\perp}$ distribution particularily interesting.
With the typical experimental spectrometers this study somes
at no cost.

Notice, that the nuclear attenuation produces a distortion of the
apparent transverse Fermi distribution too. Although this
distortion can be regarded as a {\sl mean-field effect},
it is too proportional to a strength of FSI - the mean-field
distortion and the
rescattering broadening  always come in the same package.
In the semiexclusive scattering the pure elastic rescattering
effect is observable only provided that the mean-field
distortion of the $\nu=0$ term is not overwhelming. Numerical
estimates for the rescattering CT signal
and an anlysis of the distortion background are in
progress and will be reported elsewhere [113].

One can disentangle the elastic rescattering from the
mean-field distortion experimentally. Namely, a signature
of the $\nu$-fold elastic rescattering is production of the
knock-out or recoil protons $p_{i}(recoil)$,
\beq
e+A \rightarrow e'+p(forward)+p_{1}(recoil)+...+p_{\nu}(recoil)+A^{*}
\label{eq:18.2.8}
\endeq
having the longitudinal momentum of the order of the Fermi
momentum and the transverse momentum typical of the elastic
$p-p$ scattering. The total transverse momentum of the recoil
nucleons is correlated with, and balances, the transverse momentum of
the ejected forward proton. Eq.(\ref{eq:18.2.7}) predicts
a decrease of the relative yield of the $\nu$ recoil nucleons
$\propto (\sigma_{min}(Q)/\sigma_{tot}(pN))^{2\nu}$. In fact,
in CT regime one may expect the recoil momentum which is
somewhat higher than in the free-nucleon scattering, since by the
geometrical considerations
\beq
B(Q) \approx {1\over 2}B_{pp}\left[1+
{ \sigma_{min}(Q) \over \sigma_{tot}(pN) } \right]
\label{eq:18.2.9}
\endeq
To this end, the semi-inclusive experiments with the forward
proton spectrometer complemented with the large acceptance
recoil-nucleon
detector are of great interest.


\section{Search for colour transparency in the quasielastic
$(p,p'p)$ scattering on nuclei.}

This reaction was suggested in the pioneering papers by
Mueller [3] and Brodsky [4] and the dedicated search for CT signal
was preformed at
AGS [114,115].  The theoretical expectation was that
the nuclear transparency should rise,
$Tr_{A}\rightarrow 1$, with energy $\nu$. The early
evidence for such rise, fig.24, was one of the principle
motivations behind the subsequent theoretical development
of the CT subject. However, at higher energy the CT
signal has disappeared, fig.24.

The theoretical analysis of the large-angle $pp$ scattering
is compounded for at least two reasons: \\
(1) The so-called
Landshoff amplitude [116], fig.25, is
the asymptotically leading one and
is known to be dominated by the large-size configurations in the
proton. \\
(2) The free-nucleon $90^{o}$ $pp$ scattering
exhibits irregular energy dependence precisely in the
energy range of the AGS experiment [114].
The origin of this irregular energy dependence is not yet
completely understood.

In the kinematics of the AGS experiment the projectile proton's
energy is equally shared by the scattered and the recoil protons.
Judging from the above analysis of the energy dependence of
$\sigma_{min}(Q)$, one must not expect large signal of CT.
In the energy range of the AGS experiment [114] the frozen size
approximation does not hold, moreover,  $l_{f} \lsim R_{A}$. The most
detailed analysis
of the energy dependence and ejectile's expansion problems
is due to Kopeliovich and Zakharov [41],
who have applied the path integral formalism outlined in
section 11.3. Their results are numerically
not quite realistic, though,
as they start with the pointlike ejectile at the production vertex.
The predicted
nuclear transparency rises monotonously with energy, fig.24.

Ralston and Pire have suggested that the hard-scattering and the
Landshoff amplitudes have the rapidly energy dependent relative
phase [36], which produces a nonmonotonous energy dependence
of the $90^{o}$ $pp$ scattering cross section. Then, the experimentally
observed bump in $Tr_{A}$ is attributed to nuclear distortion of
interference of the two mechanisms. Qualitatively, this is the
most plausible explanation of the AGS results.
Kopeliovich and Zakharov [41] have studied the specific
model suggested by Ralston and Pire. Their conclusion:
the Ralston-Pire model fails to explain quantitatively
a drop of the transparency to the Glauber model value.

More recently, Jennings and Kopeliovich [50] have noticed that,
in fact, the energy dependence shown in fig.24 is ficticious.
The three sets of the data points correspond to the three
experimental runs at the  fixed
energy $\nu=$ 6, 10 and 12 GeV.
The effective energy
\beq
\nu_{eff}=\nu(1-k_{z}/m_{p})
\label{eq:19.1}
\endeq
was varied selecting the Fermi momentum of the struck nucleon.
Since the formation length $l_{f}$ and the evolution rate
depend rather on the laboratory energy $\nu$, plotting
the measured values of $Tr_{A}$ vs. $\nu_{eff}$ is quite misleading
and, in fact, erroneous.
Jennings and Kopeliovich have noticed that because of the
reaction kinematics, eq.(\ref{eq:19.1}),
the small value of $\nu_{eff}$ corresponds to $x_{Bj}<1$.
Here the Fermi-bias suppresses
the relative  proton content of the ejectile wave
packet, see eq.(\ref{eq:18.1.7}) and enhances CT signal.
To the contrary, large $\nu_{eff}$ corresponds to
$x_{Bj} > 1$, the Fermi-bias suppresses the resonance admixture
in the ejectile wave packet and suppresses CT signal.

Frankel adn Frati [117] have questioned the Glauber model estimates
of $Tr_{A}(p,p'p)$ and suggested very strong enhancement
of nuclear transparency by the short-range repulsion of
nucleons in the nuclear matter. As I shall show in section 21,
an accurate evaluation of the short-reange repulsion leads
to a negligible correction to the Galuber model.

The quantitative, and even qualitative, understanding of
the AGS data [114] is as yet lacking, and much more
theoretical work is needed. I would like to
emphasize that the $90^{o}$ scaterring
experiment is a typical poor kinematics experiment, as
one artificially correlates the energy $\nu$ needed to have large
formation length $l_{f}$ with the momentum transfer $Q$, which
increases rapidly with energy, $|t| = Q^{2} \sim m_{p}\nu$,
 driving very steep decrease
of the measured cross section. In fact, the moderate and
even small, diffractive, momentum transfer experiments
could as well yield strong CT signal, which is a subject of
the next section.


\section{Colour transparency and colour opacity
in dif\-frac\-tive had\-ron-nucleus scattering}


\subsection{Anomalous nuclear attenuation in diffraction
dissociation of hadrons on nuclei}

In a broad sense, CT signal is a departure of the nuclear
transmission coefficient from the corresponding Glauber model
prediction. In this sense, CT related effects have already been
observed 20 odd years ago in the diffractive scattering
off nuclei and were much discussed in 70's
under the title of anomalous nuclear
attenuation of resonances and multiparticle states
{\sl in statu nascendi} in the nuclear matter.

The traditional description of diffraction dissociation
$hA\rightarrow h^{*}A$ on nuclei
was based on the two-channel approximation, treating the
off-diagonal $hN\rightarrow h^{*}N$ transitions to the
first order. Then, the nuclear attenuation will be
sensitive to the interaction cross section
of the produced state
$\sigma_{tot}(h^{*}N)$ [118-120]. In sections 10,\,11
 I did
much to discredit a variant of this approximation - the
Vector Dominance Model.
Still, following the historic path, I present the basics
of the formalism in vogue in the late 60's - early 70's.

The starting point is an expansion (\ref{eq:6.3.2}), in
which one makes a substitution (here for the sake of
brevity I put $|1\rangle = |h\rangle,\,\,|2\rangle = |h^{*}\rangle,
\,\, \sigma_{1}=\sigma_{tot}(hN),\,\,\sigma_{2}=\sigma_{tot}(h^{*}N)$)
\beq
\langle 2|\hat{\sigma}^{\nu}|1\rangle=
\sum_{n=0}^{\nu-1}\langle 2|\hat{\sigma}^{n}|2\rangle
\langle 2|\hat{\sigma}|1\rangle\langle 1|\hat{\sigma}^{\nu-n-1}|1\rangle
=\sigma_{21}{\sigma_{2}\,^{\nu}-\sigma_{1}\,^{\nu} \over
\sigma_{2}-\sigma_{1} } \, \, .
\label{eq:20.1.1}
\endeq
In this approximation, the nuclear profile function of
the coherent diffraction excitation
$hA\rightarrow h^{*}A$ will be equal to [120]
\beq
\langle 2|\exp\left[-{1 \over 2}\hat{\sigma}T(b)\right]|1\rangle
=
2\sigma_{21}G_{A}(\kappa_{12})
\left\{ { \exp\left[-{1\over 2}\sigma_{1}T(\vec{b})\right]
 -\exp\left[-{1\over 2}\sigma_{2}T(\vec{b})\right]
\over \sigma_{1}-\sigma_{2} }\right\}       \, .
\label{eq:20.1.2}
\endeq

In the literature there was much use of the so called
K\"olbig-Margolis
formula [119] for  nuclear transparency in the quasielastic, or
incoherent, diffraction dissociation $hA \rightarrow h^{*}A^{*}$,
when one sums over all excitations and breakup of a nucleus
\beq
Tr_{A}=
\int d^{2}\vec{b}
\left\{  { \exp\left[ -\sigma_{1}T(\vec{b})\right]
-\exp\left[ -\sigma_{2}T(\vec{b})\right]
 \over \sigma_{1}-\sigma_{2}}
\right\}                                                     \, .
\label{eq:20.1.3}
\endeq
This formula can be derived  following a probabilistic
derivation of the
Glauber formula (\ref{eq:14.2.1}), in which one introduces
different initial ($\sigma_{1}$) and final ($\sigma_{2}$)
 state attenuations.
Formulas (\ref{eq:20.1.2}) and (\ref{eq:20.1.3}) were a basis
of suggestions to
measure $\sigma_{tot}(h^{*}N)=\sigma_{2}$ for resonances and
multiparticle states in the coherent and incoherent diffraction
dissociation on nuclei [118-120].

Evidently, the probabilistic formula (\ref{eq:20.1.3}) is correct only
at $l_{f} \ll R_{A}$, although in the literature
it is widely and erroneously applied
to the high energy data too. In the high-energy limit of
$l_{f} \gg R_{A}$, more involved analysis like the one
outlined in section 14.3, gives
\beq
Tr_{A}=\int d^{2}\vec{b} T(\vec{b})
\left\{ { \sigma_{1}\exp\left[ -\sigma_{1}T(\vec{b})/2 \right]
-\sigma_{2}\exp\left[ -\sigma_{2}T(\vec{b})/2\right]
 \over \sigma_{1}-\sigma_{2} }\right\}^{2}               \, .
\label{eq:20.1.4}
\endeq
The two-channel formula (\ref{eq:20.1.4}) was first derived by
Gevorkyan, Zaimidoroga and Tarasov [121] (see also [122]), the
multi-channel formalism was developed by the author [14].

More than a decade of active experiments
followed, often with the confusing and
counterintuitive result that  for
resonances and multiparticle states $\sigma_{2}$
is much smaller than
$\sigma_{tot}(hN)$. The typical results of the high energy
experiments [123,124] are shown in fig.26.
The data on the coherent dissociation
$nA \rightarrow p\pi^{-} A$ at $E=(12-14) GeV$ give
rather $\sigma _{2} \approx 40 mb$ [125].
Different values of $\sigma_{2}$ for the same $p\pi^{-}\pi^{+}$ system
come out from the coherent and incoherent dissociation [126,127].
For the $5\pi$ system $\sigma_{2}$ is smaller than for the $3\pi$
system, for the $3\pi$ system $\sigma_{2} \lsim \sigma_{tot}(\pi N)$
and depends strongly on the spin and parity of the produced system [128].

The subject was abandoned before
the coherent dynamical theory of the diffraction dissociation
has emerged. Still, by the late 70's it was well understood
that the perturrbative
two-channel approximation is too crude, that the formulae
(\ref{eq:20.1.2})-(\ref{eq:20.1.3}) do not have any
applicability domain and the diffraction
dissociation does not measure the unstable particle cross
sections [129,130,12,14] (for the review see [15]).
An extensive anlysis of the vector-meson photoproduction
in sections 10,11 is an excellent illustration to this statement.

The above anomalies can easily be understood [8,49] in terms of
the overlap structure of moments
\beq
\langle 2|\hat{\sigma}^{\nu}|1\rangle=
\int d^{2}\vec{\rho}dz\Psi_{2}(z,\vec{\rho})^{*}\sigma(\rho)^{\nu}
\Psi_{1}(z,\vec{\rho})
\label{eq:20.1.5}
\endeq
and of the observable
\beq
\Sigma_{21}={\langle 2|\hat{\sigma}^{2}|1\rangle \over
\langle 2|\hat{\sigma}|1\rangle}
\label{eq:20.1.6}
\endeq
 (in this case higher order
observables are important too).
Here I just comment on a diffractive production of the first
radial excitation of the pion $\pi^{*}(1300)$. Evidently, as it was
discussed in section 5.3, the nonvanishing forward diffraction
production on nucleons is possible only because $\sigma(\rho)$
varies with $\rho$. Regarding $\Sigma_{21}$ and higher overlap
integrals, for the presence of a node in the  wave function
of $\pi^{*}(1300)$, the situation is very much reminiscent of
the $\rho'$ photoproduction. Because of
the $\rho$ dependence of $\sigma(\rho)$ the moments
$\langle 2|\hat{\sigma}|1\rangle$ and
$\langle 2|\hat{\sigma}^{2}|1\rangle$
are dominated by different regions of $\rho$ and there are
no good reasons for the observable $\Sigma_{12}$ to equal the
Kolbig-Margolis value
$\Sigma_{21}=\sigma_{1}+\sigma_{2}$.
The both undercompensation
and overcompensation scenarios are possible.

Unlike the case of virtual photoproduction of the $\rho^{o}$ and
$\rho'$, in the $\pi A \rightarrow \pi^{*} A$ scatering one
can not vary the scanning radius. Nevertheless, since the nuclear
filtering produces suppression of a contribution  from large
$\rho$ to the overlap integrals, the
undercompensation and the overcompensation scenarios can be
distinguished by the $A$ dependence. For instance, the
quasielastic (incoherent) diffraction dissociation
measures the matrix element of the form [14]
$\langle 2|\sigma(\rho)\exp[-{1\over 2}\sigma(\rho)T(b)]|1\rangle$,
 (see eq.(\ref{eq:10.1.2})).
The $A$ dependence of this matrix element
can qualitatively  be described  following the considerations in
section (10.4):\smallskip\\
{\bf
i) Overcompensation scenario:}\smallskip\\
One interesting possibility is that the corresponding node
in the hadronic counterpart of Fig.11 takes place at small
impact parameters and only rather heavy nuclei.
Then, the nuclear filtering must
effectively lower the nuclear overlap integral, which will mimic
stronger nuclear attenuation. In terms of the parameter $\sigma_{2}$
this filtering will amount to $\sigma_{2} > \sigma_{tot}(\pi N)$,
i.e. the same mechanism which has led to the nuclear
antishadowing $Tr(\gamma \rightarrow \Psi') > 1$ may lead
to the {\sl nuclear opacity} in the diffraction
dissociation $\pi A \rightarrow \pi^{*} A$.
However, on heavier nuclei one will encounter flattening
or even a rise of $Tr_{A}$ with A. The most likely case
is, however, the one of Figs.11,12 : the light nuclei are
the best place to search for the anomalous $A$ dependence.
\smallskip\\
 {\bf ii) Underscompensation scenario:} \smallskip\\
      Nuclear
filtering must effectively enhance the nuclear overlap
     integral, which mimics
weaker nuclear attenuation. Again, I refer to Fig.12: one may
start with the antishadowing or flat $Tr_{A} \approx 1$ in light
nuclei followed by the onset of shadowing in heavier nuclei. If such
an $A$ dependence is fitted in terms of the parameter $\sigma_{2}$
this filtering will amount to $\sigma_{2} < \sigma_{tot}(\pi N)$,
even to $\sigma_{2} < 0\,!$.
\smallskip\\
iii) {\bf Nonmonotonous A-dependence:}\smallskip\\
The discussion in section 10.4 is fully applicable here too. The regime of
overcompensation on the light nuclei can be followed by the regime of
undercompensation on heavy nuclei. More precisely, such a change
of the regime depends on the value of $T(b)$. In such a
transient regime the diffraction dissociation cross section may
exhibit nonmonotonous $A$ dependence. For instance, one might
start with the shadowing in light nuclei, which is followed by
weaker shadowing and/or antishadowing in intermediate nuclei,
and by shadowing in heavier nuclei. Another option is that
one starts with the antishadowing in light nuclei followed by
shadowing in heavier nuclei.\smallskip\\
{\bf iv) Anomalous $t$ dependence.\smallskip\\  }
Such a nonmonotonous
$A$ dependence effects must be even stronger in the coherent diffraction
dissiociation and may produce, for instance, anomalous
$A$ dependence of the diffraction slope of the forward
coherent peack. If the reduced matrix element (\ref{eq:10.4.1})
does not have a node, which is a case for the $\rho^{0}$
photoproduction,
then the forward peak of the
diffrential cross section of the coherent production
\beq
d\sigma_{coh}/dt \sim G_{A}(t)^{2}  \,\, .
\label{eq:20.1.7}
\endeq
In presence of the node the diffraction cone will exhibit anomalous
$A$ dependence, dramatically different from (\ref{eq:20.1.6}).
Specifically, a node of the reduced nuclear matrix element in
Fig.11 corresponds to anomalous reduction of the diffraction slope.
\smallskip\\

The experimental data on the diffractive production of
$\pi^{*}(1300)$ are very scanty. The K\"olbig-Margolis
parametrization of the $15 GeV$  $\pi A \rightarrow 3\pi A$ data
in the $A_{1}-\pi^{*}$ mass range gives
$\sigma(0^{-})=49_{-7}^{+9} mb$ compared to
$\sigma(1^{+})=20_{-1.5}^{+1.8} mb$ [125]. Similar analysis [131]
of the $40 GeV$ data from the Bologna-Dubna-Milano
experiment [132]
gives $\sigma_{2}(0^{-})= 32_{-5}^{+6} mb$ compared to
$\sigma_{2}(1^{+})=11\pm 1 mb$. In both cases the combined mass
spectrum is dominated by the $1^{+}$ production.

The QCD approach to diffractive transitions allows one to combine
the colour transparency ideas with the quark model spectroscopy
of resonances, and a fresh look at the
diffraction dissociation physics is called upon.
Evidently, studies of diffraction dissociation into the
low-lying states can not tell much on the detailed from of
       $\sigma(\rho)$ at $\rho \ll R_{p}$,
        but they are indispensable
for tests of our understanding of the QCD mechanism of
diffractive scattering and for the better theoretical control of the
onset of CT in reactions like $(e,e'p),\,\,(e,e'\pi)$.

{}From the pracrtical point of view, a search for the anomalous
$A$ dependence requires data taking on many nuclei, which
particular attention being paid to the light nuclei and to
taking the data on the hydrogen target.


\subsection{Signal of colour transparency in the charge-exchange
scattering on nuclei [{\sl Refs.34,41}]}

Other soft processes can be treated similarly
(for the recent review see [37]). Above I focused
on the forward or near forward production processes.
At larger scattering angles $Tr_{A}$ increases and overshoots
unity because of the conventinal multiple scattering
effects [97], see section 18.2.
Still, in certain cases the CT effects are
quite noticable.

The first quantitative evidence for CT signal was obtained [31,41],
in fact, from an analysis of the charge-exchange scattering
of pions on nuclei. The specifics of the charge-exchange
is that the elementary charge-exchange transition proceeds
on one constituent quark of the pion, so that the
change-exchange amplitude depends on the $q\bar{q}$ separation
in the pion as (the spin-flip amplitude in included)
\beq
f_{cex}(\vec{q},\vec{\rho})=C\left[1+i\beta q\vec{\sigma}\cdot\vec{n}\right]
\exp\left[{i\over 2}\vec{q}\cdot\vec{\rho} -
{1\over 2}\lambda\vec{q}\,^{2}\right]              \, .
\label{eq:20.2.1}
\endeq
The cross section of the quasielastic charge-exchange on the nucleus
contains the familiar nuclear attenuation factor [34,41]:
\arr
{d\sigma(\pi^{+}A \rightarrow \pi^{0}A') \over dt }
=
{1\over 8 \pi}
{Z \over A} \int d^{2}\vec{b} \cdot ~~~~~~~~~~~~~~~~~~~~~~~~
{}~~~~~\\
\cdot Tr\left\{
\left< f_{cex}(\vec{q},\vec{\rho}\,')^{+}
\exp\left[-{1\over 2}\sigma(\rho')T(\vec{b})\right]\right>_{\vec{\rho}\,'}
\left< f_{cex}(\vec{q},\vec{\rho})
\exp\left[-{1\over 2}\sigma(\rho)T(\vec{b})\right]\right>_{\vec{\rho}}
\right\}                     \, .\nonumber
\label{eq:20.2.2}
\endarr
The trace in (\ref{eq:20.2.2}) refers to the spin variables.
Since
$T(b) \sim R_{A}n_{A}(0) \sim A^{1/3}/(60 mb)$, the $\rho$ dependent
absorption factor $\exp\left[-\sigma(\rho)T(b)/2\right]$ varies
significantly at $\rho \sim R_{\pi}$ and modifies strongly the
$q^{2}$-dependence of the nuclear-absorbed amplitude
(\ref{eq:20.2.1}) already at the realitively small $q^{2}$.
Compared with the standard Glauber model formula in
which $\sigma(\rho)=\sigma_{tot}(\pi N)$,
the CT formula (\ref{eq:20.2.2}) leads
to a much more rapid rise of the nuclear
transparency with $|t|=\vec{q}^{2}$,
fig.27, which overshoots unity at $q^{2} \sim 0.4 (GeV/c)^{2}$,
and agrees with the experiment [133,134] better than the Glauber
model prediction.
The
bump at $q^{2} \approx 0.6 (GeV/c)^{2}$
is due to a dip in the charge exchange on free nucleons and a
contribution of the higher order quasielastic
rescattering of pions in a nucleus (for more
details see [34,41]).
Similar considerations of the colour transparency effects in the
antiproton charge exchange $\bar{p}p\rightarrow \bar{\Lambda}\Lambda$
on nuclei are contained in [43].

In the still earlier 1982 paper by Zamolodchikov et al.
[33] it was argued that the Regge
rise of the diffraction slope in the coherent
$K_{S}$-regeneration on nuclei
$K_{L}A \rightarrow K_{S}A$ enhances a contribution of the large-$\rho$,
i.e., $\rho \gsim R_{K}$,
components of the kaon wave function to the regeneration amplitude.
In interaction with nuclear targets this results in an attenuation
stronger than estimated with the free $KN$ total cross section, i.e.,
in the {\sl colour opacity}. The experimental data
[135] on the $K_{S}$-regeneration on
nuclei give an evidence for precisely such an enhanced nuclear opacity.
If the $K_{L}N \rightarrow K_{S}N$ regeneration
amplitude were structureless and $\rho$-independent, then
inelastic shadowing corrections would have rather predicted
reduced nuclear opacity, see eq.(\ref{eq:4.3.1}).


\subsection{Colour transparency
and diffractive production of jets on nuclei [60]}

Extension of considerations of section 12.4 to the diffraction
dissociation of hadrons on nuclei is straightforward.
Consider diffraction dissociation of pions
\beq
\pi \rightarrow jet(q)+ jet(\bar{q})
\label{eq:20.3.1}
\endeq
where
\beq
\left.{d \sigma_{D}(\pi \rightarrow X) \over dt }\right|_{t=0}=
{1\over 16\pi}
\int dz d^{2}\vec{\rho}\, |\Psi_{\pi}(z,\rho)|^{2}\sigma(\rho)^{2}
\label{eq:20.3.2}
\endeq
The principle difference from the diffraction dissociation of
virtual photons is that the pion wave function
$\Psi_{\pi}(z,\rho=0)$ is finite at $\rho \rightarrow 0$
compared to the singular wave function of the photon
$\varepsilon K_{1}(\varepsilon \rho) \propto 1/\rho$.
Application to (\ref{eq:20.3.2}) of the substitutions
(\ref{eq:12.4.5}),\,(\ref{eq:12.4.6}) gives
\beq
\left.{d \sigma_{D}(\gamma^{*} \rightarrow jet_{1}+jet_{2})
\over dt dM^{2} dk^{2}}\right|_{t=0}
\propto
{1 \over R_{\pi}^{2}}{1 \over M^{4}} {dk^{2} \over k^{6}} \,\, .
\label{eq:20.3.3}
\endeq
If the $k^{2}$ distribution (\ref{eq:20.3.3}) is reinterpreted
in terms of the quasi-two-body pion-pomeron collision, then the
differential cross section of reaction
\beq
\pi+\Pom \rightarrow jet(q)+ jet(\bar{q})
\label{eq:20.3.4}
\endeq
decreases $\propto 1/k^{6}$ compared to the $\propto 1/k^{4}$ law
in the photon-pomeron collision.
Notice a lack of the pion or nucleon form factor suppressions at
large momentum transfer squared $k^{2} \gg 1/R_{\pi}^{2}$,
which should have appeared if high-$\vec{k}$ were due to
the intrinsic transverse momentum of (anti)quark in the pion.
The reason is that very much alike to the
diffraction dissociation of photons, the large-$\vec{k}$ jets
originate from the two quark and antiquark of the pion which
have exhanged hard gluons with the same quark of the target
nucleon. In fact, derivation of the specific $1/k^{6}$ law requires
certain assumptions on the wave function of the pion, which
hold with the QCD (Culomb-dominated) asymptotics of the wave
function discussed in section 13 (recall, that in the electromagnetic
formfactor the large momentum transfer $Q$ does not automatically
imply $\rho \sim 1/Q$, one needs a special QCD dictated form
of the wave function in the nomentum representation).

The $A$ dependence of diffraction dissociation of pions and photons
on nuclei is similar, and one predicts vanishing
nuclear attenuation in diffraction dissociation of the pion into the
two high-$\vec{k}$ jets [60]. Notice a distinction between the pion
and proton interactions: In the latter case for the three-quark
structure of the proton the two high-$\vec{k}$ jets will be
accompanied by the spectator quark jet aligned along the projectile's
momentum. Unless the third jet too is produced sideways, the nuclear
dissiociation of protons will be attenuated. The same attenuation
will persist in all cases when the jet aligned along the projectile is
present in the final state, for instance, in the triple-pomeron
regime.


\section{Beyond the dilute-gas approximation: correlation effect
                       in the nuclear transparency}


\subsection{Correlations and Glauber's opti\-cal po\-ten\-tial
 [{\sl Refs.20,136}]}

The
Glauber model predictions used as a reference value for the
nuclear transparency, are usually derived in IPM
(the dilute-gas approximation)
with the factorized nuclear density (\ref{eq:6.1.7}).
They must be corrected for the strong repulsive correlations
at small internucleon distances ([20,95,112], see also textbooks
[136,137]).
I present here major results from the recent discussion of
the correlation effect [6,49].

The generic $\nu$-fold scattering amplitude corresponds to
(see eqs.(\ref{eq:6.3.4}),\,(\ref{eq:6.3.5}))

\beq
{A! \over (A-\nu)!\nu!}\langle A|
\prod_{i=1}^{\nu}\Gamma(\vec{b}-\vec{c}_{i}|A\rangle =
{A! \over (A-\nu)!\nu!}
\int \prod^{\nu}d^{3}\vec{r}_{i}
\prod^{\nu}\Gamma(\vec{b}-\vec{c}_{i})\,
n^{(\nu)}(\vec{r}_{\nu},....,\vec{r}_{1})
\label{eq:21.1.1}
\endeq
The $\nu$-particle nuclear density $n^{(\nu)}$
is normalized by
eqs.(\ref{eq:6.3.5}),(\ref{eq:6.3.6}), and satisfies
\beq
\int d^{3}\vec{r}_{\nu}n^{(\nu)}(\vec{r}_{\nu},....,\vec{r}_{1})=
n^{(\nu-1)}(\vec{r}_{\nu-1},....,\vec{r}_{1})
\label{eq:21.1.2}
\endeq

The leading correction to IPM comes from the
two-body correlations, defined by ($n^{(1)}(\vec{r})=n(\vec{r})$)
\beq
n^{(2)}(\vec{r}_{2},\vec{r}_{1})=
n(\vec{r}_{2})n(\vec{r}_{1})
\left[1-C(\vec{r}_{2},\vec{r}_{1})\right]
\label{eq:21.1.3}
\endeq
Keeping the terms linear in
$C(\vec{r}_{i},\vec{r}_{k})$ from a manifestly
positive defined correlation
factor of the form $\prod_{i>k}[1-C(\vec{r}_{i},\vec{r}_{k})]$
(for the detailed discussion of the correlation functions see
Glauber's lectures [20] and the monographs [136,137]), we have
\beq
n^{(A)}(\vec{r}_{A},....,\vec{r}_{1})=
\left[1-\sum_{i>k}C(\vec{r}_{i},\vec{r}_{k})\right]
\prod^{A} n(\vec{r}_{j})
\label{eq:21.1.6}
\endeq
With the nuclear density (\ref{eq:21.1.6})
the nuclear matrix element of
$\Gamma_{A}(\vec{b},\vec{c}_{A},...,\vec{c}_{1})$ will take the form
\arr
\langle A| \prod_{i=1}^{A} \left[1-\Gamma(\vec{b}-\vec{c}_{i})\right]
|A\rangle =  ~~~~~~~~~~~~~~~~~~~~~~~~~~~~~~~~\\
\langle 1|\left[1-\Gamma(\vec{b}-\vec{c}_{1})\right]|1\rangle^{A}
-{A(A-1)\over 2}
\langle 1|\left[1-\Gamma(\vec{b}-\vec{c}_{1})\right]|1\rangle^{A-2}
\cdot \nonumber\\
\cdot \int d^{3}\vec{r}_{2}d^{3}\vec{r}_{1}
n(\vec{r}_{2})n(\vec{r}_{1})
C(\vec{r}_{2},\vec{r}_{1})
\Gamma(\vec{b}-\vec{c}_{1})\Gamma(\vec{b}-\vec{c}_{2}) \nonumber
\label{eq:21.1.7}
\endarr

The salient feature of nuclear forces is a strong short-range
repulsion
with the correlation radius $r_{c} \sim 0.5 fm$ ([138] and
references therein).
 Since $B \ll R_{A}^{2}$, the nucleonic profile functions
$\Gamma(\vec{b}-\vec{c}_{1}),\, \Gamma(\vec{b}-\vec{c}_{2})$
enforce $\vec{c}_{1,2} \approx \vec{b}$.
The short-range repulsion radius $r_{c}$ is
of the same order in magnitude as
the $hN$ interaction radius, and an accurate evaluation of the
impact parameter integrals gives
\arr
\int d^{3}\vec{r}_{2}d^{3}\vec{r}_{1}
n(\vec{r}_{2})n(\vec{r}_{1})
C(\vec{r}_{2},\vec{r}_{1})
\Gamma(\vec{b}-\vec{c}_{1})\Gamma(\vec{b}-\vec{c}_{2})\nonumber \\
\approx {1\over A^{2}}{\sigma_{tot}(hN) \over 2}
\int dz n_{A}(\vec{b},z)^{2}
{\sigma_{tot}(hN) \over 2}
{r_{c}^{2} \over r_{c}^{2}+2B }
2\int_{0}^{\infty} dr C(r) \,\, .
\label{eq:21.1.10}
\endarr
The result is a density-dependent correction factor to Glauber's
exponential attenuation (\ref{eq:6.1.8}):
\arr
\langle A| \prod_{i=1}^{A} \left[1-\Gamma(\vec{b}-\vec{c}_{i})\right]
|A\rangle = ~~~~~~~~~~~~~~~~~~~~~~~~~~~~~~~~~~~~~~~~\nonumber\\
\left[1-\left({\sigma_{tot}(hN)\over 2}\right)^{2}
l_{cor}\int_{-\infty}^{+\infty} dz n_{A}(\vec{b},z)^{2}\right]
\exp\left[-{\sigma_{tot}(hN) \over 2}
\int_{-\infty}^{+\infty}dzn_{A}(\vec{b},z)\right]
\label{eq:21.1.11}
\endarr
where
\beq
l_{cor}={r_{c}^{2} \over r_{c}^{2}+2B}\int_{0}^{\infty} drC(r)
\approx \sqrt{{\pi\over 2}}r_{c}{r_{c}^{2} \over r_{c}^{2}+2B}
\label{eq:21.1.12}
\endeq

After the higher-order  corrections
$\sim C(\vec{r}_{k},\vec{r}_{l})C(\vec{r}_{j},\vec{r}_{i})$
 are included, the correlation factor
in (\ref{eq:21.1.11}) can be exponentiated:
\arr
\langle A| \prod_{i=1}^{A} \left[1-\Gamma(\vec{b}-\vec{c}_{i})\right]
|A\rangle = ~~~~~~~~~~~~~~~~~~~~~~~~~~~~~~~\nonumber\\
\exp\left\{-{1 \over 2} \sigma_{tot}(hN)\int dz n_{A}(\vec{b},z)
\left[1+{1\over 2}\sigma_{tot}(hN)
l_{cor}n_{A}(\vec{b},z)\right] \right\}
\label{eq:21.1.13}
\endarr
The correlation-corrected optical potential
differs from the conventional
Glauber potential
$V_{opt}^{(Gl)}(\vec{r})=
{i \over 2}\hbar v \sigma_{tot}(hN)n_{A}(\vec{r})$ by
the density-dependent factor
(see the normalization of the forward scattering amplitude
(\ref{eq:6.1.3}))
\beq
V_{opt}(\vec{r})=V_{opt}^{(Gl)}(\vec{r})
\left[1+{1 \over 2i} f_{el}(0) n_{A}(\vec{r})l_{cor}\right]
\label{eq:21.1.14}
\endeq

At high energies the $hN$ interaction radius is rather large,
$2B \approx 1 fm^{2}$ [51], and the factor
$r_{c}^{2}/(r_{c}^{2}+2B) \sim 1/5$
dilutes significantly the possible correlation effect.
With the typical cross section of $\sim 30 mb$
the second order correction to the optical potential is $\sim 2\%$
and can safely be neglected. This also
shows that the correlation expansion
converges rapidly.
The sign of the correlation effect
can easily be understood: the short-range
repulsion lowers the relative screening of nucleons, enhances the
nuclear attenuation and enhances the
total nuclear cross section.
The observation that the finite radius of
$hN$ interaction dilutes the
correlation effect is contained already in Glauber's lectures [20].


\subsection{The hole and spectator effects in the quasielastic
$(e,e'p)$ scattering
[{\sl Refs.6,49}]}

The effect of the two-nucleon correlations in the semiinclusive
$(e,e'p)$ cross section has its own specifics.
The generic form of the $\nu$-fold scattering contribution to
the quasielastic cross section is
\beq
\int \prod_{i=1}^{\nu}d^{3}\vec{r}_{i}
n^{(\nu)}(\vec{r}_{\nu},....,\vec{r}_{1})
\prod_{i>1}^{\nu}\Gamma_{in}(\vec{b}-\vec{c}_{i})
\label{eq:21.2.1}
\endeq
Let us decompose the
$A$-body density (\ref{eq:21.1.6}) as
\beq
n^{(A)}(\vec{r}_{A},....,\vec{r}_{1})=
\left[1-\sum_{i>1}C(\vec{r}_{i},\vec{r}_{1})-
\sum_{i>k>1}C(\vec{r}_{i},\vec{r}_{k})\right]
\prod_{j=1}^{A} n(\vec{r}_{j})
\label{eq:21.2.2}
\endeq
The terms $\propto \sum_{i>1}C(\vec{r}_{i},\vec{r}_{1})$ describe
correlation between the struck nucleon and spectator nucleons
(the effect of the `hole' in the nuclear matter), whereas
the terms
$\propto \sum_{i>k>1}C(\vec{r}_{i},\vec{r}_{k})$ describe correlations
between the spectator nucleons.
We find cancellations between the
hole and spectator effects.

The spectator effect can be estimated following the above
section 21.1.
There are the two minor changes [6,49]:
\begin{enumerate}
\item
In the density dependent correction factor in eq.(\ref{eq:21.1.13})
$\sigma_{tot}(hN)/2$ should be substituted for by $\sigma_{in}(hN)$.
\item
The diffraction slope of the elastic amplitude $B$ should be substituted
for by the slope $B_{in}$ of the inelastic profile function
$\Gamma_{in}(\vec{b})$. For all the practical
purposes, $B_{in} \approx B$.
\end{enumerate}
The resulting correction to the nuclear transparency can be estimated
as follows:
\arr
\delta_{sp}Tr_{A}
\approx
-{1 \over A}\int d^{2}\vec{b} dz_{1} n_{A}(\vec{b},z_{1})
\sigma_{in}(pN)^{2}
l_{cor}\cdot \nonumber\\
\cdot \int_{z}^{\infty} dz_{2}n_{A}(\vec{b},z_{2})^{2}
\exp\left[-\sigma_{in}(pN)\int_{z_{1}}^{\infty}dz n_{A}(\vec{b},z)
\right]
\label{eq:21.2.3}
\endarr
In the uniform  density approximation
$\int_{z_{1}}^{\infty} dz_{2}n_{A}(\vec{b},z_{2})^{2}
\approx n_{A}(0)\int_{z_{1}}^{\infty} dz n_{A}(\vec{b},z)
$
and
\beq
\delta_{sp}Tr_{A} \approx
-Tr_{A}(r_{c}=0)\sigma_{in}(pN)l_{cor}n_{A}(0)\beta\,  ,
\label{eq:21.2.4}
\endeq
where
\beq
\beta \approx 1-{1 \over ATr(r_{c}=0)}\int d^{2}\vec{b}T(\vec{b})
\exp\left[-\sigma_{in}(pN)T(\vec{b})\right] < 1
\label{eq:21.2.5}
\endeq
The spectator effect {\sl enhances}
the nuclear attenuation and {\sl lowers} the nuclear transparency
$Tr_{A}$.


The hole effect in the nuclear transparency  equals
\arr
\delta_{hole}Tr_{A}= ~~~~~~~~~~~~~~~~~~~~~~~~~~~~~~~\nonumber\\
{1 \over 2A}\int dz_{1} dz_{2}d^{2}\vec{c}_{1}d^{2}\vec{c}_{2}
n_{A}(\vec{r}_{2})n_{A}(\vec{r}_{1})
C(\vec{r}_{2},\vec{r}_{1})
\Gamma_{in}(\vec{b}-\vec{c}_{2})
\langle 1|\left[1-\Gamma_{in}(\vec{b}-\vec{c})\right]
|1\rangle^{A-2} \nonumber\\
\approx
{1\over A}\int d^{2}\vec{b} dz_{1}n_{A}(\vec{b},z_{1})
\exp\left[-\sigma_{in}(pN)
\int_{z_{1}}^{\infty}dzn_{A}(\vec{b},z)\right]\cdot~~~~~
{}~~~~~~~~~~~~~\nonumber\\
\cdot\sigma_{in}(pN)
\left( {r_{c}^{2} \over r_{c}^{2} + B_{in}} \right)
\int_{z_{1}}^{\infty} dz_{2}C(z_{2}-z_{1})
n_{A}(\vec{b},z_{2}) ~~~~~~~~~~~~~~~~~~~~~~~
\label{eq:21.2.6}
\endarr
When combined with the uncorrelated cross
section, the correction (\ref{eq:21.2.6}) enhances the integrand
by the factor of the form
\beq
1+ {1\over 2}\sigma_{in}(pN)
\left({r_{c}^{2} \over r_{c}^{2} + B_{in}} \right)
\int_{z_{1}}^{\infty}dz\,C(z-z_{1})
n_{A}(\vec{b},z)
\label{eq:21.2.7}
\endeq
The hole effect {\sl does not exponentiate}, but inasmuch it is
small, one can do so.
After the exponentiation, the hole effect can be
cast in the form of renormalization of the nuclear density in
the exponent of the attenuation factor, which accentuates
the hole effect:
\arr
n_{A}(\vec{b},z)\rightarrow n_{A}(\vec{b},z)
\left[1- {r_{c}^{2} \over r_{c}^{2} + B_{in}}C(z-z_{1}) \right]
\nonumber\\
=n_{A}(\vec{b},z)\left[1-{1 \over \sigma_{in}(hN) }
\int d^{2}\vec{b} \Gamma_{in}(\vec{b})C(\vec{b},z-z_{1})\right]
\label{eq:21.2.8}
\endarr
After this approximate exponentiation, our formula
for the hole effect becomes  similar to that
of Benhar et al. [112].

In the uniform density approximation, the estimate for the hole
effect is
\beq
\delta_{hole}Tr_{A}=Tr_{A}(r_{c}=0)
\left({r_{c}^{2}+2B \over r_{c}^{2}+B_{in}}\right)
\sigma_{in}(pN)l_{cor}n_{A}(0)
\label{eq:21.2.9}
\endeq
The spectator and hole
corrections (\ref{eq:21.2.4}) and (\ref{eq:21.2.9}),
tend to cancel each other, although the hole effect is numerically
larger.

Numerical results [6,49] for the spectator and hole effects
from eqs.(\ref{eq:21.2.3}) and (\ref{eq:21.2.6})
with $\sigma_{in}(pN)= 32 mb$, $B=0.5 f^{2}$ and $r_{c}=0.5f$
are shown in fig.28.
The reduction of the correlation
effect because of $r_{c}^{2} \ll 2B$ and cancellation
of the hole and spectator
corrections makes the total correlation effect negligibly
small.
The hole effect was previously discussed by Benhar et al. [112], who
quote the larger effect (apparently, the numerical estimates
in Ref.112 are given for $B \ll r_{c}^{2}$; in this limit
our estimate of the hole effect is close to that of Benhar et al.).
Benhar et al.
did not consider the spectator effect, which enhances the
nuclear attenuation and lowers the nuclear transparency.
 In the estimate of the repulsion
effect in $(e,e'p)$ scattering by Frankel and Frati [117]
neitherthe dilution factor nor the spectator effect were
included. Besides, Frankel and Frati use the overestimated
$r_{c}=1.25 \, fm$.

I conlcude that the correlation effect is negligible
and the dilute-gas Glauber model calculations
have a sufficiently high accuracy.


\section{Use and misuse of the multiple scattering theory}

Above we have repeatedly seen, how important both
quantitatively and qualitatively are the {\sl quantum interference}
and {\sl coherence} effects. In the recent literature on CT physics
there is a
wide-spead use of the time-dependent cross-section,
\beq
\sigma(z)=\left\{
\sigma_{hard}+
{z \over l_{f}}
\left[
\sigma_{soft}-\sigma_{hard}
\right]
\right\}
\theta(l_{f}-z)+\sigma_{soft}
\theta(z-l_{f})
\label{eq:22.1}
\endeq
introduced by Farrar, Frankfurt, Liu and Strikman [70].
The cross section (\ref{eq:22.1}) rises with the distance
$z$ from the hard scattering
vertex, saturates at the free-nucleon cross section
$\sigma_{soft}$ at
large distances, and is meant to describe the
expansion of the evloving ejectile wave packet.
Farrar et al. apply (\ref{eq:22.1}) to calculation of the
{\sl classical probability} that the observed hadron
is not absorbed in a nucleus:
\beq
Tr={1 \over A} \int d^{2}\vec{b}dz n_{A}(b,z)
\exp\left[-\int_{z}^{\infty}dz'\sigma(z'-z)n_{A}(b,z')\right]
\label{eq:22.2}
\endeq
Typical of predictions of Farrar et al., asymptotically vanishing
attenuation in the photoproduction of the $J/\Psi$,
is shown in fig.13  and is ruled out by the $NMC$ data [69].

The model also predicts much stronger attenuation
for the  $\Psi'$ compared to the $J/\Psi$,
and in all aspects is quite orthogonal to predictions
with correct treatment of the quantal coherence and
interference. The quantum interference obviously does
not fit the Procrustean bed of the classical probability
calculus!

The VDM model is another classic
(but not classical!) model I have questioned
above. Unlike the model of Farrar et al., VDM is the
perfect quantum-mechanical model.
It fails, $\Sigma_{V}\neq \sigma_{tot}(VN)$,
because of too crude a
reduction of the coupled-channel problem
to the two-channel, $\gamma+V$, problem. The perils of such
reduction were well understood long ago, in the context of
diffractive production on nuclei [127,128,12,14], the $\Psi'$
photoproduction is just an excellent demonstration case.
In certain cases, like $J/\Psi$ photoproduction,
basically because $J/\Psi$ is a ground state of charmonium and
its wave function does not have a node, such a
reduction produces reasonable results by a numerical conspiracy,
which still does not justify it.

Jennings and Miller [78], in their analysis of quasielastic
knock-out of protons off nuclei, start with the
Karmanov-Kondratyuk [10] formulation of the
coupled-channel problem. Then, they reduce it first to the
two-channel, $N-N^{*}$ problem. Next, they assume that
one particular superposition of $N$ and $N^{*}$ has a
vanishing cross section, although it is not true even within
their own  model of the nucleonic resonances.
Further, they truncate the multiple scattering expansion to
the single- and double-scattering terms. What they do next
amounts to reinterpretation of the double-scattering term,
evaluated with very specific $N-N^{*}$ mixing, as the leading
term of the shadowing of the effective single-channel problem
with the complex cross section
\beq
\sigma_{eff}(z) = \sigma_{soft}\left[1-\exp(-iz/l_{c})\right]
\label{eq:22.3}
\endeq
Then, this complex cross section (\ref{eq:22.3}) is used
for evaluation of the
higher order terms of the multiple scattering
theory and for calcualtion of the attenuation amplitude
following Farrar et al., eq.(\ref{eq:22.2}).

Recall that, by the optical
theorem of Bohr, Peierls and Placzek, i.e., by the unitarity
relation, the total cross
section is a {\sl real} and {\sl positive-defined}
observable, given by
the imaginary part of the forward scattering amplitude.
 The questionable
approximations made in ref.78 and the unphysical "cross section"
(\ref{eq:22.3})
are unnecessary, because the required
coupled-channel formalism at $l_{f} \lsim R_{A}$ was already
developed to much detail in 70's by Karmanov and Kondratyuk [10] and
Kopeliovich and Lapidus [11] (for the early work see
also the review by Tarasov [21]). Evidently, none of the numerical
predictions from such a mock-up multiple scattering model,
which conflicts the unitarity relation, can
be trusted. Our novel technique of the effective diffraction
operator gives a full solution of the multiple-scattering
problem with allowance for the coherency constraint.


\section{Conclusions}

A brief summary of CT physics was already given in the Introduction.
I emphasize here few more points:
\begin{itemize}

\item
CT leads to {\sl weak, nonexponential,}
 attenuation of the {\sl strongly} interacting hadronic waves
 in a nuclear matter.
This nonexponentiality is a useful heart of CT.
The underlying QCD physics is a filtering of the
{perturbative} Fock components of hadrons in the
non-perturbative soft scattering, i.e., earlier applicability
of the perturbative QCD in nuclear interactions.

\item  QCD observables of
CT experiments allow a unique scanning of the non-per\-tur\-ba\-tive
wave functions of hadrons with the well-controlled
scanning radius. To this end, for the exceptionally rich
anomalous shadowing and antishadowing phenomena, the
virtual photoproduction of
vector mesons is outstanding as an ideal testing ground of
CT ideas. Compared to other candidate reactions, in this case
one can easily have high counting rates, as all the interesting
phenomena take place at already moderate values of
$Q^{2} \sim m_{\rho}^{2}$. We have a solid case for the
vector meson physics at SLAC and EEF.

\item
CT in the quasielastic $(e,e'p)$ scattering derives from the same
QCD mechanism, which predicts the power-law asymptotics
of the electromagnetic form factor. CT observables of
this experiment will shed a new light on the transition
from the nonperturbative to the perturbative regimes of
scattering. In view of the coherency constraint high
precision experiments with light nuclear targets deserve
a particular attention. This is a solid case for the
high-luminosity, continuous-beam acceleartor like EEF.

\item
CT experiments on electroproduction of baryon resonances
are of great interest. Here the high luminosity of EEF is
indispensable.

\item
Quasielastic $(e,e'p)$ scattering in the CT regime is a clean
measure  of the true Fermi momentum distribution.
Separation of the mean-field and elastic-rescattering effects
in the transverse momentum distribution of ejected protons
is of great interest for the nuclear matter theories, since
that will give a handle on the nuclear density matrix.
The forward spectrometers of future $(e,e'p)$ experiments
must be complemented with the large-acceptance recoil
proton detectors.

\item
Studies of the Fermi-bias in the signal of CT are a must as
the Fermi-bias gives a very interesting handle on the composition
of the ejectile wave packet.

\item
The $Q^{2}$- and energy-dependence of CT signal, in particular
the onset of CT regime, give a unique information on the
quark-hadron duality in the diffractive scattering. Diffractive
vector meson studies will contribute decisively to the poorly
known spectroscopy of light vector mesons.

\item
A fresh look at hadronic diffractive interactions from the
QCD point of view is called upon. This is a vast area of research
which was not touched on yet. The hadronic beams available or
to be available at BNL, KAON, internal target experiments at
RHIC, fixet target experiments at FNAL can shed much light on
the CT physics. Of particular interest here is a search for
the nonmonotonous A-dependence. Both the coherent and
incoherent diffraction production are equally imporatnt,
although anomalous $A$ dependence might be stronger in the
coherent production. Diffraction dissociation
studies are a must for the quantitative interpretation of
CT $(e,e'p)$ experiments. Partial-wave analysis of the
diffractively produced states is important to establish the
spin properties of the QCD diffraction scattering - this
topic too is in its infancy.

\item
Diffractive production of jets in nuclear interactions of
hadrons and leptons is particularly interesting. The
high energy experiments at FNAL or jet target experiments
at RHIC would allow to study the {\sl coherent} diffraction
dissociation into the two-jet states. The coherent
diffraction studies require detection of the recoil nucleus
or separation of the very sharp forward coherent peak
(for instance, see [68]), which
will be next to impossible in the jet production experiments.
The potential of the incoherent production is as high as
of the coherent
production, and might be easier to study experimentally.

\end{itemize}

At last, but not the least, a unique combination of tests of
the perturbative and nonperturbative QCD in a single experiment
makes Colour Transparency physics a solid case for the
high-luminosity electron and hadron facilities (EEF, KAON)
in the 15-50 GeV energy range. Of the candidate reactions
discussed in these lectures, I would single out the virtual
photoproduction of vector mesons for the theoretically well
understood ejectile state, for the rich shadowing and
antishadowing phenomena, for the well controlled scanning of
wave function of the vector mesons, for testing the
quark-hadron duality via energy dependence, and for the relatively
high counting rates. Furthermore, the most
exciting predictions could easily be tested at SLAC and EEF.
The quasielastic scattering of electrons is noteworthy for
unique combination of different sources of weak FSI in a
single process: QCD mechanism of electromagnetic form factors
at large $Q$, colour transparency sum rules, quark-hadron
duality and the coherency constraint.

CT physics is a unique example of very sophisticated interplay
of the hard and soft, diffractive, scattering ideas
unparallelled in other areas of the high energy and nuclear
physics. We are heading towards the revival of the diffraction
production physics as a valid QCD physics. Much theoretical
work is needed. An extensive experimental program
which stretches far beyond 2000 is lying ahead. Let's do it!
\bigskip\\

{\bf \sl \Large Acknowledgements.\smallskip\\}
Thanks are due to A.A.Anselm and M.I.Eides for invitation to the
Winter School of the St-Petersburg Nuclear Physics Institute
and to A.B.Kaidalov and Yu.A.Simonov for invitation to the Winter
School of the Institute for Theoretical and Experimental Physics.
I have made much use of the lecture notes [49]  prepared for
the RCNP Kikuchi School on Spin Physics at Intermediate
Energies, held on 16-19 September  at the Research Center for
Nuclear Physics of the Osaka University. Thanks are due to
M.Fujiwara and T.Suzuki for invitation to the RCNP Kikuchi
School.

\pagebreak

\pagebreak

\begin{table}
\center\begin{tabular}{|c||c|c|c|c|c|c||} \hline
$(n_{r},L)$ & $(1,0)$ & $(2,0)$ &  $(1,2)$ & $(3,0)$
& $(2,2)$ & $(1,4)$\\
\hline \hline
$(1,0)$   &  32.0  & -13.1  &  -9.2   &  -5.8  &  -4.9  &  -2.3 \\
\hline
$(2,0)$   & -13.1  &  42.7  &   7.5   & -11.9  &  -4.0  &   4.8  \\
\hline
$(1,2)$   &  -9.2  &  7.5   &   37.3  &  1.7   &  -7.1  &   -12.9 \\
\hline
$(3,0)$   &  -5.8  &  -11.9 &   1.7   &  46.9  &  7.2   &   -2.6  \\
\hline
$(2,2)$   &  -4.9  &  -4.0  &   -7.1  &  7.2   &  40.4  &  6.5 \\
\hline
$(1,4)$   &  -2.3  &  4.8   &  -12.9  & -2.6   &  6.5   &  44.7 \\
\hline \hline
\end{tabular}
\caption[x]{\sl The diffraction matrix (matrix elements are
in millibarns).}
\label{matrix}
\end{table}

\begin{table}
\center\begin{tabular}{|c|c|c|c|} \hline
Number of the & Number of the& Excitation & $\sigma_{min}$ \\
coherent shells, & frozen out states,  & energy  &  \\
 $K$ &  $N_{eff}$ & (GeV) & (mb) \\
\hline
 1 & 1 & 0.94 & 32.0\\
 2 & 3 & 1.64 & 22.1 \\
 3 & 6 & 2.34 & 17.1 \\
 4 & 10 & 3.04 & 14.0 \\
 5 & 15 & 3.74 & 11.8 \\
 6 & 21 & 4.34 & 10.3\\
 7 & 28 & 5.04 &  9.1\\
 8 & 36 & 5.74 &  8.2\\
 9 & 45 & 6.44 &  7.5 \\
10 & 55 & 7.14 &  6.9 \\
\hline
\end{tabular}
\caption[x]{\sl The minimal eigenvalue of the diffraction matrix
vs. the number of the conspiring shells $K$
($\Delta m = 0.7 GeV/c^{2}$).}
\label{sigmin}
\end{table}
\phantom{.}\pagebreak
{\bf \LARGE Figure captions:\bigskip\\}

\begin{itemize}

\item[Fig.1 - ]
A subset of the
two-gluon exchange diagrams for $ab$ scattering amplitude.\\

\item[Fig.2 - ]
Universal $\sigma(\rho)$ vs. $\rho$ for the
nucleon target [39,44]. Different processes
probe this universal cross section at different $\rho$.  \\

\item[Fig.3 - ]
Compilations of the experimental data on {\sl (left)} $nA$ [22]
and {\sl (right)} $K_{L}A$ [23] total
cross sections vs. energy. The dashed curve
is the Glauber calculation
with only elastic rescatterings,  i.e., with the exponential
attenuation. The solid curve includes Gribov's
inelastic shadowing, i.e., the nonexponential
attenuation, and gives much better description of the energy
dependence of $\sigma_{tot}(nA)$ and $\sigma_{tot}(K_{L}A)$.\\

\item[Fig.4 - ]
 Hadron-nucleus multiple scattering diagrams: (a) elastic
rescatterings, (b) Gribov's inelastic rescatterings. \\

\item[Fig.5 - ]
Deep inelastic scattering on the radiatively generated
sea, or photoabsorption mediated by conversion of photons into
$q\bar{q}$ pairs.\\

\item[Fig.6 - ]
 Structure function of $\mu D$ scattering calculated in the
CT approach [46] vs. the compilation of the experimental data.
\\

\item[Fig.7 - ]
The qualitative pattern of the the $Q^{2}$-dependent scanning
of the wave functions of the ground state $V$ and
the radial excitation
$V'$ of the vector meson [7,8,47,48]. The scanning distributions
$\sigma(\rho)\Psi_{\gamma^{*}}(\rho)$ shown
by the solid and dashed
curve have the scanning radii $\rho_{Q}$ differing by a factor 3.
All wave functions and radius $\rho$ are in arbitrary units.\\

\item[Fig.8 - ]
The predicted $Q^{2}$ and $\nu$-dependence of the nuclear
transparency in the virtual phtotoproduction
of the heavy quarkonia [48].
The qualitative pattern is the same from the light to heavy nuclei.
Notice the $Q^{2}$ dependence of the threshold energy.  \\

\item[Fig.9 - ]
The predicted $Q^{2}$-dependence of the nuclear transparency in the
virtual photoproduction {\sl (top)} of the $\Psi'$ (the undercompensation
scenario) and {\sl (bottom)} of the $\rho'$ (the overcompensation scenario)
[48]. \\

\item[Fig.10 - ]
The predicted $Q^{2}$ and $\nu-$dependence of the nuclear transparency
in the virtual photoproduction of the $\rho^{0}$-mesons [48].
\\

\item[Fig.11 - ]
The impact parameter dependence of the reduced nuclear matrix
element $M(T)$ for different nuclei [68]. The node of $M(T)$
takes place on the diffuse edge of nuclei.

\item[Fig.12 - ]
The anomalous $A$ dependence of (virtual) photoproduction of
the $\rho'$ meson compared to the smooth $A$ dependence for
thr $\rho^{0}$ meson.

\item[Fig.13 - ]
 QCD prediction by Benhar et al. [47], {\sl (solid line)} and
prediction of the classical attenuation model
by Farrar et al. [70] {\sl (dotted line)} for the
ratio of transparency in quasielastic $J/\Psi$ photoproduction
on tin and carbon as a function of energy versus the NMC data for the
200 and 280 GeV primary muons [68].\\

\item[Fig.14 - ]
QCD predictions [46] for the nuclear shadowing vs.
the NMC data [74].\\

\item[Fig.15 - ]
CT predictions [51] for the atomic number dependence of
$R=\sigma_{L}/\sigma_{T}$. \\
{\sl The top box:} $R_{Cu}(x)-R_{N}(x)$ and $R_{Pb}(x)-R_{N}(x)$
in muon and neutrino scattering compared with the SLAC
result for $R_{Fe}(x)-R_{N}(x)$ [75] and the NMC result
for $R_{D}(x)-R_{p}(x)$ [76].\\
{\sl The bottom box:} $R_{Ca}(x)-R_{C}(x)$
in muon scattering compared with the NMC data  [76].\\


\item[Fig.16 - ]
 The $\nu-$fold scattering contribution
to photoabsorption on nuclei: one quark-antiquark loop is
shared by all $\nu$ participating nucleons.\\

\item[Fig.17 - ]
 Correlation of the asymptotic behaviour of the electromagnetic
 form factor with the strength of final state interaction [45]:
 top - $Q^{2}F_{em}(Q)$ , bottom - the strength of final
state interaction measured by
$\langle \rho^{2}(Q)\rangle /\langle \rho^{2}(0)\rangle$
vs. $Q^{2}$. The order of magnitude of the asymptotic normalization
suggested by perturbative QCD is indicated by the arrow.
The curves correspond to:
\subitem (a) - the Gaussian wave function, $m=150 (MeV/c)^{2}$;
\subitem (b) - the Gaussian+Coulomb wave function,
$m=150 (MeV/c)^{2}$, $\alpha=0.1$;
\subitem (c) - the Gaussian+Coulomb wave function,
$m=150 (MeV/c)^{2}$, $\alpha=0.01$;
\subitem (d) - the Gaussian+Coulomb wave function,
$m=300 (MeV/c)^{2}$, $\alpha=0.01$;
\subitem (e) - the monopole form factor.\\

\item[Fig.18 - ]
 The minimal eigenvalue of the effective QCD diffraction matrix
[5]: \\
{\sl (top)}
as a function of energy for different nuclei, \\
{\sl (bottom)} as a function of the reduced energy
$\nu R_{ch}(C)/R_{ch}(A)$.\\

\item[Fig.19 - ]
 The least possible interaction cross section $\sigma_{min}$
of the wave packet made of the conspiring states of $K$
positive-parity shells as a function of mass $M$ of the highest
shell [5]. The $1/M^{2}$ behaviour expected from the
naive scaling considerations is shown by the straight line.\\

\item[Fig.20 - ]
 The energy dependence of the nuclear
transparency $Tr_{A}$ in the quasielastic $(e,e'p)$ scattering on
different nuclei.
The low energy predictions
from the Glauber model  are
shown separately.\\


\item[Fig.21 - ]
 Nuclear transparency eq.(\ref{eq:14.2.1})
for different nuclei is plotted as a function of the scaling
variables $x$ defined by
eq.(\ref{eq:17.4.1}).\\

\item[Fig.22 - ]
Test of the scaling law (\ref{eq:10.2.1}) in
the virtual photoproduction of the
$J/\Psi$ and $\rho^{0}$ mesons: \\
{\sl top}  - the $\rho^{0}$ production at fixed energy. The dashed
straight line corresponds to the $1/(Q^{2}+m_{\rho}^{2})$ dependence.\\
{\sl bottom } - at the asymptotic energy,
the dotted straight line corresponds to the $1/(Q^{2}+m_{J/\Psi})^{2}$
dependence. \\

\item[Fig.23 - ]
 The smoothed mass
spectrum in the diffraction dissociation of protons,
predicted by the QCD diffraction operator [5],
compared with the experimental data
compiled in [100]($\nu \approx 270 GeV$,  $|t|=0.025 (GeV/c)^{2}$).
\\

\item[Fig.24 - ]
 The Kopeliovich-Zakharov predictions [41] of the
nuclear transparency in the quasielastic
$(p,p'p)$ scattering on aluminum as a function of energy
$\nu$ vs. the AGS data [114] plotted vs. the effective
c.m.s. energy $\nu_{eff}$. The solid curve
is a prediction for the
pointlike ejectile wave function at the production vertex.
The dashed curve corresponds to the Ralston-Pire
wave function [36]
with admixture of the Landshoff mechanism. The sets of the
data points
marked by $\triangle$, $\nabla$ and $\bigcirc$ correspond to
the beam momentum of 6,~10 and 12 $GeV/c$, respectively. \\

\item[Fig.25 - ]
 The Landshoff mechanism of the large-angle $pp$ elastic
scattering.\\


\item[Fig.26 - ]
 Cross section $\sigma_{2}(N^{*}N)$
for $p\pi^{-}$ and $p\pi^{-}\pi^{+}$
systems determined in the K\"olbig-Margolis approximation from
{\sl (top)} diffraction
dissociation on deuterons [123] and {\sl (bottom)}
 from the diffractive
dissociation $n\rightarrow p\pi^{-}$ on nuclei [124].\\

\item[Fig.27 - ]
 The nuclear transparency in the pion charge exchange on carbon
at 40 (GeV/c) [133,134]. The solid curve is computed with $\sigma(\rho)
\propto \rho^{2}$, the dashed curve is the Glauber model calculation
with $\sigma(\rho) =\sigma_{tot}(\pi N)$ [34,41].     \\

\item[Fig.28 - ]
 Correlation effect in the nuclear transparency as a function
of the atomic number $A$ [6].

\end{itemize}

\end{document}